\documentclass[12pt,letterpaper]{article}
\pdfoutput=1
\usepackage{epsfig,graphicx,bm,amsmath}
\usepackage{amssymb}
\numberwithin{equation}{section}
\usepackage{float}

\def\pa{\partial}

\renewcommand{\(}{\left(}
\renewcommand{\)}{\right)}
\renewcommand{\[}{\left[}
\renewcommand{\]}{\right]}

\newcommand\blfootnote[1]{%
  \begingroup
  \renewcommand\thefootnote{}\footnote{#1}%
  \addtocounter{footnote}{-1}%
  \endgroup
}

\begin{document}

\setlength{\unitlength}{1mm}

\thispagestyle{empty}
\vspace*{2cm}

\begin{center}
{\bf \Large Thermodynamically stable asymptotically flat hairy black holes with a dilaton potential}\\
\vspace*{2cm}

{\bf Dumitru Astefanesei,$^{1}$ David Choque,$^{2,3}$ Francisco G\'omez,$^{1}$ and Ra\'ul Rojas$^{1}$} \\
\vspace*{0.4cm}

{\it $^1$Pontificia Universidad Cat\'olica de
	Valpara\'\i so, Instituto de F\'{\i}sica,
 \\ Av. Brasil 2950, Valpara\'{\i}so, Chile.}\\
{\it $^2$Universidad  Adolfo Ib\'{a}\~{n}ez, Dept. de Ciencias,
	Facultad de Artes Liberales \\ Av. Padre
	Hurtado 750, Vi\~{n}a del Mar, Chile.}\\
{\it $^3$Universidad Nacional de San Ant\'onio Abad del Cusco, \\ Av. La Cultura 733, Cusco, Per\'u.}
\vspace*{1.4cm}

{\bf ABSTRACT}
\end{center}

We present a detailed analysis of the thermodynamics of exact asymptotically flat hairy black holes in Einstein-Maxwell-dilaton theory. We compute the regularized action, quasilocal stress tensor, and conserved charges by using a `counterterm method' similar to the one extensively used in the AdS-CFT duality. In the presence of a non-trivial dilaton potential that vanishes at the boundary we prove that, for some range of parameters, there exist thermodynamically stable black holes in the grand canonical and canonical ensembles. To the best of our knowledge, this is the first example of a thermodynamically stable asymptotically flat black hole, without imposing artificial conditions corresponding to embedding in a finite box. 
\vfill \setcounter{page}{0} \setcounter{footnote}{0}
\blfootnote{E-mail: {\tt dumitru.astefanesei@pucv.cl}}\\
\blfootnote{E-mail: {\tt brst1010123@gmail.com}}\\
\blfootnote{E-mail: {\tt francisco.gomez.serrano96@gmail.com}}\\
\blfootnote{E-mail: {\tt raulox.2012@gmail.com}}
\newpage

\tableofcontents
\newpage
\section{Introduction} \label{intro}
The thermodynamic character of gravity becomes apparent in the context of black hole physics.  When considering the quantum mechanical effects, the thermodynamic significance of physical quantities, which were originally considered purely geometrical \cite{Bekenstein:1973ur}, arises naturally \cite{Hawking:1974sw,Hawking:1976de}. This subtle connection between thermodynamics and black hole physics provides one of the most important features of any proposed theory that attempts to unify quantum mechanics and general relativity. Particularly, the relation between the black hole entropy and its event horizon area could be an important clue that the `holographic principle' \cite{tHooft:1993dmi, Susskind:1994vu}  is fundamental for constructing such a theory \cite{Bigatti:1999dp}.

A black hole in thermal equilibrium with its surroundings is clearly relevant for understanding the relation between thermodynamic quantities and their geometric counterparts. The conventional wisdom is that the asymptotically flat black holes are not thermodynamically stable \cite{Hawking:1976de}. This is a direct consequence of the fact that Schwarzschild black hole in flat spacetime, without imposition of further conditions, has negative specific heat  and can not be in equilibrium with an indefinitely large reservoir of energy. That is, a thermal fluctuation can break the equilibrium between the rate of absorption of thermal radiation and rate of emission of Hawking radiation, which in turn will lead to the evaporation of the black hole or its indefinite growth (depending on whether the initial fluctuation made the black hole a bit hotter or a bit cooler than the heat bath, respectively). This implies that the canonical ensemble is not suitable in this special case.  The specific heat for an asymptotically flat neutral black hole can  be rendered positive in the canonical ensemble by giving it an electric charge (or by rotating the black hole), but this does not in itself ensure full thermodynamic stability as the other response function becomes negative in the same range of parameters.

In this paper, we shall revisit this issue by investigating black hole solutions in a more general theory with a scalar field and its non-trivial self-interaction. The advantage of considering Einstein-Maxwell-dilaton theory with a dilaton potential is that there exist exact asymptotically flat static hairy black hole solutions \cite{Anabalon:2013qua} and we do not have to rely on numerical methods. The main assumptions for the no-hair theorem proofs are the asymptotic flatness and nature of the energy-momentum tensor and so, at first sight, the existence of regular hairy asymptotically flat black hole solutions  could be surprising.  However, their existence was conjectured in \cite{Nucamendi:1995ex} and some numerical evidence was presented.\footnote{A different class of (spinning) hairy black holes was found in \cite{Herdeiro:2014goa, Herdeiro:2015gia}, for a review see, e.g., \cite{Herdeiro:2015waa}.} The idea behind this conjecture is that in a theory with a non-trivial  potential of the scalar field, its parameters can be adjusted so that the effective cosmological constant could cancel out. The dilaton potential considered in \cite{Anabalon:2013qua}, which vanishes at the boundary, is reminiscent of the general potential that was obtained for a one scalar field truncation of $\omega$-deformed supergravity \cite{DallAgata:2012mfj, Tarrio:2013qga, Anabalon:2013eaa} (see, e.g., \cite{Trigiante:2016mnt} for a nice review) and, also, $N= 2$  gauged  supergravity  with  an  electromagnetic Fayet-Iliopoulos term \cite {Faedo:2015jqa, Anabalon:2017yhv}. We are going to elaborate further and obtain the relevant thermodynamic quantities by using a relatively recent development, the so called `counterterm method in asymptotically flat spacetime', which was motivated by similar work \cite{Henningson:1998gx, Balasubramanian:1999re, Skenderis:2000in, deHaro:2000vlm} in the AdS-CFT duality \cite{Maldacena:1997re}. Interestingly, we are going to prove that the self-interaction of the scalar field is the key for obtaining asymptotically flat hairy black holes, which are thermodynamically stable. 

Since a gravitational system is intrinsically non-linear, the conserved quantities are remarkable analytic tools to investigate its behaviour.  However, due to the equivalence principle, it became clear that a local definition for gravitational energy can not be possible.  Nevertheless, in general relativity, for asymptotically flat
spacetimes, conserved quantities associated with asymptotic symmetries have been defined at spatial and null infinity. Early work on total energy associated with an asymptotic geometry is due to Arnowitt, Deser, and Misner ($ADM$) \cite{Arnowitt:1960es, Arnowitt:1960zzc, Arnowitt:1961zz, Arnowitt:1962hi}, which led to a  well-defined construction of the Hamiltonian and  global definitions of energy, linear and angular momenta. The infinite systems are idealizations of more realistic physical situations, and it is desirable to have available a similar analysis in the case of physical systems of finite extent. That is the notion of `quasilocal energy', which is currently the most promising description of energy in the context of general relativity and it is also relevant for our work.  This approach, pioneered by Brown and York \cite{Brown:1992br}, can be characterized simply as follows: the gravitational energy is associated with closed spacelike two-surfaces in spacetime, not with a point. Once the closed spacelike surface is pushed to the boundary, the results match the ones obtained by the $ADM$ formalism.

One obvious problem that appears when computing the action (and quasilocal stress tensor) at the spatial infinity is the existence of infrared divergences associated with the infinite volume of the spacetime manifold. The initial approach of dealing with this issue was to use a background subtraction whose asymptotic geometry matches that of the solutions. However, such a procedure causes the resulting physical quantities to depend on the choice of reference background. Furthermore, it is not possible  to embed the boundary surface into a background spacetime even for simple solutions, e.g. when matter fields are present or for spinning black holes. Unexpectedly, the rescue came by a completely different route, namely the AdS-CFT duality in string theory. The observation that the infrared divergencies of the (super)gravity in the bulk are equivalent with the ultraviolet divergencies of the dual field theory was at the basis of the counterterm method in AdS. The procedure consists of adding suitable boundary terms (such that  the bulk equations of motion are not altered), referred to as `counterterms', to regularize the action. The duality imposes the restriction that these counterterms are built up only with curvature invariants of the boundary metric and not with quantities extrinsic to the boundary like in the case of the Gibbons-Hawking term. It was observed that, even in flat spacetime, one can obtain a regularized quasilocal stress tensor \cite{Astefanesei:2005ad} and the method was successfully applied to study the thermodynamics of spinning black holes and black rings (particularly, obtaining the correct first law including the dipole charge of \cite{Emparan:2004wy}). This was an important hint that, indeed, the counterterm method is also suitable for flat spacetimes and, no longer after, a  general covariant method was proposed in \cite{Mann:2005yr}. Subsequently, this method was used for many concrete examples, e.g. \cite{Mann:2006bd,Astefanesei:2006zd,Herdeiro:2010aq,Astefanesei:2009wi,Compere:2011db,Compere:2011ve} and, also, to prove that the scalar charges, contrary to what was claimed before, can not enter in the first law of thermodynamics \cite{Astefanesei:2018vga}.

The remainder of the paper is organized as follows: In section \ref{sec:sols}, we present the exact asymptotically flat hairy charged black hole solutions of \cite{Anabalon:2013qua} and some of their geometric properties relevant for the thermodynamic stability analysis. In section \ref{sec:counter}, we give a short overview of the quasilocal formalism and counterterm method for asymptotically flat spacetimes. We compute the thermodynamic quantities for Reissner-Nordstr\"om black hole and the hairy black holes presented in the previous section. In section \ref{sec:therm1}, we briefly review the thermodynamic stability conditions in canonical and grand canonical ensembles and study the Reissner-Nordstr\"om black hole, showing explicitly that it is thermodynamically unstable. We then proceed in section \ref{sec:therm2} to perform the local thermodynamic stability analysis of the solutions with $\gamma=1$ presented in section \ref{sec:sols} and show that, due to the presence of the scalar field and its non-trivial self-interaction, they are thermodynamically stable in some region of the parameter space. We close in section \ref{disc} with a detailed discussion and outlook. We  leave a similar analysis for hairy black holes when the scalar self-interaction is turned off ($\alpha=0$) for Appendix (\ref{A}) and when $\gamma=\sqrt{3}$, but $\alpha \neq 0$, for Appendix (\ref{B}).

\section{Exact hairy black hole solutions}\label{sec:sols}

In this section, we describe relevant features of the exact asymptotically flat hairy charged black hole of interest \cite{Anabalon:2013qua}. 
We then consider the following action of Einstein-Maxwell-dilaton theory:\footnote{We should emphasize that our conventions are slightly different from the ones in \cite{Anabalon:2013qua}.}
\begin{equation}
I\left[g_{\mu\nu},A_\mu,\phi\right]
=\frac{1}{2\kappa}\int_{\mathcal{M}}
{d^{4}x\sqrt{-g}}\left[
R-\text{e}^{\gamma\phi}F^2
-\frac{1}{2}(\pa\phi)^2-V(\phi)\right]
\label{action1}
\end{equation}
where $F^2=F_{\mu\nu}F^{\mu\nu}$, $(\pa\phi)^2=\pa_\mu\phi\,\pa^\mu\phi$, $V(\phi)$ is the dilaton potential, and $c = G_N = 4\pi \epsilon_0=1$ such that $\kappa=8\pi$. 
With our conventions, the equations of motion are
\begin{align}
R_{\mu\nu}-\frac{1}{2}g_{\mu\nu}R&=
T_{\mu\nu}^{\phi}+T_{\mu\nu}^{EM}
\label{eins} \\
\pa_{\mu}\(\sqrt{-g}e^{\gamma\phi}F^{\mu\nu}\)
&=0 \label{maxw} \\
\frac{1}{\sqrt{-g}}\partial_{\mu}
\(\sqrt{-g}g^{\mu\nu}\partial_{\nu}\phi\)
&=\frac{dV(\phi)}{d\phi}+\gamma{}\text{e}^{\gamma\phi}F^2
\label{klein}
\end{align}
where 
$T_{\mu\nu}^{\phi}
\equiv\frac{1}{2}\partial_{\mu}\phi\partial_{\nu}\phi
-\frac{1}{2}g_{\mu\nu}
\[\tfrac{1}{2}(\partial\phi)^2+V(\phi)\]$ and
$T_{\mu\nu}^{EM}\equiv2e^{\gamma\phi}
\left(F_{\mu\alpha}F_{\nu}^{\,\,\alpha}
-\tfrac{1}{4}g_{\mu\nu}F^2\right)$ are the dilaton and electromagnetic energy-momentum tensors. The non-trivial coupling between the dilaton and gauge field gives rise to a new contribution in the right hand side of the equation of motion for the dilaton and that is why the hair is secondary, there is no independent integration constant associated to the scalar field. The dilaton potential considered in \cite{Anabalon:2013qua}, which vanishes at the boundary, matches the supergravity potential that was obtained for a one scalar field truncation of  $N= 2$  gauged  supergravity  with  an  electromagnetic Fayet-Iliopoulos term \cite {Anabalon:2017yhv}, when the cosmological constant vanishes.

In a series of papers \cite{Anabalon:2012ta, Acena:2012mr, Anabalon:2013sra, Acena:2013jya,  Anabalon:2017eri},  by  using  a  specific  ansatz  \cite{Anabalon:2012ta}, a new procedure was developed for obtaining exact regular hairy black hole solutions for a general  scalar  potential.\footnote{The properties of these hairy black holes were carefully studied in related work \cite{Anabalon:2014fla, Anabalon:2015vda, Anabalon:2015xvl}.} In flat spacetime, the field equations should be solved by employing a similar special ansatz for the metric and dilaton. For simplicity and clarity, we are going to focus on two particular cases for which the exponent coefficient of the dilaton coupling with the gauge field takes the values $\gamma=1$ and $\gamma=\sqrt{3}$, but more details about general solutions can be found in \cite{Anabalon:2013qua}.
%
\subsection{Black hole solution for $\gamma=1$}
\label{s1}
We are interested in a theory with the following scalar field potential:

\begin{equation}
\label{dilaton1}
V(\phi)=2\alpha (2\phi+\phi\cosh{\phi}-3\sinh{\phi})
\end{equation}
where $\alpha$ is an arbitrary parameter. For completness, the behaviour of the scalar field potential is depicted below, in Fig. \ref{pot1}, and we note that it vanishes for $\phi=0$.
\begin{figure}[H]
\centering
\includegraphics[height=4.5cm]{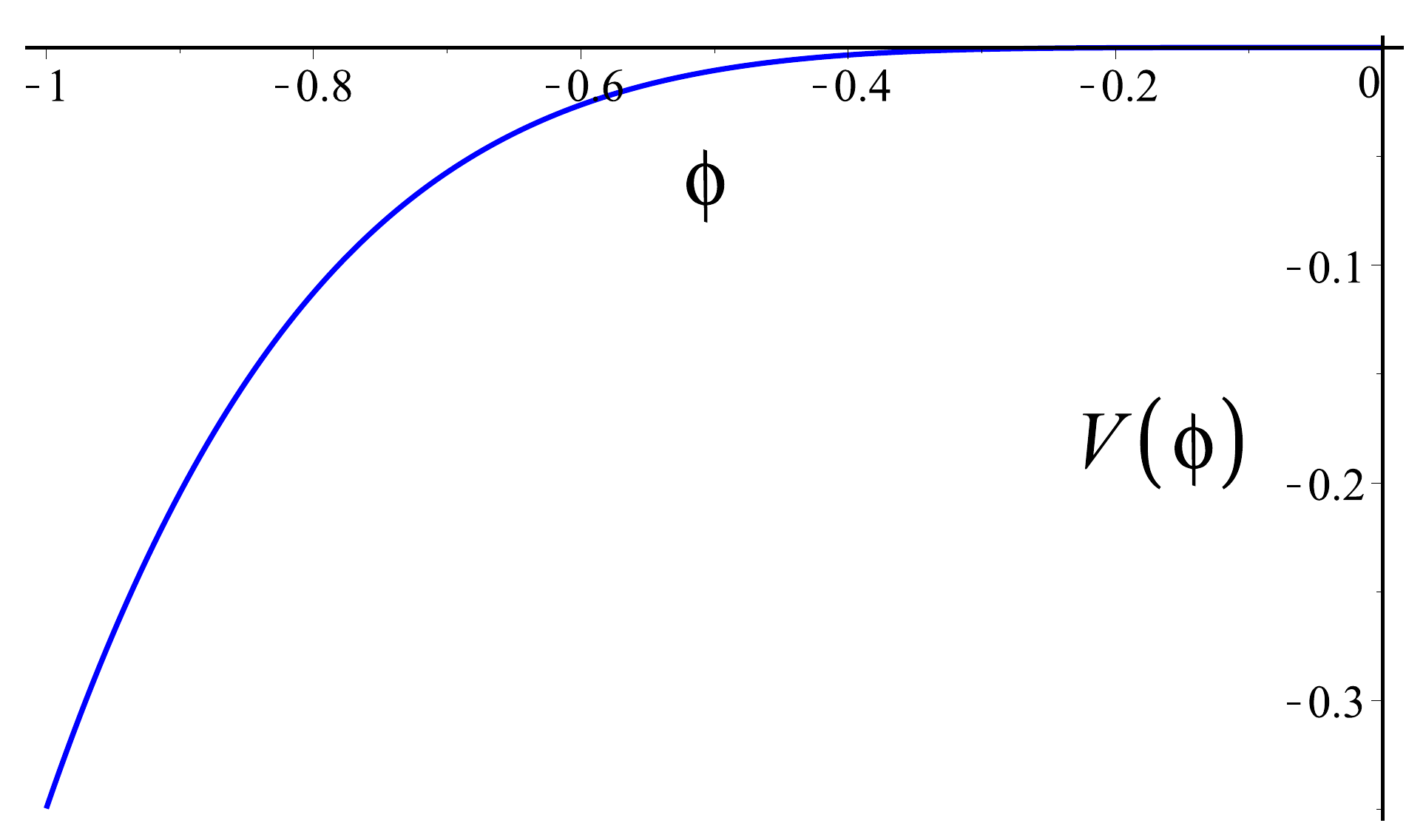}\hspace{0.5cm}
\includegraphics[height=4.5cm]{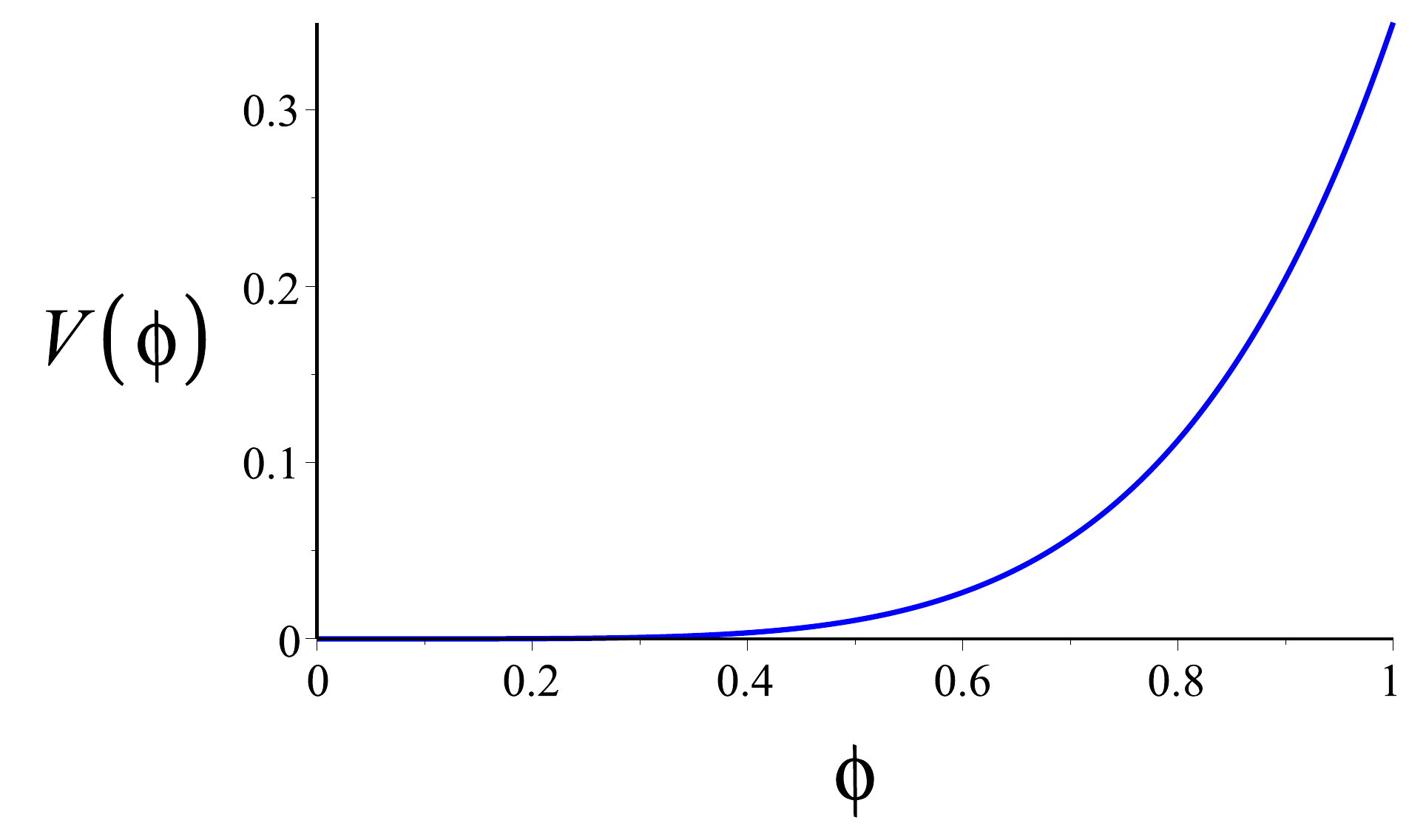}
\begin{picture}(0,0)(0,0)
\end{picture}
\vspace{0.1cm}
\caption{\small The dilaton potential for the negative branch (left hand side) and positive branch (right hand side), both for $\alpha=10$. }
\label{pot1}
\end{figure}

The suitable ansatz for the static hairy black hole (the metric\footnote{The ansatz slightly differs from the usual general ansatz \cite{Anabalon:2012ta} by the appearance of $1/x^2$ in the coefficient of $dx^2$. The reason is that, to get this solution, one should consider a specific limit procedure for which the details can be found in \cite{Anabalon:2013qua}.} and gauge field) when $\gamma=1$ is
\begin{align}
ds^{2}&=\Omega(x)
\[-f(x)dt^{2}+\frac{\eta^{2}dx^{2}}{x^{2}f(x)}+d\Sigma^2\]
\label{sol1} \\
F&=\frac{1}{2}F_{\mu\nu}dx^\mu\wedge dx^\nu
=\frac{qe^{-\phi}}{x} \; dt\wedge dx
\label{sol2}
\end{align}
where $\eta$ and $q$ are independent parameters of the solution that are going to be related to the mass and charge of the black hole, the spherical line element is $d\Sigma^2=d\theta^{2}+\sin^2\theta{}d\varphi^{2}$, and the coordinate $x$ is restricted to be positive, $x\in [0, \infty)$. It can be assumed, without losing generality, that $\eta > 0$.

By choosing a conformal factor of the form
\begin{equation}
\Omega(x)=\frac{x}{\eta^{2}\left(x-1\right)^{2}}
\end{equation}
it is straightforward to check that the equations of motion are satisfied by the metric, gauge potential, and dilaton with the following expressions:
\begin{equation}
\label{1gamma}
f(x)=\alpha
\[\frac{x^{2}-1}{2x}-\ln(x)\]+\frac{\eta^{2}(x-1)^2}{x}
\left[1-\frac{2q^2(x-1)}{x}\right]
\end{equation}
\begin{equation}
\label{1gammaA}
A=\left(\frac{q}{x}-\frac{q}{x_+}\right)dt
\; , \qquad \phi=\ln(x) 
\end{equation}
where $x_+$ is the location of black hole's event horizon, given by $f(x_+)=0$, and the additive constant in the gauge field is fixed in order to make $A=0$ at the horizon.

A detailed analysis reveals that this solution corresponds, in fact, to two disconnected black hole branches. To understand that, first we note that the conformal factor blows up in the limit $x=1$, which is the boundary where both the scalar field and its potential vanish.  To get a better understanding of this fact, let us provide the change to canonical coordinates, namely the radial coordinate $r$.  The change of coordinates 
\begin{equation}
\label{expansion}
\Omega(x)=r^2+\mathcal{O}(r^{-2})
\end{equation}
corresponds asymptotically, near $x=1$, to two distinct cases:
\begin{align}
x&=1-\frac{1}{\eta r}+\frac{1}{2\eta^2{r}^2}
-\frac{1}{8\eta^3{}r^3}+\mathcal{O}\left(r^{-5}\right)
\label{change1} \\
x&=1+\frac{1}{\eta r}+\frac{1}{2\eta^2{r}^2}
+\frac{1}{8\eta^3{}r^3}+\mathcal{O}\left(r^{-5}\right)
\label{change2}
\end{align}
Therefore, the boundary $x=1$ can be reached either from left hand side or from right hand side, which splits the solutions into two branches. The change of coordinate (\ref{change1}) restricts $0<x\leq 1$, while (\ref{change2}) restricts $1\leq x<\infty$. Since the former case implies that the scalar field acquires negative values, this branch is called `negative branch'. The latter, following the same reasoning, is called `positive branch'. Since important differences could appear in their thermodynamic behaviour, each branch must be studied separately. At $x=0$ and $x=\infty$, the scalar field is blowing up and one can check that these points correspond to real singularities of the spacetime.

We would also like to point out that the existence of regular black hole solutions depends on the parameter $\alpha$. Since we are interested in static solutions for which the Killing horizon is defined by the Killing vector $\xi^\mu=\partial / \partial t$, by imposing the condition for the existence of an event horizon, $f(x_+)=0$, we obtain
\begin{equation}
\alpha\left[1-x_+^2+2x_+\ln(x_+)\right]=
2\eta^2\left(x_{+}-1\right)^2
\left[x_{+}-2q^2(x_{+}-1)\right]
\label{condition}
\end{equation}
For the negative branch, $0<x_+\leq 1$, the right hand side term is obviously positive. One can also prove that the combination $1-x_+^2+2x_+\ln(x_+)$ that multiplies $\alpha$ on the left hand side is a positive defined function. Therefore, for the negative branch, only theories with $\alpha>0$ support black holes. No restriction for $\alpha$ is found within the positive branch.

\subsection{Black hole solution for $\gamma=\sqrt{3}$}

In this subsection, we present another exact hairy black hole solution for a different  coupling of the dilaton with the gauge field, namely $\gamma=\sqrt{3}$. To obtain exact solutions, we consider now the following dilaton potential:
\begin{equation}
\label{dilaton2}
V(\phi)=\alpha
\left[\sinh(\sqrt{3}\phi)+9\sinh\left(\frac{\phi}{\sqrt{3}}\right)
-4\sqrt{3}\,\phi \cosh\left(\frac{\phi}{\sqrt{3}}\right)
\right]
\end{equation}
that is depicted below in Fig. \ref{pot3} and has a similar behaviour as in the case $\gamma=1$.
\begin{figure}[H] 
\centering
\includegraphics[height=4.5cm]{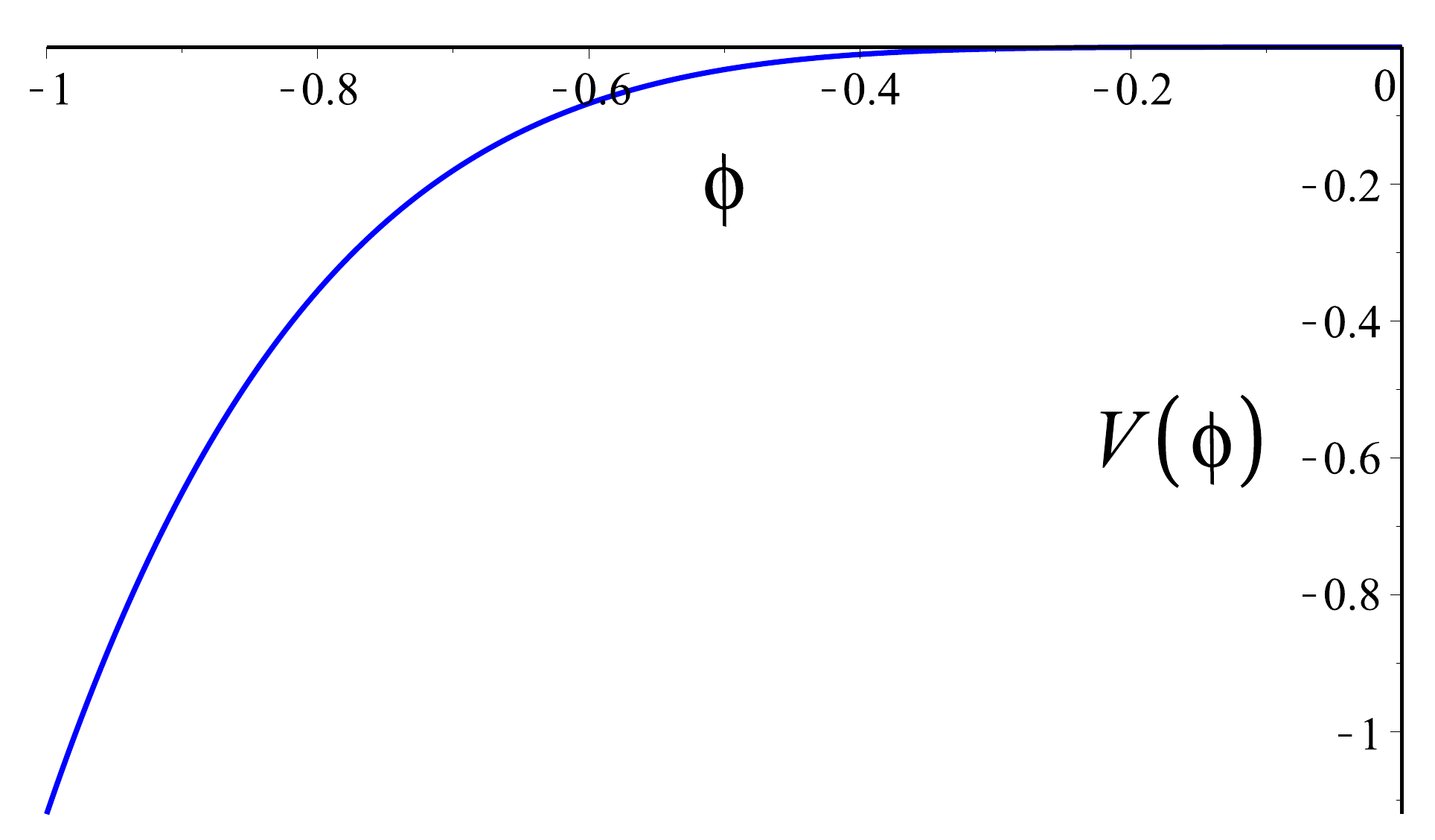}\hspace{0.5cm}
\includegraphics[height=4.5cm]{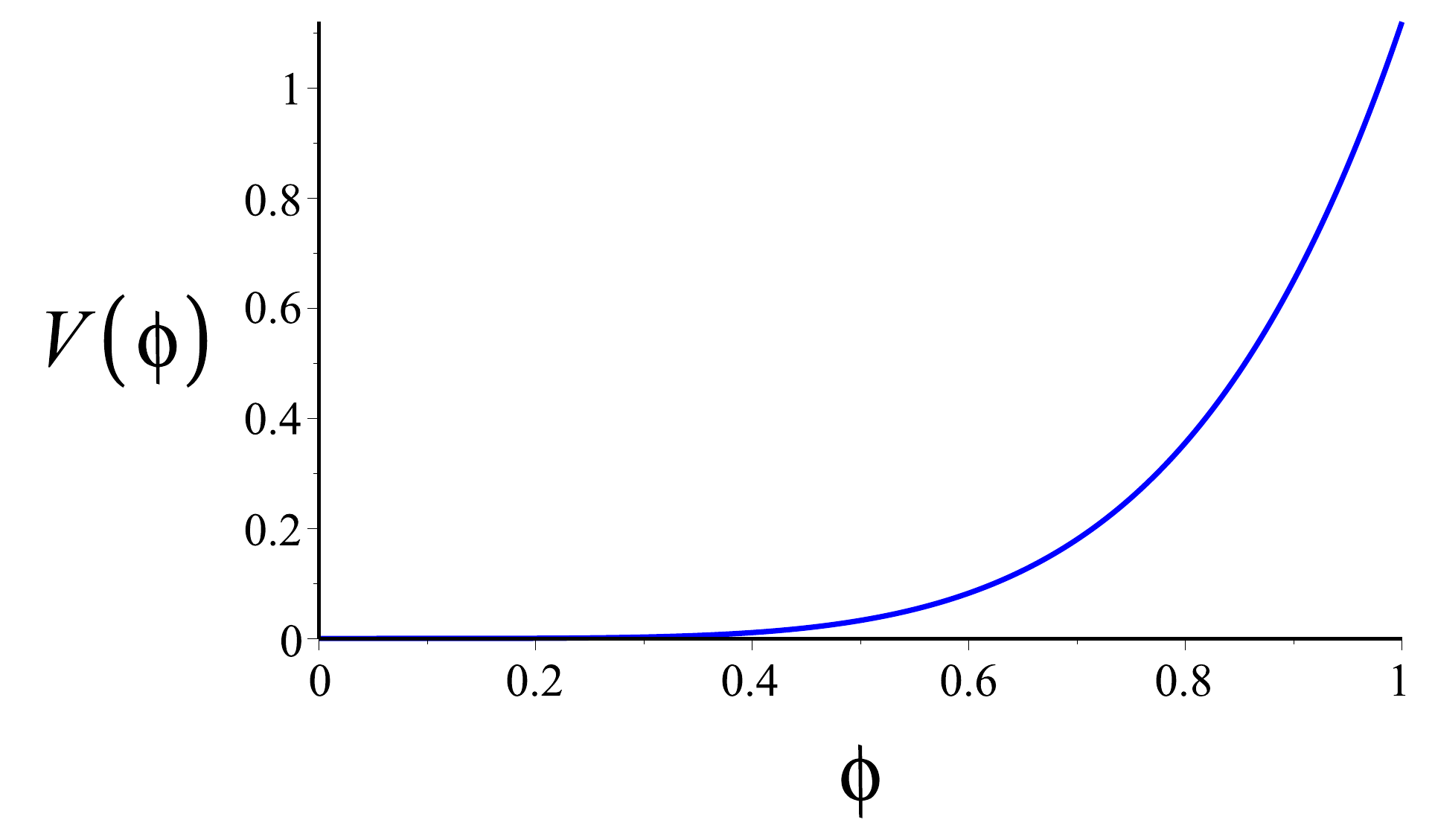}
\begin{picture}(0,0)(0,0)
\end{picture}
\vspace{0.1cm}
\caption{\small Behaviour of scalar field potential for negative branch (left), and positive branch (right side), both for $\alpha=10$.}
\label{pot3}
\end{figure}

In this case, we use the usual general ansatz \cite{Anabalon:2013qua}
\begin{equation}
ds^2=\Omega(x)\[
-f(x)dt^2+\frac{\eta^2dx^2}{f(x)}
+d\Sigma^2\]
\end{equation}
and, by peaking the conformal factor 
\begin{equation}
\Omega(x)=\frac{4x}{\eta^{2}\left(x^2-1\right)^{2}}
\end{equation}
we obtain an exact regular hairy black hole solution with the following expressions for the metric, gauge potential, and dilaton:
\begin{equation}
f(x)=\alpha
\left[\frac{x^4}{2}-2x^2+\frac{3}{2}+2\ln(x)\right]
+\frac{\eta^2(x^2-1)^2}{4}\[1-\frac{q^2(x^2-1)}{x^2}\]
\label{3gamma}
\end{equation}
\begin{equation}
A=\left(\frac{q}{2x^2} - \frac{q}{2x_+^2}\right)\,dt 
\; , \qquad \phi=\sqrt{3}\ln(x)
\end{equation}
where $x_+$ is the location of black hole's event horizon, given by $f(x_+)=0$, and the additive constant in the gauge potential is again fixed in order to make $A=0$ at the horizon.

As before, we can also prove that, in fact, there are two distinct family of solutions labeled by a scalar field that has positive values for the first branch and negative values for the second one, respectively. The corresponding changes of coordinates near the boundary are
\begin{align}
x&=1-\frac{1}{\eta r}+\frac{1}{8\eta^3 r^3}+\frac{1}{8\eta^4 r^4}
+\mathcal{O}(r^{-5}) \label{change3} \\
x&=1+\frac{1}{\eta r}-\frac{1}{8\eta^3 r^3}+\frac{1}{8\eta^4 r^4}
+\mathcal{O}(r^{-5}) \label{change4}
\end{align}
As in the previous case, we can make a discussion on the existence of the regular solutions when the parameter $\alpha$ takes positive or negative values. At the horizon, $f(x_+)=0 $ implies that
\begin{equation}
\alpha\left[-\frac{(1-x_+^2)(3-x_+^2)}{2}-2\ln(x_+)\right]
=\frac{\eta^2}{4x_+^2}
\left[x_+^2-(x_+^2-1)\,q^2\right](x_+^2-1)^2
\end{equation}
and so the restriction imposed on the parameter $\alpha$ for the existence of regular hairy black holes within the negative branch is again $\alpha>0$. Instead, no restriction on $\alpha$ for the positive branch is found.


\section{Counterterm method}
\label{sec:counter}

In this section, we present a brief review of quasilocal formalism and counterterm method for asymptotically flat spacetimes. Then, we are going to compute the thermodynamic quantities for the Reissner-Nordstr\"{o}m (RN) black hole and hairy black holes presented in section \ref{sec:sols} and prove that the first law of thermodynamics and quantum statistical relation are satisfied. 

We consider four-dimensional asymptotically flat static black hole solutions and so the spatial infinity, which is the part of infinity reached along spacelike geodesics, is the one relevant for our analysis. Brown and York proposed a surface stress-energy-momentum tensor, referred to as the `quasilocal stress tensor' in what follows, for the gravitational field \cite{Brown:1992br} that is obtained by varying the action with respect to the induced metric on the boundary of the quasilocal region. The quasilocal energy obtained in this way has the correct Newtonian limit and matches the value of the Hamiltonian that generates unit magnitude proper-time translations on the boundary. A concrete expression for the {\it regularized}
boundary stress-energy tensor when the spatial boundary is pushed to infinity was given in \cite{Astefanesei:2005ad}:
\begin{equation}
\label{stress}
\tau_{ab}=\frac{2}{\sqrt{-h}}
\frac{\delta I}{\delta h^{ab}} = \frac{1}{\kappa}\left[
K_{ab}-h_{ab}K-\Psi
\left(\mathcal{R}^{(3)}_{ab}-\mathcal{R}^{(3)}h_{ab}
\right)-h_{ab}\Box \Psi+\Psi_{;ab}\right]
\end{equation}
where $\Psi=\sqrt{{2}/{\mathcal{R}^{(3)}}}$. It was obtained from varying the action supplemented with the gravitational counterterm \cite{Lau:1999dp,Mann:1999pc,Kraus:1999di} in four dimensions:
\begin{equation}
\label{actcounterterm}
I=I_{\text{bulk}} + I_{\text{GH}} + I_{\text{ct}}=I_{\text{bulk}}
+\frac{1}{\kappa}\int_{\pa\mathcal{M}}{d^3x\sqrt{-h}K}
-\frac{1}{\kappa}\int_{\pa\mathcal{M}}
{d^3x\sqrt{-h}\sqrt{2\mathcal{R}^{(3)}}}
\end{equation}
where $K$ is the trace of extrinsic curvature $K_{ab}$, $h_{ab}$ is the induced metric on the boundary hypersurface $\mathcal{\pa\mathcal{M}}$, and $\mathcal{R}^{(3)}$ is the trace of the Ricci tensor $\mathcal{R}^{(3)}_{ab}$ of the metric $h_{ab}$. Also, with our notation, $I_{\text{bulk}}$ is the action in the bulk (\ref{action1}), the next term, $I_{\text{GH}}$, is the  Gibbons-Hawking boundary term, and the last one, $ I_{\text{ct}}$, is the gravitational counterterm for asymptotically flat spacetime. In this way, the difficulties associated with the choice of a reference background are avoided.

Once the quasilocal stress tensor is known, the conserved quantities can be obtained provided the quasilocal surface has an isometry generated by a Killing vector $\xi^{\mu}$. If the Killing vector is $\xi=\partial/\partial t$, the energy of the gravitational system is \cite{Brown:1992br}
\begin{equation}
\label{energy1}
E=\oint_{s_\infty^2}
{d^2\sigma\sqrt{\sigma}n^a\xi^b\tau_{ab}}
\end{equation}
where $n^a$ is the unit normal to the $t=constant$ surface, $s_\infty^2$, at spatial infinity, $\xi^a$ is the time Killing vector due to the time translational symmetry of the metric tensor, and $\sigma$ is the determinant of the induced metric on $s_\infty^2$. From a physical point of view, the existence of the isometry of the hypersurface with the induced metric $h_{ab}$ means that a collection of observers on that hypersurface, all measure the same value for the quasilocal energy.

Before presented concrete examples, we would like to emphasize that the quasilocal stress tensor can be computed on the Lorentzian section. On the other hand, since we need a finite range for the time coordinate to get a regularized action, the computation of the action is always done on the Euclidean section. Its definition differs from the usual action by changed signs of the kinetic terms, which are bilinear in time derivatives, and the overall sign. The temperature is computed by eliminating the conical singularity of the Euclidean instanton and the Euclidean action, $I^E$, is related to the thermodinamic potential of the grand canonical ensemble, which corresponds to $\Phi=constant$, by the quantum statistical relation
\begin{equation}
\label{freeenrgy1}
\mathcal{G}(T, \Phi)=\frac{I^E}{\beta}=M-TS-\Phi Q \,,\qquad 
d\mathcal{G}=-SdT- Qd\Phi
\end{equation}
where $\beta$ is the periodicity of the Euclidean time $\tau=it$ and $\Phi$ is the conjugate potential of the electric charge $Q$. The last observation is that the thermodynamic potential of the canonical ensemble ($Q=constant$) can be obtained by a Legendre transform of (\ref{freeenrgy1}):
\begin{equation}
\label{freeenrgy3}
\mathcal{F}(T, Q)=\frac{I^E}{\beta} = M-TS \,,\qquad d\mathcal{F}(T,Q)=-SdT+\Phi dQ
\end{equation}
Geometrically, it corresponds to adding an extra boundary term to the action of the form \cite{Hawking:1995ap}
\begin{equation}
\label{canterm}
I_A=\frac{2}{\kappa}\int_{\pa\mathcal{M}}
{d^3x\sqrt{-h}\,n_\nu F^{\mu\nu}A_\nu}
\end{equation}
for RN black hole. We found that for the hairy black holes of interest, a similar term (though, depending also of the scalar field) should be added to correctly define the canonical ensemble:
\begin{equation}
\label{canterm2}
I_A=\frac{2}{\kappa}\int_{\pa\mathcal{M}}
{d^3x\sqrt{-h}e^{\gamma\phi}\,n_\nu F^{\mu\nu}A_\nu}
\end{equation}

\subsection{Reissner-Nordstr\"{o}m black hole}
As a warm-up exercise, we are going to obtain the regularized Euclidean action, quasilocal stress tensor, and conserved charges for the RN black hole. With our conventions, the metric and gauge potential are
\begin{align}
\label{rnmetric}
ds^2&=-f(r)dt^2+f(r)^{-1}dr^2+r^2\(d\theta^2+\sin^2\theta d\varphi^2\) \\
\label{rnpotential}
A&=\left(\frac{q}{r}-\frac{q}{r_+} \right)dt
\end{align}
with $f(r)=1-2m/r+q^2/r^2$ and the radius of black hole outer horizon, $r_+$, satisfies $f(r_+)=0$. 

The gauge field is $F=dA$ and its equation of motion is $d(\star F)=0$. The electric charge is obtained by the standard Gauss law considering the sphere $s_\infty^2$ at the boundary:
\begin{equation}
Q
=\frac{1}{4\pi}\oint_{s_\infty^2}{\star F}
=\frac{1}{4\pi}
\oint_{s_\infty^2}{\frac{1}{4}\sqrt{-g}\epsilon_{\mu\nu\alpha\beta}
	F^{\mu\nu}dx^\alpha\wedge dx^\beta}
={q}
\end{equation}
where {$\epsilon$} is the totally antisymmetric Levi-Civita symbol, with $\epsilon_{tr\theta\varphi}=1$. 

The thermodynamic quantities associated with the RN black hole are the temperature $T$, entropy $S$, and chemical potential $\Phi$:
\begin{equation}
\label{thermod}
T=\frac{f'(r_+)}{4\pi}
=\frac{1}{4\pi r_+}\(1-\frac{q^2}{r_+^2}\) , \quad
S=\frac{\mathcal{A}}{4}
=\pi r_+^2  , \quad
\Phi=\frac{Q}{r_+}
\end{equation}
where $\mathcal{A}$ is the area of the black hole horizon. 

We choose the spacetime foliation with spherical hypersurfaces $r=constant$. The unit normal to these hypersurfaces, extrinsic curvature, and its trace are 
\begin{equation}
n_{\mu}=\frac{\delta_{\mu}^{r}}{\sqrt{g^{rr}}}, \qquad K_{\mu\nu}=\nabla_\mu n_{\nu}, \qquad 
K=g^{\mu\nu}K_{\mu\nu}
\end{equation}
Then, we obtain the following non-trivial components of the quasilocal stress tensor:
\begin{align}
\tau_{tt}&=-\frac{2M}{\kappa r^2}
+\frac{3M^2+q^2}{\kappa r^3}+\mathcal{O}(r^{-4})\\
\tau_{\theta\theta}=
\frac{\tau_{\phi\phi}}{\sin^2\theta}
&=-\frac{M^2-q^2}{2\kappa r}
-\frac{M(M^2-q^2)}{\kappa r^{2}}+\mathcal{O}(r^{-3})
\end{align}
and is a straightforward computation to check that it is, indeed, covariantly conserved.

In order to compute the energy (\ref{energy1}), which is the conserved quantity associated to the Killing vector $\xi=\pa/\pa t$, we note that 
$\sigma_{ab}=r^2 (d\theta+\sin^2\theta d\varphi^2)$, and the (time-like) normal 
to the surface $t=constant$ is $n_a=\delta_a^t/\sqrt{-g^{tt}}$. We obtain
\begin{equation}
\label{massRN}
E =\frac{8\pi M}{\kappa} + \mathcal{O}\(r^{-1}\)
=M+ \mathcal{O}\(r^{-1}\)
\end{equation}
that matches the $ADM$ mass of the black hole computed by expanding the $g_{tt}$ component of the metric at spatial infinity. Using the thermodynamic quantities (\ref{thermod}),
quasilocal mass (\ref{massRN}), and horizon equation, $f(r_+)=0$, one can prove the first law of thermodynamics for the RN black hole:
\begin{equation}
dM=T dS+\Phi\,dQ
\end{equation}

For the grand canonical ensemble the conjugate potential is fixed, $\Phi=constant$, and the Euclidean action, computed on-shell, satisfies the quantum statistical relation
\begin{equation}
\mathcal{G}=\frac{I^E}{\beta}= -\frac{Q^2}{2r_+}+\frac{1}{2}M = M-TS-\Phi Q
\end{equation}
Similarly, one can compute on-shell the Euclidean action for the canonical ensemble that corresponds to a fixed charge, $Q=constant$. However, in this case, there is an extra contribution (\ref{canterm}) that provides the right Legendre transform of the thermodynamic potential:
\begin{equation}
\mathcal{F}=\frac{I^E}{\beta} = -\frac{Q^2}{2r_+}+\frac{1}{2}M+\frac{Q^2}{r_+}
=M-TS 
\end{equation}

\subsection{Hairy black holes}
In what follows, we use the counterterm method to obtain the thermodynamic quantities for the hairy black holes in the theories with $\gamma=1$ and $\gamma=\sqrt{3}$ and prove that the first law of thermodynamics and quantum statistical relation are satisfied.

We study the hairy black hole solution given by (\ref{1gamma}) and (\ref{1gammaA}). The conserved charge can be obtained by integrating the equation of motion $d(\star e^{\phi} F)=0$,
\begin{equation}
Q=\frac{1}{4\pi}\oint_{s_\infty^2}{\star  e^{\phi}F}
=\frac{1}{4\pi}\oint_{s_\infty^2}
{e^{\phi}\(\frac{1}{4}\sqrt{-g}\epsilon_{\mu\nu\alpha\beta}
	F^{\mu\nu}dx^\alpha\wedge dx^\beta\)}=\frac{q}{\eta}
\label{charge1}
\end{equation}
and the other thermodynamic quantities associated to this solution, the Hawking temperature $T$, entropy $S$, and conjugate potential $\Phi$ are 
\begin{align}
\label{quant1}
T&=\frac{x_+}{4\pi\eta}f'(x_+) 
=\frac{(x_{+}-1)^2}{8\pi\eta x_+}
\[\alpha-\frac{4\eta^2q^2(x_{+}+2)}{x_+}
+2\eta^2\(\frac{x_{+}+1}{x_{+}-1}\)\]\notag\\
S&=\pi\,\Omega(x_{+})=\frac{\pi x_{+}}{\eta^{2}(x_{+}-1)^{2}}, \qquad
\Phi=A_{t}(x_{+})-A_{t}(x=1)=-\frac{Q\eta (x_{+}-1)}{x_{+}}
\end{align}
In order to compute the quasilocal stress tensor and Euclidean action, we should consider the foliation $x=constant$ with the induced metric
\begin{equation}
ds^2=h_{ab}\,dx^{a}dx^{b}=\Omega(x) \[-f(x)dt^{2}+d\theta^{2}+\sin^{2}{\theta}d\varphi^{2}\,\]
\label{xcte1}
\end{equation}
and using the equation (\ref{stress}) we obtain
\begin{align}
\tau_{tt}&
=\frac{12\eta^{2}q^{2}-\alpha}{6\eta\kappa}\,(x-1)^{2}
+\mathcal{O}\[(x-1)^{3}\] \\
\tau_{\theta\theta}&=\frac{\tau_{\theta\theta}}{\sin^2\theta}
=\frac{\(\alpha-12\eta^{2}q^{2}\)^{2}-36\eta^{4}\(4q^{2}-1\)}
{288\kappa\,\eta^{5}}\,(x-1)+\mathcal{O}[(x-1)^{2}]
\end{align}
The quasilocal stress tensor is covariantly conserved, which also hints that the solution is regular \cite{Astefanesei:2009mc}. To compute the quasilocal energy, we should use the eq. (\ref{energy1}) and the Killing vector $\xi=\pa/\pa t$. To check that this is the correct normalized Killing vector we have to use, let us expand asymptotically the metric near the boundary $x=1$, 
\begin{align}
ds^2&=g_{tt}dt^2+g_{xx}dx^2+g_{\theta\theta} d\Sigma^2 
\notag \\
&=-dt^2
+\frac{dx^2}{\eta^2(x-1)^4}
+\frac{x\,d\Sigma^2}{\eta^2(x-1)^2}+\mathcal{O}(x-1)dt^2
+\mathcal{O}\[(x-1)^{-3}\]dx^2
\end{align}
Now, by changing the coordinates 
\begin{equation}
\frac {dx^2}{\eta^2(x-1)^4}=dr^2 
\quad\Longrightarrow\quad r=\pm \frac{1}{\eta(x-1)}
\end{equation}
at the boundary, in the limit $r\rightarrow\infty$ one can see that the metric becomes Minkowski spacetime in spherical coordinates $ds^2=-dt^2+dr^2+r^2d\Sigma^2$, with the time coordinate proper properly normalized. We obtain the following expression for the quasilocal energy:
\begin{equation}
E = \frac{\alpha-12\eta^2q^2}{12\eta^3}+\mathcal{O}\(x-1\)
=M+\mathcal{O}\(x-1\)
\label{mass}
\end{equation}
The $ADM$ mass can be read off as the monopole in the expansion of $g_{tt}$ component of the metric in the canonical coordinates, with $r$ given in the eq. (\ref{expansion}),
\begin{equation}
-g_{tt}=1-\frac{2M}{r}+\mathcal{O}\(r^{-2}\) = 1-\frac{2}{r}\, \(\frac{\alpha-12\eta^{2}q^{2}}{12\eta^{3}}\)
+\mathcal{O}\(r^{-2}\) 
\end{equation}
and matches the quasilocal energy. Using the thermodynamic quantities (\ref{charge1}), (\ref{quant1}), quasilocal mass (\ref{mass}), and the horizon eq. $f(x_+)=0$, one can verify that, indeed, the first law of thermodynamics is satisfied.

The last step in our analysis is to also verify the quantum statistical relation. For that, we have to compute the Euclidean action. The bulk and Gibbons-Hawking contributions are
\begin{align}
\label{ctq3}
I^{E}_{bulk}+I_{GH}^{E}=&\beta(-TS-\Phi Q)+\frac{4\pi\beta}{\kappa}
\(-\frac{xf\Omega^{'}}{\eta}\)\bigg|_{x\rightarrow 1} \\ \notag
=& \beta(-TS-\Phi Q)
+\frac{4\pi\beta}{\kappa}\left[\frac{2}{\eta(x-1)}
+\frac{\alpha-12\eta^{2}q^2+3\eta^2}{3\eta^{3}}+\mathcal{O}(x-1)\right] \notag 
\end{align}
and contain a divergent term of order $\mathcal{O}\left[(x-1)^{-1}\right]$. This divergence is canceled by the gravitational counterterm
\begin{eqnarray}
\label{ctgrav}
I_{ct}^{E}=\frac{4\pi\beta}{\kappa}
\(2\Omega \sqrt{f}\)\bigg|_{x\rightarrow 1}
=\frac{4\pi\beta}{\kappa}
\[-\frac{2}{\eta(x-1)}
-\frac{\alpha-12\eta^{2}q^{2}+6\eta^{2}}{6\eta^{3}}
+\mathcal{O}(x-1)\]\;\;\;
\end{eqnarray} 
that contributes also with a finite part. Adding (\ref{ctq3}) and (\ref{ctgrav}), we obtain the following expression for the (finite) Euclidean action at the limit $x=1$: 
\begin{equation}
I^E=I^{E}_{bulk}+I_{GH}^{E}+I_{ct}^{E}=\beta(-TS-\Phi Q)+\beta\left(\frac{\alpha-12\eta^{2}q^{2}}{12\eta^{3}}\right)
\end{equation}
where the last term is exactly the quasilocal energy multiplied by the periodicity of the Euclidean time. Therefore, the finite on-shell action satisfies the quantum statistical relation for the grand canonical ensemble. Now, in order to obtain the thermodynamic potential for the canonical ensemble,  we should also consider the Euclidean contribution of the boundary term $I_A$ given by the eq. (\ref{canterm2}),
\begin{equation}
I^E_A=-\frac{2}{\kappa}\int_{\pa\mathcal{M}}
{d^3x\sqrt{-h}e^\phi n_\mu F^{\mu\nu}A_\nu}=\beta \Phi Q
\end{equation} 
and so the on-shell action for the canonical ensemble is $I^E=\beta(M-TS)$, which is again consistent with the corresponding quantum statistical relation.

Let us now perform the same analysis for the theory $\gamma=\sqrt{3}$. Since the procedure is basically the same than the one done before, we present here the equivalent results without other details. 

The electric charge is
\begin{equation}
Q=\frac{1}{4\pi}\oint_{s_\infty^2}
{\star e^{\sqrt{3}\phi}F}
=\frac{q}{\eta}
\end{equation}
and the remaining relevant thermodynamic quantities are
\begin{align}
\label{quant2}
T&=\frac{f'(x_+)}{4\pi\eta}
=\frac{(x_+^2-1)^2}{2\pi\eta\,x_+}\[
\alpha-\frac{\eta^2q^2(2x_+^2+1)}{4x_+^2}
+\frac{\eta^2x_+^2}{2(x_+^2-1)}\] \\
S&=\pi\Omega(x_+)=\frac{4\pi x_{h}}{\eta^{2}(x^{2}_{h}-1)^{2}} \,,\qquad
\Phi=A_{t}(x_+)-A_{t}(x=1)
=-\frac{Q\eta (x_{+}^{2}-1)}{2x_{+}^{2}}
\end{align}
The quasilocal energy matches the $ADM$ mass and has the following expression:
\begin{equation}
E=\frac{8\alpha+3\(1-2q^{2}\)\eta^2}{6\eta^3}
\end{equation}
and, as expected, satisfies the first law of thermodynamics.

The regularized Euclidean action is
\begin{align}
I^{E}_{bulk}+I_{GH}^{E}+I_{ct}^{E}=
&\beta(-TS-\Phi Q)+\frac{4\pi\beta}{\kappa}
\(2\Omega \sqrt{f}
-\frac{f\Omega^{'}}{\eta}\)\bigg|_{x\rightarrow 1} \\ \notag
=& \beta(-TS-\Phi Q)+\frac{4\pi\beta}{\kappa}
\[\frac{8\alpha+3\(1-2q^{2}\)\eta^{2}}{3\eta^{3}}\] 
+\mathcal{O}(x-1)\notag 
\end{align}
and, since with our conventions $\kappa=8\pi$, we obtain the quantum statistical relation in the grand canonical ensemble $I^{E}=\beta(M-TS-\Phi Q)$. A similar analysis can be done for the canonical ensemble, but we do not present the details here.


\section{Thermodynamic stability} 
\label{sec:therm1}

In the previous section, we have provided a thermodynamic description of various hairy black holes. The key idea of our analysis was to consider boundary conditions  for constructing the appropriate ensemble. However, black hole thermodynamics as deduced above will only make sense if a black hole can be in locally stable equilibrium in the corresponding ensemble.

Thermal  stability  in  an  ensemble  with  a  black   hole  must  apply  to  the  entire  system,  because  such  systems  obviously  cannot  be  subdivided  into  spatially  separate parts  as  is usually  done  in  treating  questions  of  thermodynamic  stability.  The  response  functions  relevant  to  the  thermal  stability  of  a given  type  of  ensemble,  therefore,  are  those  which  can  be  obtained  by  variation  of  the  thermodynamic parameters  that  are  not  fixed   by  the  boundary  conditions  defining  the  given  ensemble. In the canonical ensemble description of an asymptotically flat neutral black hole, the specific heat can be made positive if the black hole is put into a box. The boundary conditions must be specified, i.e. in the Schwarzschild case one can fix the temperature of the box and its radius. It follows then that stability can be achieved only for a sufficiently small cavity, leading to a well-defined canonical ensemble. 

After a review of the thermodynamic stability conditions, the first model that we shall briefly describe is the case of RN black hole. 

\subsection{Thermodynamic stability conditions}
\label{stabilitycond}
In order to analyse the thermodynanic stability of the solutions presented so far, it is important to distinguish between local and global stability. The former is related with how an equilibrium configuration responds under a small fluctuation in thermodynamic variables, while the later is related with a global maximum in the  entropy (or global minimum in the energy). We shall briefly describe the local stability below, however more details can be found in \cite{Landau} (see, also, \cite{Callen}).

Stationary black holes are localized thermal equilibrium states of the quantum gravitational field and so should respect the ordinary laws of thermodynamics. We are interested in a static charged black hole with the conserved charges $M$ and $Q$ for which the first law of thermodynamics can be written as
\begin{equation}
\label{masss}
dM=TdS+\Phi dQ \,\, \Longrightarrow \,\,\,\,\, T=\(\frac{\pa M}{\pa S}\)_Q \,,\,\,\,\,
\Phi=\(\frac{\pa M}{\pa Q}\)_S
\end{equation}
All irreversible processes in isolated system which lead to equilibrium are  governed by an increase in entropy and the equilibrium will be reestablished only when the entropy will assume its maximum value. This is the second law of thermodynamics, $dS \geqslant 0$. Since first we work in the grand canonical ensemble, it is more convenient to use the corresponding thermodynamic potential to study local stability and not the condition of maximum entropy. 

In the grand canonical ensemble for which the electrostatic potential $\Phi$ is constant, we have at equilibrium  
\begin{equation}
\label{freeenrgy2}
\mathcal{G}_e(T_e, \Phi_e)=M-T_eS-\Phi_e Q \,,\qquad 
d\mathcal{G}_e = -(SdT+Qd\Phi)_e = 0
\end{equation}
where the subscript `e' indicates the values at the thermodynamic equilibrium. Consider now a small deviation, then the conditions for stable local equilibrium are
\begin{equation}
\left(\frac{\partial^2 M}{\partial Q^2}\right)_{S}\left(\frac{\partial^2 M}{\partial S^2}\right)_{Q}-\left[\left(\frac{\partial}{\partial S}\right)_{Q}\left(\frac{\partial M}{\partial Q}\right)_{S}\right]^{2}= \frac{T}{C_{Q}}
\(\frac{1}{\epsilon_{S}}-\frac{T\alpha_{Q}^{2}}{C_{Q}}\)>0
\label{mixcondM}
\end{equation}
and
\begin{equation}
\(\frac{\pa^2 M}{\pa S^2}\)_{Q}
=\frac{T}{C_{Q}} > 0 \,\,\,\, \text{or} \,\,\, \(\frac{\pa^2 M}{\pa Q^2}\)_{S}
=\frac{1}{\epsilon_{S}} > 0
\label{indcondM}
\end{equation}
The expressions above were also written in terms of the usual physical thermodynamic quantities, the heat capacity ($C_{Q}$),  electric permittivity $(\epsilon_{S})$, and $\alpha_Q \equiv (\pa \Phi/\pa T)_Q$ (see, e.g., \cite{Chamblin:1999hg}).

We would like to obtain similar relations for the fluctuations in grand canonical and canonical ensembles, but using the second derivatives of the corresponding thermodynamic potential. For that, we have to use the following relations between thermodynamic quantities of interest:
\begin{equation}
\label{relations}
C_{\Phi}=C_{Q}+T\epsilon_T\alpha_Q^2
,\qquad 
\epsilon_S=\epsilon_T-\frac{T\alpha_\Phi^2}{C_{\Phi}}
,\qquad 
\alpha_{\Phi}=-\epsilon_{T}\alpha_{Q}
\end{equation}
Without presenting the details here, we can write the conditions for local stability 
in grand canonical in canonical ensembles in the following compact form:
\begin{equation}
\epsilon_S>0, \qquad C_\Phi>0 \qquad \text{grand canonical ensemble}
\end{equation}
\begin{equation}
\epsilon_T>0, \qquad C_Q>0 \qquad \,\,\,\,\,\,\,\,\,\,\,\,\,\,\,\, \text{canonical ensemble}
\end{equation}
These relations are consistent with the general stability criterion that states that the thermodynamic potentials are convex functions of their extensive variables and concave functions of their intensive variables (see, e.g., \cite{Callen}).

\subsection{Reissner-Nordstr\"{o}m black hole}
\label{RN}
If a black hole is to be in thermodynamic  equilibrium  at  some  temperature  $T$,  then  it  must  be  surrounded  by  a  heat  bath  at  the  same  temperature. 
The Schwarzschild black hole, the simplest static spherically symmetric solution of Einstein equations, is thermodynamically unstable. This can be easily obtained by observing that the temperature is inverse proportional with respect to the black hole mass:
\begin{equation}
T=\frac{1}{8\pi}\frac{\hbar c^3}{k_B G}\frac{1}{M} \,,\,\,\,\, S=\frac{k_B c^3}{4\hbar G}A_h = \frac{4\pi k_B G}{\hbar c}M^2 \,\,\, \Longrightarrow  C=T\,\frac{\pa S}{\pa T} =-\frac{8\pi k_BG}{\hbar c}M^2
\end{equation}
Since the heat capacity is negative, the  black hole heats up as it radiates energy. The immediate consequence is that the canonical ensemble is not well defined, e.g. the root-mean-square energy fluctuations  are imaginary: $\langle(\Delta E)^2\rangle = CT^2$. One way to go around this problem was proposed by York in \cite{York:1986it}, namely to consider the black hole in a box of radius $r_B$. The principle of equivalence requires that the temperature measured locally by a static observer is blue-shifted with respect to the usual temperature that is determined at asymptotically flat spatial infinity:
\begin{equation}
\label{t}
T(r_B) = T_{\infty} |g_{tt}(r_B)|^{-1/2} =\frac{\hbar}{8\pi GM}\(1-\frac{2GM}{r_B} \)^{-1/2}
\end{equation} 
where, as in \cite{York:1986it}, we have used the conventions in which $c=k_B=1$. The heat capacity at a fixed value of $r_B$ can now be computed using the temperature (\ref{t}) and the result is
\begin{equation}
C=T\,\frac{\pa S}{\pa T} = \frac{8\pi G M^2}{\hbar} \(1-\frac{2GM}{r_B}\)
\(\frac{3GM}{r_B} - 1\)^{-1}
\end{equation}
It is clear now that, when $2M\leq r_B < 3M$, the heat capacity is positive (the energy fluctuations are real) and so the canonical ensemble is well defined. 

A natural question is if can it be possible to obtain a positive heat capacity when the electromagnetic field is added in the problem without embedding the system in box? We are going to see that the answer is negative.  However, as a warm-up exercise,  we are also going to present in this section the thermodynamic stability of the RN asymptotically flat black hole. The metric and gauge potential that solve the corresponding equations of motion are given in (\ref{rnmetric}) and (\ref{rnpotential}), and the thermodynamic quantities are given in (\ref{thermod}). The existence of the black hole is conditioned by the inequality $Q\leq M $, otherwise the solution is a naked singularity. In the extremal case, $M=Q$, the temperature vanishes and the horizon radius becomes $r_+=M=Q$. This is going to impose a restriction on the values of the electrostatic potential, $\Phi=Q/r_+ \leq 1$. 

\begin{figure}[H]
	\centering
	\includegraphics[width=7 cm]{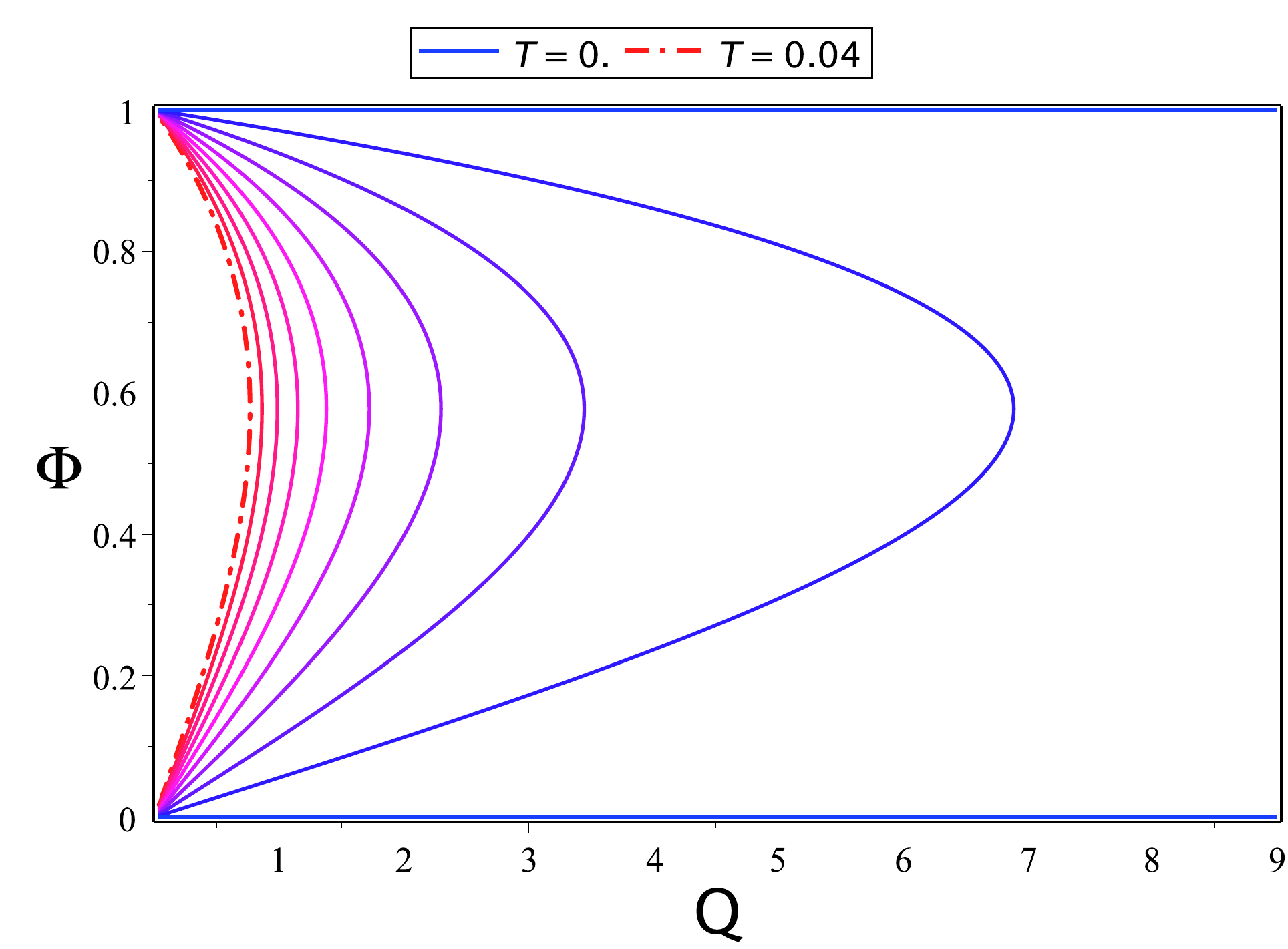} \quad
	\includegraphics[width=7 cm]{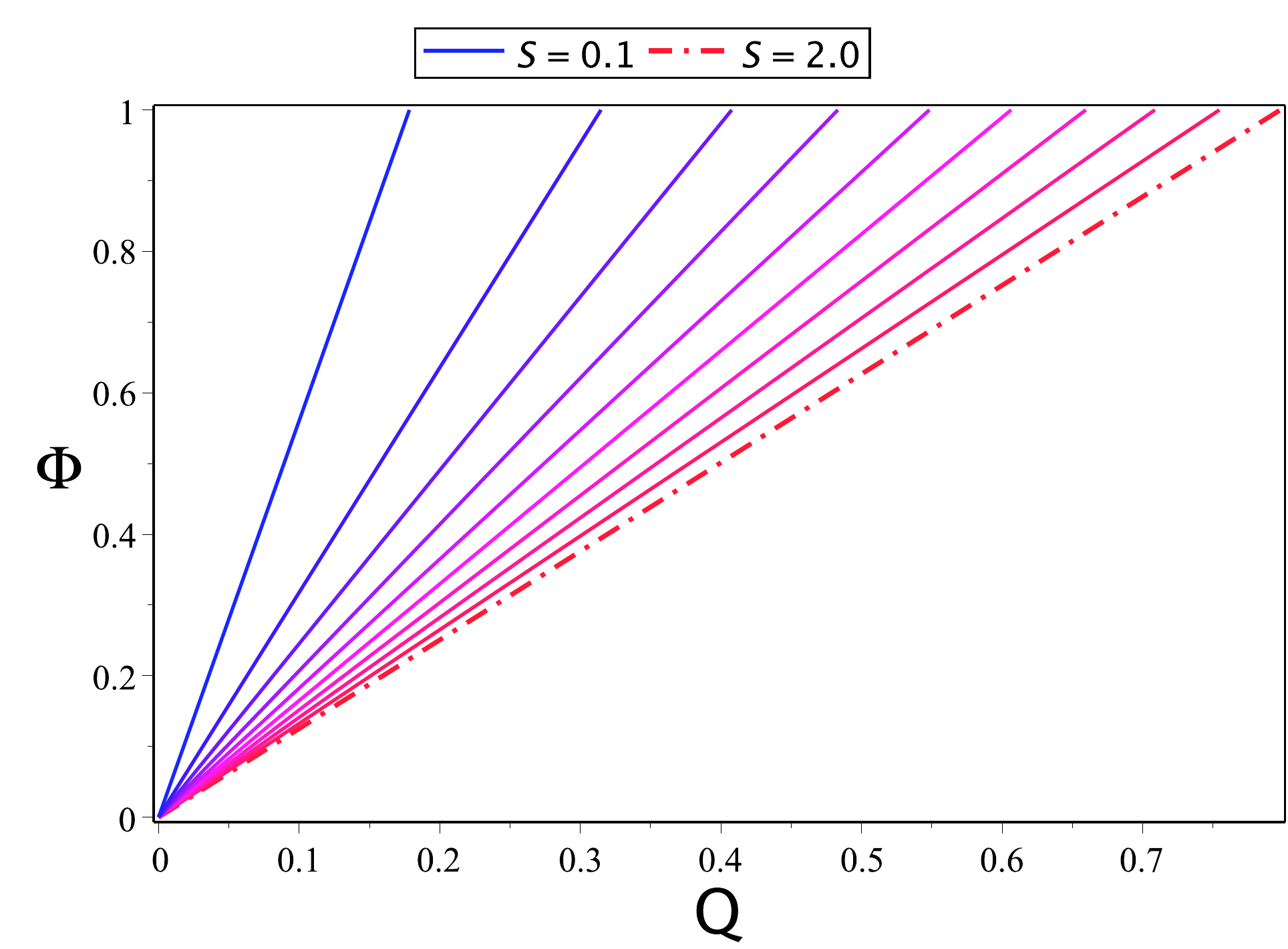}
	\caption{Left Hand Side: Equation of state of RN black hole. Horizontal (blue) curve at $\Phi=1$ represents extremal charged black holes. Right Hand Side: Isentropic curves.}
	\label{RN1}
\end{figure}

The electric permittivity is a measure of the electric fluctuations. The black holes will be electrically unstable under electric fluctuations if the electrical permittivity is negative. This happens if the potential of the black
hole decreases as a result of placing more charge on it. The potential should
of course increase, in an attempt to make it harder to move the system from equilibrium. The equation of state gives important information on stability against fluctuation of either electric charge $Q$ or conjugate potential $\Phi$, at a fixed temperature. By combining the first and the last equation  in (\ref{thermod}) to eliminate $r_+$, we obtain the equation of state of a RN black hole,
\begin{equation}
4\pi TQ+\Phi\, \left(\Phi^2-1\right)=0
\label{Eqstatern}
\end{equation}
from where the electric permittivity at fixed temperature can be read off. By combining the first two equations in (\ref{thermod}) instead, we obtain
$\pi Q^2-S\Phi^2=0$,
from where the electric permittivity at fixed entropy can be read off. These equations are depicted in Fig. \ref{RN1}. It can be easily proven that only black holes with $\Phi<1/\sqrt{3}$ are electrically stable, since by differenciating eq. $(\ref{Eqstatern})$ with respect to $\Phi$ at fixed $T$, one gets
\begin{equation}
\epsilon_T=\(\frac{\pa Q}{\pa\Phi}\)_{T}=\frac{1-3\Phi^2}{4\pi T}
\end{equation}

Let us now briefly present the stability under the mixed fluctuations, according to the discussion in the previous subsection. In grand canonical ensemble, the thermodynamic stability is guaranteed by the simultaneous positivity of the response functions $C_\Phi$ and $\epsilon_S$. It is clear that
\begin{equation}
\epsilon_S=\frac{1-\Phi^2}{4\pi T}>0
\end{equation}
because $\Phi \leq 1$. On the other hand, we can obtain 
\begin{equation}
\mathcal{G}(r_+,\Phi)=\frac{\(1-\Phi^2\)r_+}{4},
\quad 
T(r_+,\Phi)=\frac{1-\Phi^2}{4\pi r_+}
\quad\Longrightarrow\quad
\mathcal{G}(T,\Phi)
=\frac{\(1-\Phi^2\)^2}{16\pi T}
\end{equation}
and so the heat capacity
\begin{equation}
C_\Phi=
-\frac{1}{8\pi}\(\frac{1-\Phi^2}{T}\)^2<0
\end{equation}
In other words, $\mathcal{G}(T,\Phi)$ does not change its concavity with respect to $T$, at $\Phi$ fixed. This result implies that there is no thermodynamic stable configuration in this ensemble.

In the canonical ensemble, we shall investigate if $C_Q$ and $\epsilon_T$ can be simultaneously positive definite. The electric permittivity at fixed $T$ can be expressed as
\begin{equation}
\epsilon_T=\frac{\(r_+^2-3Q^2\)r_+}{r_+^2-Q^2}
\end{equation}
Note that, since $r_+^2-Q^2$ is a positive quantity, the region where $\epsilon_T>0$ corresponds to those configurations satisfying $r_+^2-3Q^2>0$ (this is equivalent to $\Phi<1/\sqrt{3}$).
On the other hand, the thermodynamic potential and temperature can be written as
\begin{equation}
\mathcal{F}(r_+,Q)=\frac{r_+^2+3Q^2}{4r_+},
\qquad T(r_+,Q)=\frac{r_+^2-Q^2}{4\pi r_+^3}
\end{equation}
and thus the heat capacity,
\begin{equation}
C_Q=-\frac{2\pi r_+^2\(r_+^2-Q^2\)}{r_+^2-3Q^2}
\end{equation}
has positive values only if $r_+^2-3Q^2<0$. Then, $C_Q$ and $\epsilon_T$ are not both positive, which confirms that there are also no thermodynamic stable configurations in canonical ensemble.

\section{Thermodynamic stability of hairy black holes}
\label{sec:therm2}

In this section, we are going to obtain our main result, namely that the asymptotically flat hairy black holes in theories with a non-trivial self interaction for the scalar field can be thermodynamically stable and so the embedding in a box is not necessary. We would like to point out that, when $\alpha=0$, there also exist exact asymptotically flat regular hairy black hole solutions, but they are not thermodynamically stable. The case of interest is when the dilaton potential is non-trivial, but for completeness we also present the case without potential in Appendix (\ref{A}). For simplicity, we are going to include in this section only the case $\gamma=1$ with both branches, positive and negative, but in Appendix (\ref{B}) we are going to explicitly show that the  results are very similar for the case $\gamma=\sqrt{3}$, which hints to a general conclusion on the local stability of hairy black holes. That is, the self-interaction of the scalar field plays a similar role as the one played by the `box' for non-hairy black holes and, therefore, it is the key ingredient for the thermodynamic stability of asymptotically flat hairy black holes.

\subsection{Positive branch with $\gamma=1$ and $\alpha\neq 0$}
\label{main}
In what follows, we investigate the local thermodynamic stability of the black holes presented in section \ref{sec:sols} in the positive branch, when $\phi>0$ or, equivalently, $x\in (1, +\infty)$, for $\gamma=1$ and $\alpha>0$.

\subsubsection{Grand canonical ensemble}
\label{gcpb}
\begin{figure}[H]
\centering
\includegraphics[width=7 cm]{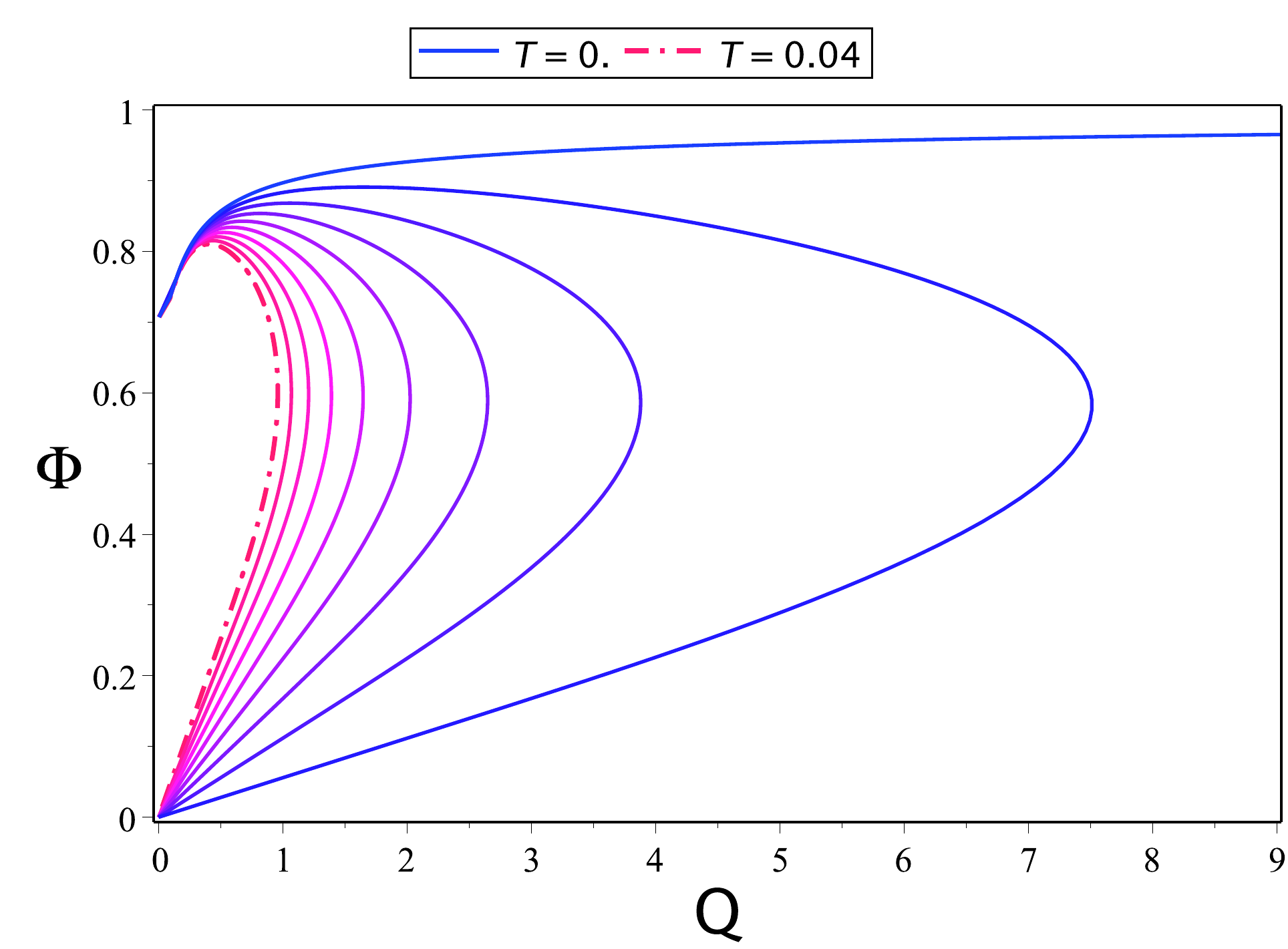}
\includegraphics[width=7 cm]{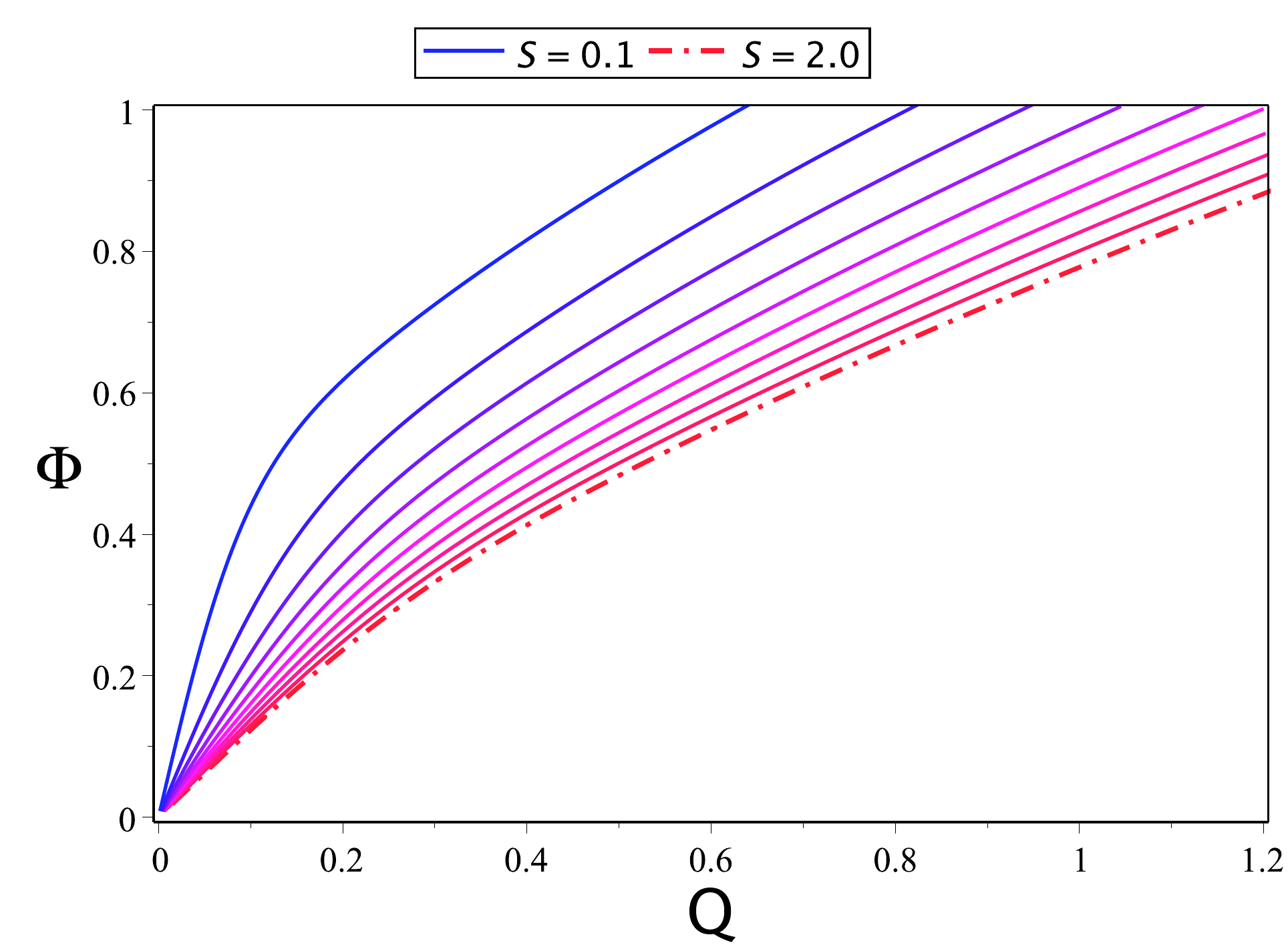}
\\\caption{Left Hand Side: isotherm curves $Q$--$\Phi$. Right Hand Side: isentropic curves $Q$---$\Phi$. $\gamma=1$ and $\alpha=10$.}
\label{state_hairy1}
\includegraphics[width=7 cm]{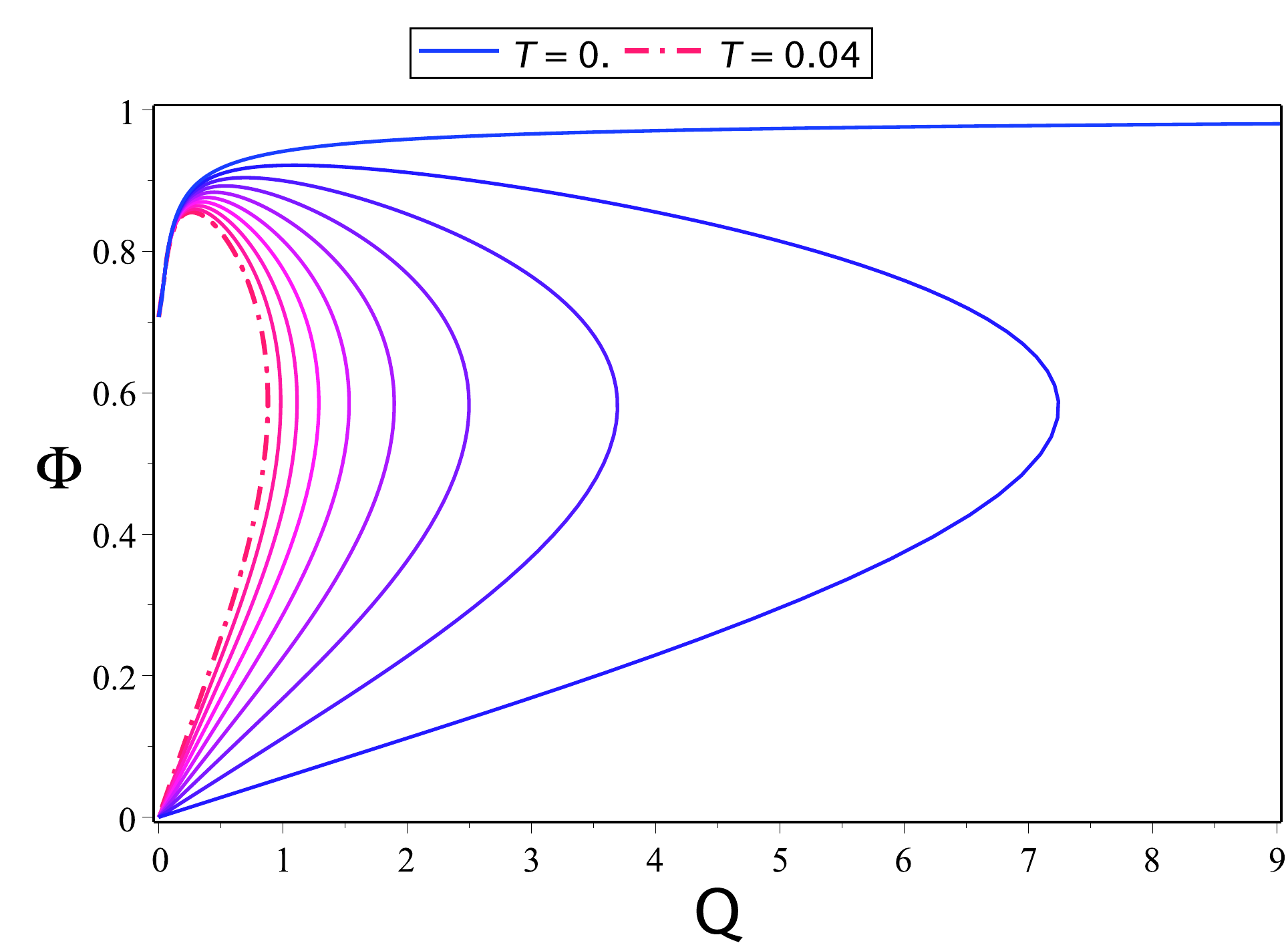}
\includegraphics[width=7 cm]{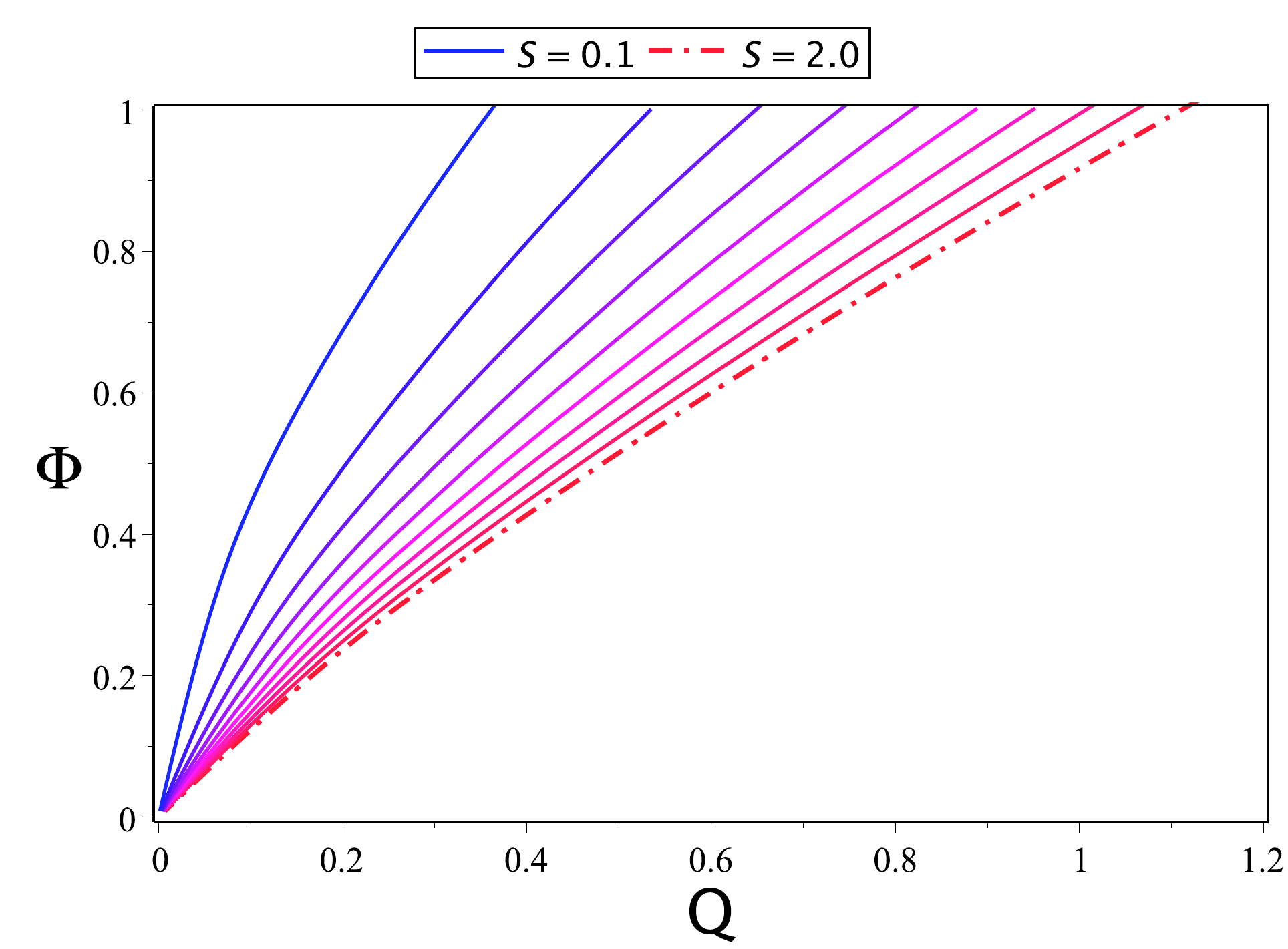}
\caption{Isotherms and isentropic curves $Q$--$\Phi$, for $\gamma=1$ and $\alpha=100$. By comparing with the case $\alpha=10$, we observe that the general behaviour does not depend on $\alpha$.}
	\label{state_hairy2}
\end{figure}

In this case, it is not possible to isolate $x_+$ from the horizon eq. (\ref{1gamma}) and we shall work with parametric equations. In order to obtain the equation of state, $(Q, \Phi)$ at fixed $T$, it is necessary to express both the electric charge $Q$ and conjugate potential $\Phi$ as functions of $(x_+,T)$. To complete this plan, let us first isolate the negative root\footnote{In this way we work with positive definite $Q$ and $\Phi$.} of $q$ from the horizon equation,
\begin{equation}
q=-\frac{\sqrt{x_{+}\(2\,\eta^2x_+^2
		-2\alpha x_{+}\ln x_{+}+\alpha{x_+}^2
		-4\,\eta^2x_{+}+2\eta^2-\alpha\)}}
{2\eta\(x_{+}-1\)^{3/2}}
\label{q1}
\end{equation}
Next, let us replace $q$ in the expression of the temperature (\ref{quant1}) to get 
\begin{equation}
\label{tempbun}
T(x_+,\eta)=\frac{\(x_+-1\)\eta}{4\pi x_+}
-\frac{\(2x_+^2\ln{x_+}+4x_+\ln{x_+}-5x_+^2+4x_++1\)\alpha}
{8\pi x_{+}\(x_{+}-1\)\eta}
\end{equation}
and
\begin{equation}
\eta(x_+,T)=\frac{4\pi T x_{+}+ w(x_+,T)}
{2(1-x_+)}
\label{eta1}
\end{equation}
where $w(x_+,T)=\sqrt{ 16\pi^2T^2x_+^2
	+\alpha(4x_+^2\ln{x_+}+8x_+\ln{x_+}-10x_+^2+8x_{+}+2)}$. Finally, by replacing (\ref{eta1}) back into (\ref{q1}) to obtain $q=q(x_+,T)$, it is straightforward to get the concrete expressions for $Q=Q(x_+,T)$ and $\Phi=\Phi(x_+,T)$, by using (\ref{charge1}) and (\ref{quant1}). 
We obtain the following results:
\begin{align}
\Phi&=\frac{\sqrt{x_+-1}
	\sqrt{32\pi^2T^2x_+^2+ 16\pi T x_+w-8\alpha x_+\ln{x_+}+4\alpha x_+^2+2w^2-4\alpha}}
{2\sqrt{x_+}\(4\pi x_+T+ w\)}
\label{hairychem1} \\
Q&=\frac{\sqrt{x_+(x_+-1)}
	\sqrt{32\pi^2T^2x_+^2+ 16\pi T x_+w-8\alpha x_+\ln{x_+}+4\alpha x_+^2+2w^2-4\alpha}}
{\(4\pi x_+T+ w\)^2}
\label{hairycharg1}
\end{align}
A similar procedure can be used to obtain the expressions for $\Phi(x_+,S)$ and $Q(x_+,S)$,
\begin{align}
\Phi&=\frac{\sqrt{{x_+}-1}
	\sqrt{\alpha{S}\(x_+^2-1-2x_+\ln{x_+}\)+2\pi x_+}}
{2\pi\sqrt{x_+}}
\\
Q&=\frac{\sqrt{({x_+}-1)S}
	\sqrt{\alpha{S}\(x_+^2-1-2x_+\ln{x_+}\)+2\pi x_+}}
{2\pi\sqrt{x_+}}
\end{align}
In Figs. \ref{state_hairy1} and \ref{state_hairy2} it is shown $\Phi$ vs $Q$ at fixed temperature and entropy, for two different order of magnitude values of $\alpha$.
From this graphic representation, one can simply conclude that the isentropic permittivity is positive, $\epsilon_S>0$. However, this result can also be obtained by computing it explicitly
\begin{equation}
\epsilon_S=\frac{\alpha S
	\(2x_+^3-2x_+^2\ln{x_+}-3x_+^2+2{x_+}-1\)
	+2\pi x_+^2}
{\alpha S\(x_+^3-2x_+\ln{x_+}-2x_+^2+3{x_+}-2\)
	+2\pi x_+} \(\frac{Sx_+}{\pi}\)^{\frac{1}{2}}
\end{equation}
and by noticing that both expressions, the ones at the numerator and denominator, are positive definite in the positive branch.
The positiveness of this response function is directly related with the stability in grand canonical ensemble. If some configurations with $\epsilon_S>0$ were also characterized by $C_\Phi>0$, then those ones would represent thermodynamic stable black holes, according with the discussion in the previous section.

\begin{figure}[H]
	\centering
	\includegraphics[width=11 cm]{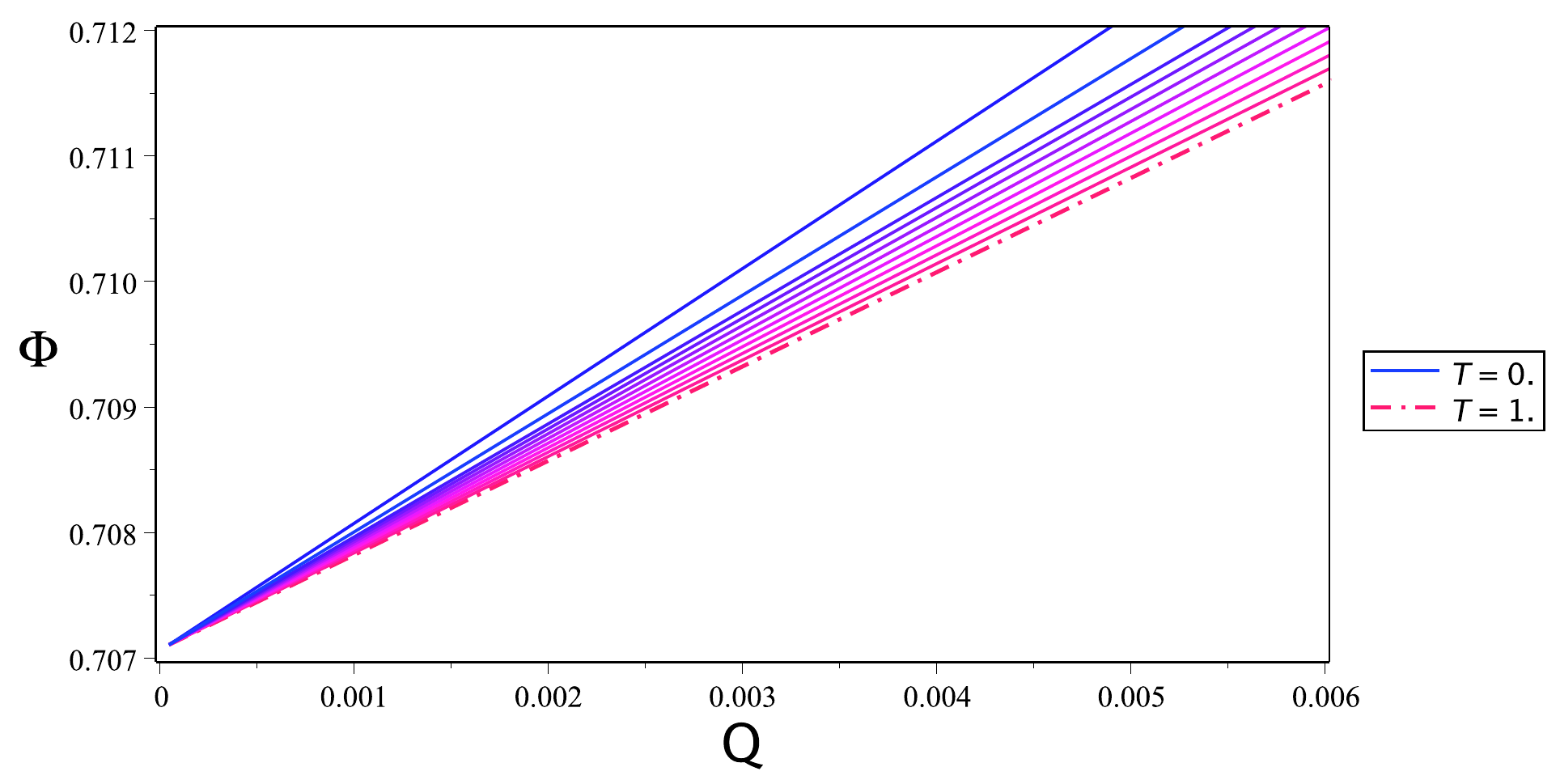}
	\caption{Zoom in the equation of state, near $Q=0$ for $\Phi=-1/2q$. Temperature ranges from $T=0$ to $T=1$. }
	\label{zoomeqestate}
\end{figure}

The equation of state, depicted on the left hand sides in Figs. \ref{state_hairy1} and \ref{state_hairy2}, reveals two separated regions where $\epsilon_T>0$. Also, notice that extremal black holes are electrically stable in this case. These two features are different compared with RN black hole. Another interesting aspect of the equation of state is that all isotherms start at $Q=0$, $\Phi=0$ (like RN black hole) but end at $Q=0$, $\Phi=1/\sqrt{2}<1$. To understand why,
we observe that the physical charge is $Q=-q/\eta$ (we consider $q$ negative in our analysis) and it vanishes not only when $q=0$, but also in the limit when $\eta \rightarrow \infty$ for a finite $q$. When considering the horizon equation (\ref{1gamma}), the term proportional with $\eta^2$ is going to be the relevant one and so we obtain the horizon value as:
\begin{equation}
\frac{\eta^{2}(x_+ - 1)^2}{x_+}\left[1-\frac{2q^2(x_+ - 1)}{x_+}\right] = 0 \,\,\,\Longrightarrow \,\,\, 
\frac{x_+}{x_+ - 1} =2q^2
\end{equation}
By replacing  this result in the expression of the electric potential (\ref{quant1})
\begin{equation}
\Phi=A_{t}(x_{+})-A_{t}(x=1)=-\frac{Q\eta (x_{+}-1)}{x_{+}} \,\,\,\Longrightarrow \,\,\, \Phi=-\frac{1}{2q}
\end{equation}
So, it  seems that, indeed, we can consider a zero physical charge for two distinct values of the conjugate electric potential $\Phi$. We would like then to obtain the isotherm curves for any temperature given by the equation (\ref{tempbun}), which it is not possible when the horizon value $x_+$ is very small. In this case, for large values of $\eta$ we obtain large temperatures. However, we can get a full range of values for the temperatures when the horizon value is large (that means, small black holes, because $x$ is inverse proportional with the canonical normal coordinate $r$), which in turn implies that $q=-1/\sqrt{2}$ in this limit (as can be seen in the Figs. \ref{state_hairy1}, \ref{state_hairy2}, and \ref{zoomeqestate}.).

Let us then examine the other relevant response function, the heat capacity at fixed $\Phi$, by means of the thermodynamic potential. By solving $\eta=\eta(x_+,\Phi)$ from the horizon equation, we get the following parametric expressions
\begin{align}
M&=-\frac{\alpha}{12\eta^3}
+\frac{x_+^2\Phi^2}{\eta(x_{+}-1)^2}, \quad
Q=\frac{x_{+}\Phi}{\eta(x_{+}-1)}, \quad
S=\frac{\pi x_+}{\eta^2(x_{+}-1)^2},\\
T&=\frac{(x_{+}-1)^2}{8\pi\eta\,x_+}
\[-\alpha+4\Phi^2\eta^2x_+\,\frac{x_{+}+2}{(x_{+}-1)^2}
-2\eta^2\,\frac{x_{+}+1}{x_+-1}\]
\end{align}
and the thermodynamic potential
\begin{equation}
\mathcal{G}(x_+,\Phi)=\frac{\alpha}{24\eta^3}
-\frac{\Phi^2x_+^2}{2\eta(x_+-1)^2}
+\frac{x_{+}+1}{4\eta(x_+-1)}
\end{equation}
The heat capacity 
$C_\Phi$ can now be directly computed by obtaining the corresponding second derivative of the thermodynamic potential.
Concretely, the conditions for stable local equilibrium in grand canonical ensemble are
\begin{align}
-\(\frac{\pa^2\mathcal{G}}{\pa T^2}\)_\Phi&>0
\label{firstcondition}
\\
\(\frac{\pa^2\mathcal{G}}{\pa T^2}\)_\Phi
\(\frac{\pa^2\mathcal{G}}{\pa\Phi^2}\)_T
-\[\(\frac{\pa}{\pa\Phi}\)_T
\(\frac{\pa\mathcal{G}}{\pa T}\)_\Phi\]^2&>0
\label{secondcondition}
\\
-\(\frac{\pa^2\mathcal{G}}{\pa T^2}\)_\Phi
\left\{\(\frac{\pa^2\mathcal{G}}{\pa T^2}\)_\Phi
\(\frac{\pa^2\mathcal{G}}{\pa\Phi^2}\)_T
-\[\(\frac{\pa}{\pa\Phi}\)_T
\(\frac{\pa\mathcal{G}}{\pa T}\)_\Phi\]^2\right\}&>0
\label{thirdcondition}
\end{align}
According to the discussion in section (\ref{stabilitycond}), it is consistent that only two of them are required, since the other one automatically holds. To be more explicit, inequalities (\ref{firstcondition}) and (\ref{thirdcondition}) are equivalent to the statements $C_\Phi>0$ and $\epsilon_S>0$, respectively, while (\ref{secondcondition}) is equivalent to $C_\Phi\epsilon_S>0$. 

\begin{figure}[H]
	\centering
\includegraphics[width=4.2 cm]{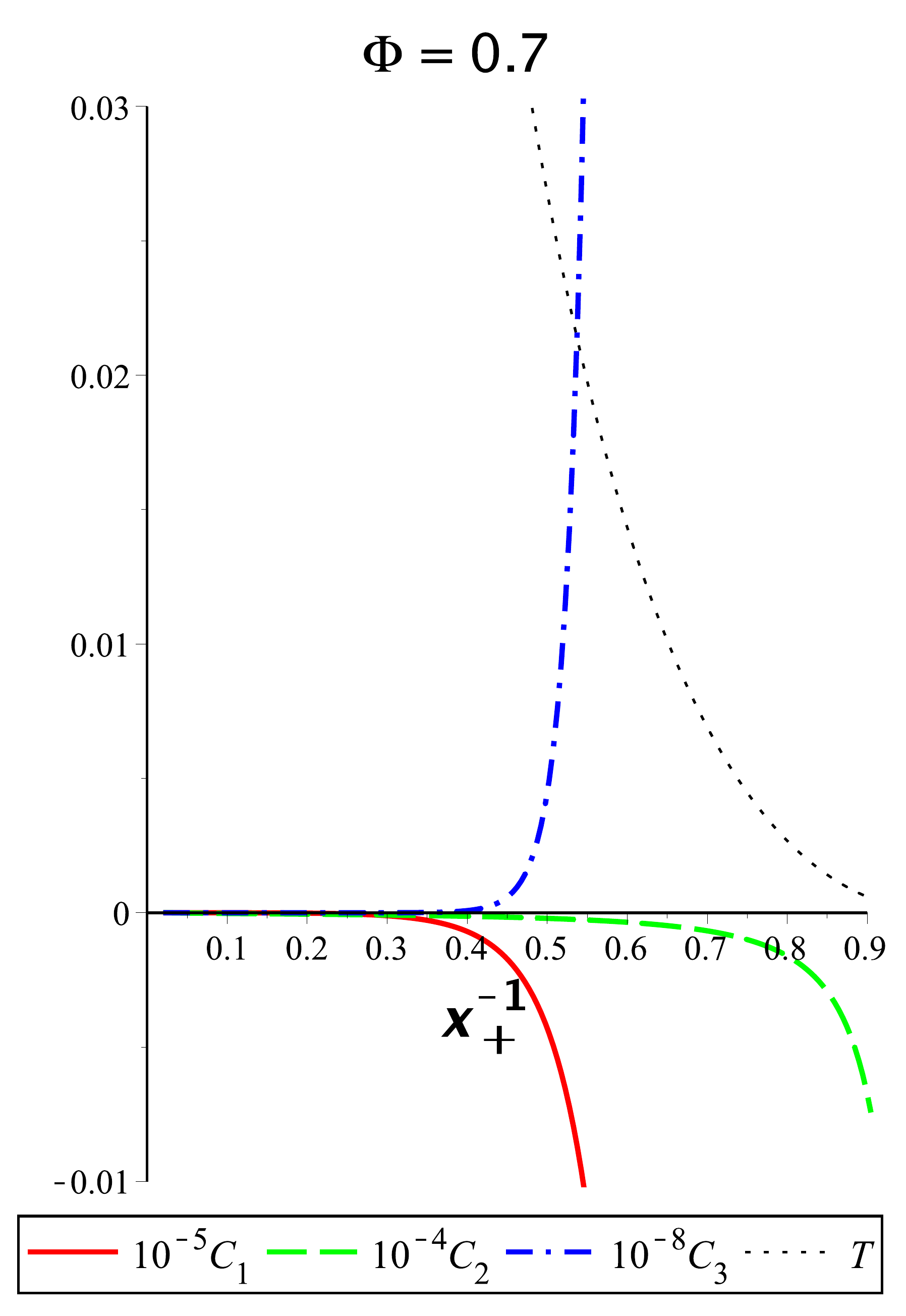}
\quad
\includegraphics[width=4.2 cm]{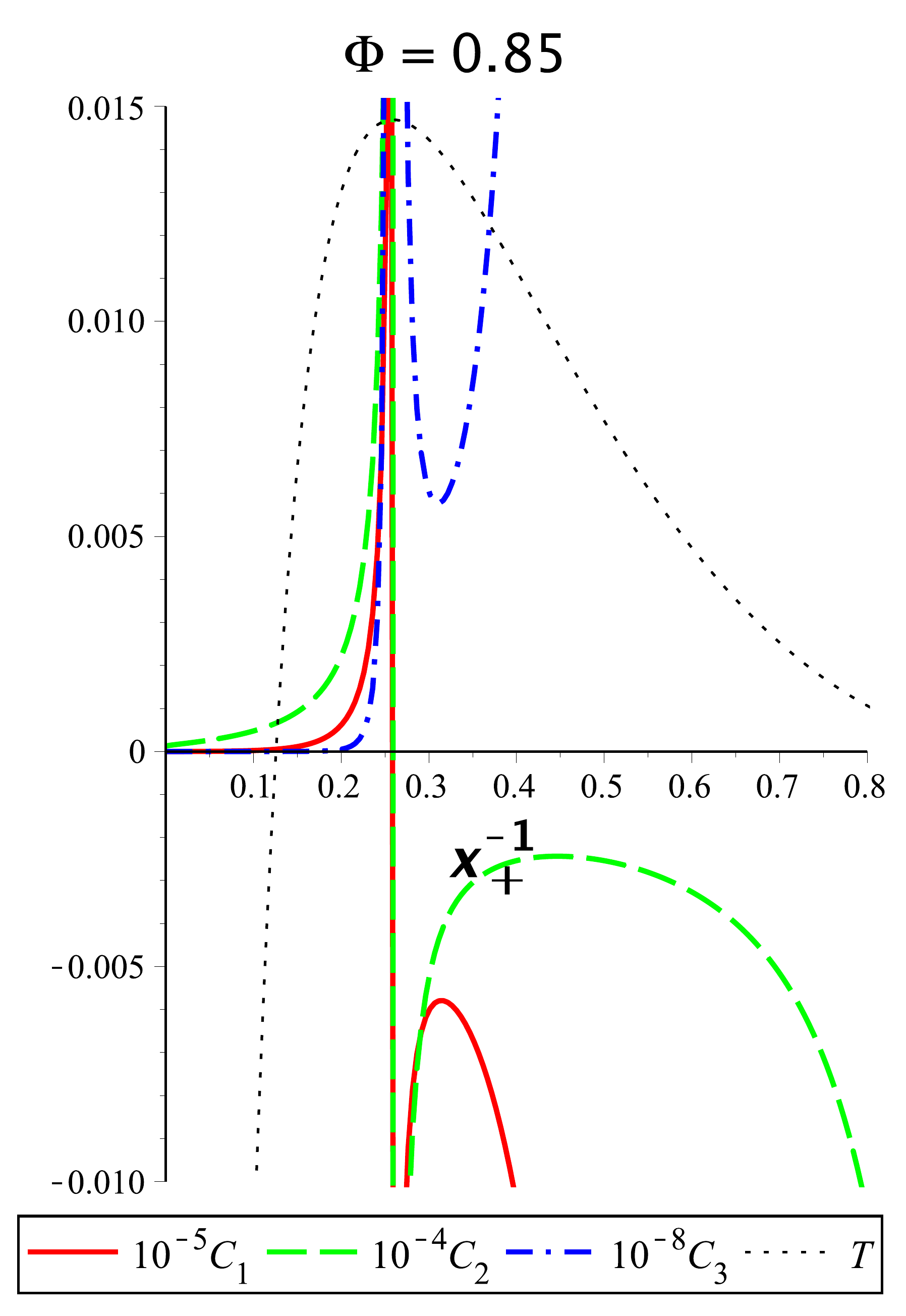}
\quad
\includegraphics[width=4.2 cm]{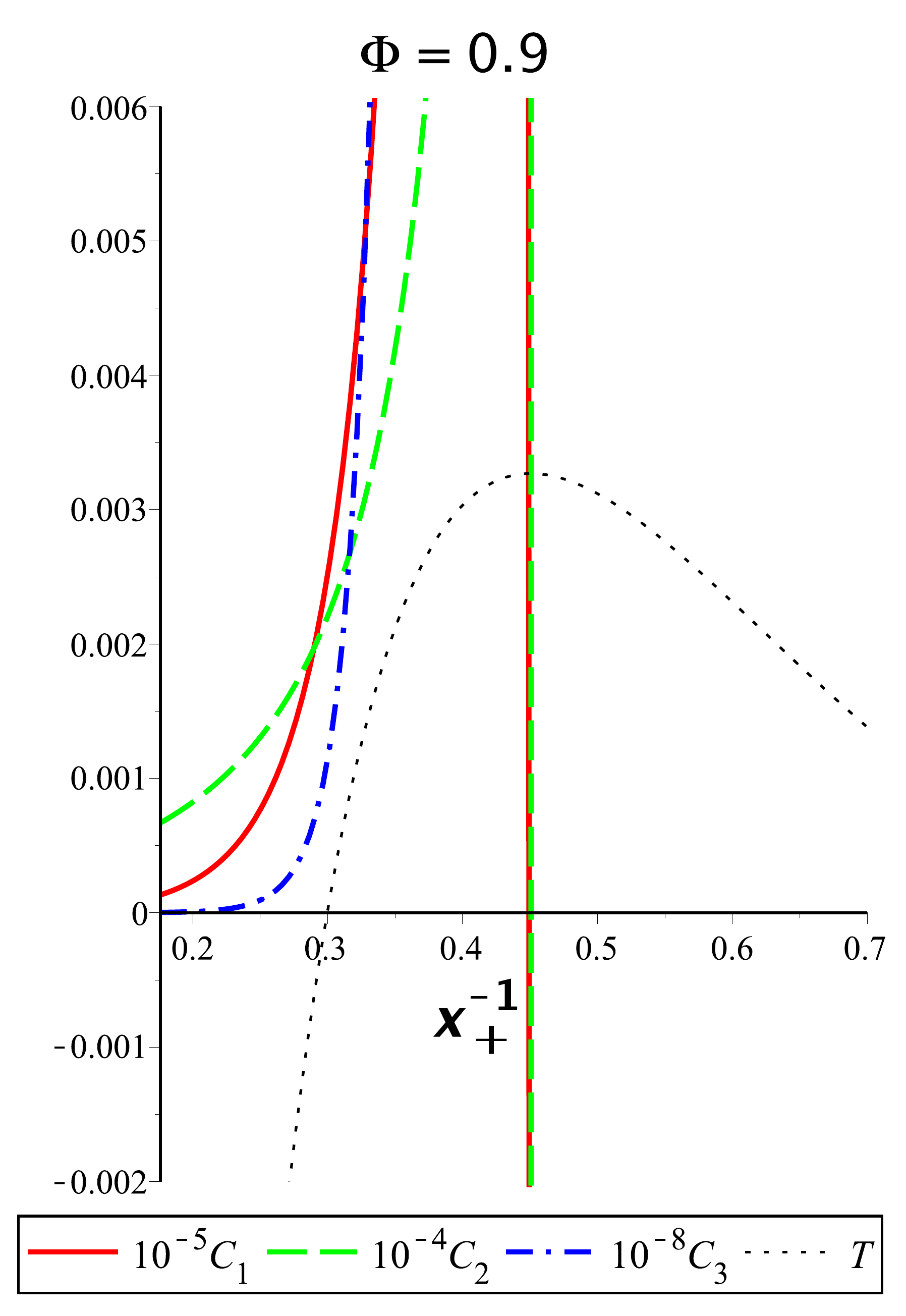}
\caption{Response functions in terms of the second derivatives of the thermodynamic potential, for $\gamma=1$ and $\alpha=10$. Three different values of the conjugate potential $\Phi$ is considered.}
	\label{resp10}
\end{figure}

Below, in Fig. \ref{resp10}, it is shown that the stability criteria hold for a set of black hole configurations. Concretely, we plot the following quantities (up to a constant factor chosen by convenience)
\begin{align}
C_1
&:=-\(\frac{\pa^2\mathcal{G}}{\pa T^2}\)_\Phi ,\qquad
C_2
:=-\(\frac{\pa^2\mathcal{G}}{\pa\Phi^2}\)_T
\label{cc2}\\
C_3
&:=-\(\frac{\pa^2\mathcal{G}}{\pa T^2}\)_\Phi
\left\{\(\frac{\pa^2\mathcal{G}}{\pa T^2}\)_\Phi
\(\frac{\pa^2\mathcal{G}}{\pa\Phi^2}\)_T
-\[\(\frac{\pa}{\pa\Phi}\)_T
\(\frac{\pa\mathcal{G}}{\pa T}\)_\Phi\]^2\right\}
\label{cc3}
\end{align}
As commented, $C_1>0$ and $C_3>0$ are sufficient for full stability, however, for completeness, we additionally plot $C_2$ to show explicitly that everything is consistent.\footnote{Since, for stability, the second derivative of the free energy with respect to intensive variables (such as $\Phi$ and $T$) is negative.}

\begin{figure}[H]
\centering
\includegraphics[width=13 cm]{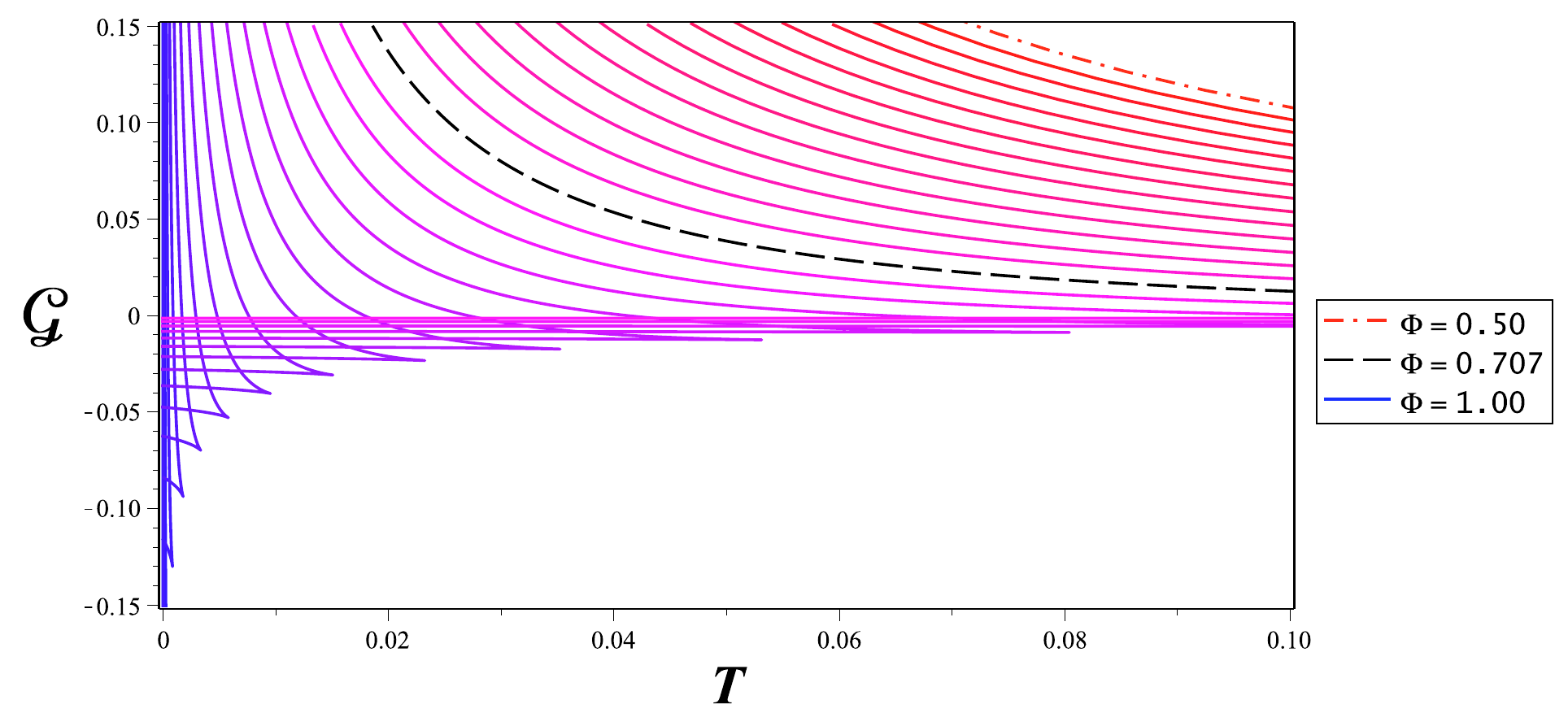}
\caption{Thermodynamic potential $\mathcal{G}$ vs $T$ in positive branch, for $\gamma=1$ and $\alpha=10$. Given a $\Phi>1/\sqrt{2}$, it develops a sector with negative concavity, that is, $C_\Phi>0$.}
	\label{GT}
\end{figure}

In Fig. \ref{resp10}, we observe that, for a given $\Phi>{1}/{\sqrt{2}}$ (the second and third graphs), $C_1$, $C_2$ and $C_3$ develop a divergence and are simultaneously positive inside the physical region ranging from $T=0$ to $T=T_{max}$ (the location of those divergences). This novel region is characterized by both $\epsilon_S>0$ and $C_\Phi>0$. In Fig. \ref{GT}, below, it can be observed that these stable black holes appear provided $\Phi>1/\sqrt{2}$ and are characterized by $\mathcal{G}<0$ and by $\left(\pa^2\mathcal{G}/\pa{T}^2\right)_\Phi<0$, as expected for $C_\Phi>0$. We would like to emphasize that, in the case of RN black hole, there are no physical configuration with positive heat capacity at $\Phi$ fixed.

A detailed discussion on the physical interpretation of these results and comparison with the AdS stable black hole is going to be presented in Discussion, section \ref{disc}.

\subsubsection{Canonical ensemble}
\label{cpb}
\begin{figure}[H]
	\centering
	\includegraphics[width=5.6 cm]{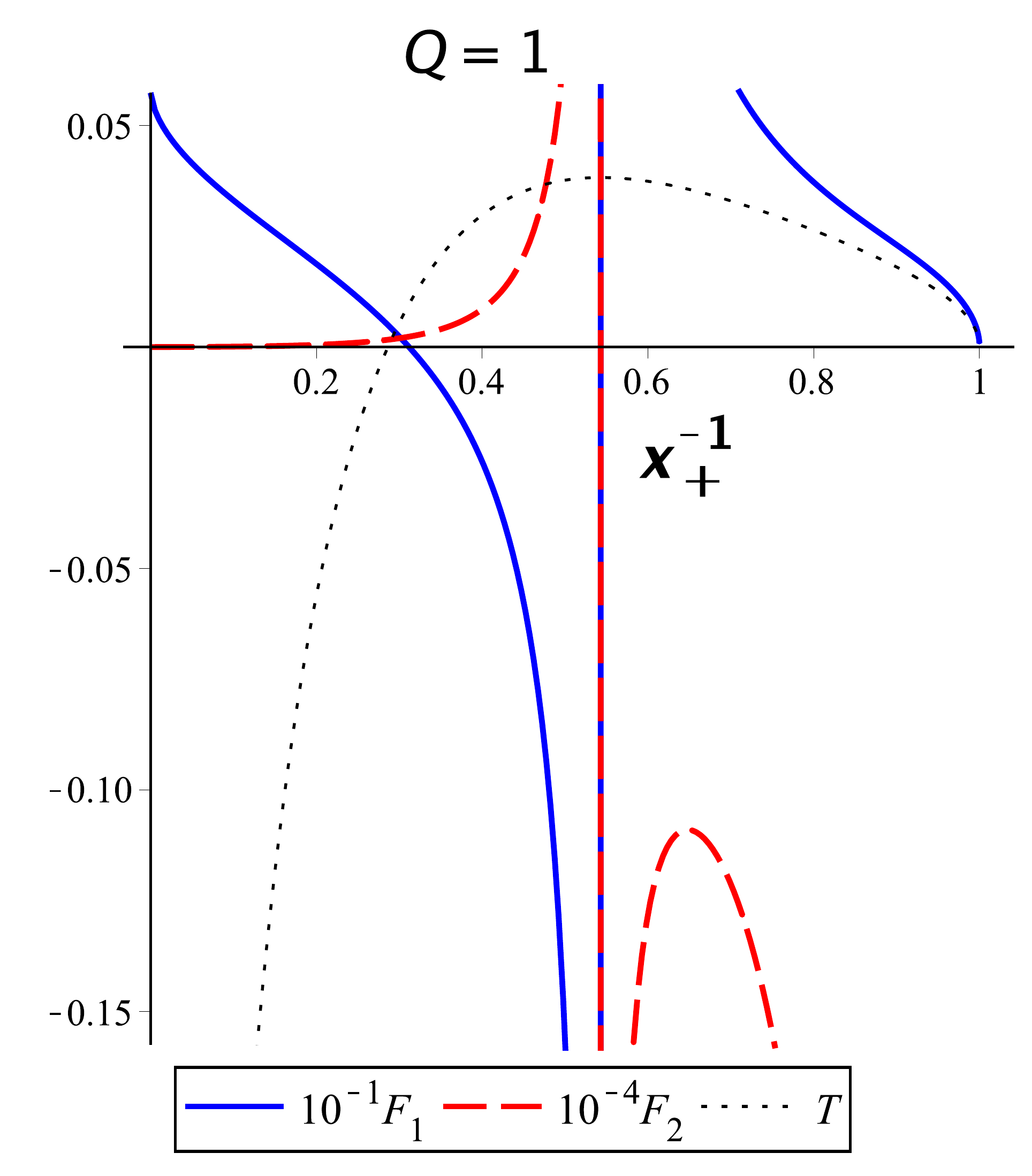} \qquad
	\includegraphics[width=5.6 cm]{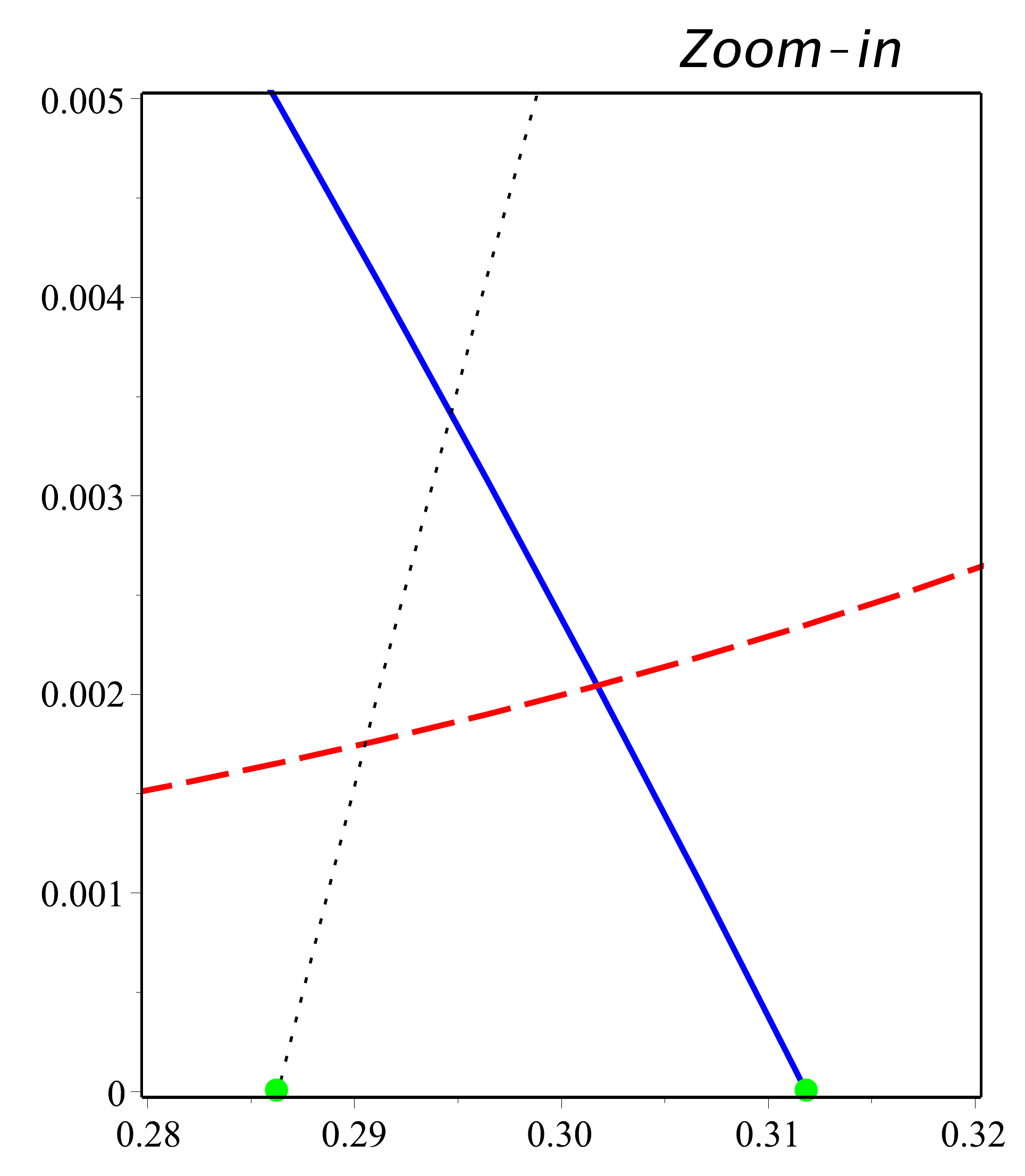}
	\caption{Second derivatives of the thermodynamic potential, given by $F_1$ and $F_2$, for $\gamma=1$, $\alpha=10$ and the charge was fixed to $Q=1$. Black dotted line represents temperature. The region for thermodynamic stability starts at $T=0$ and finishes at the zero of $F_1$ ($\epsilon_T=0$).}
	\label{resp4}
\end{figure}

\begin{figure}[H]
	\centering
	\includegraphics[width=5.6 cm]{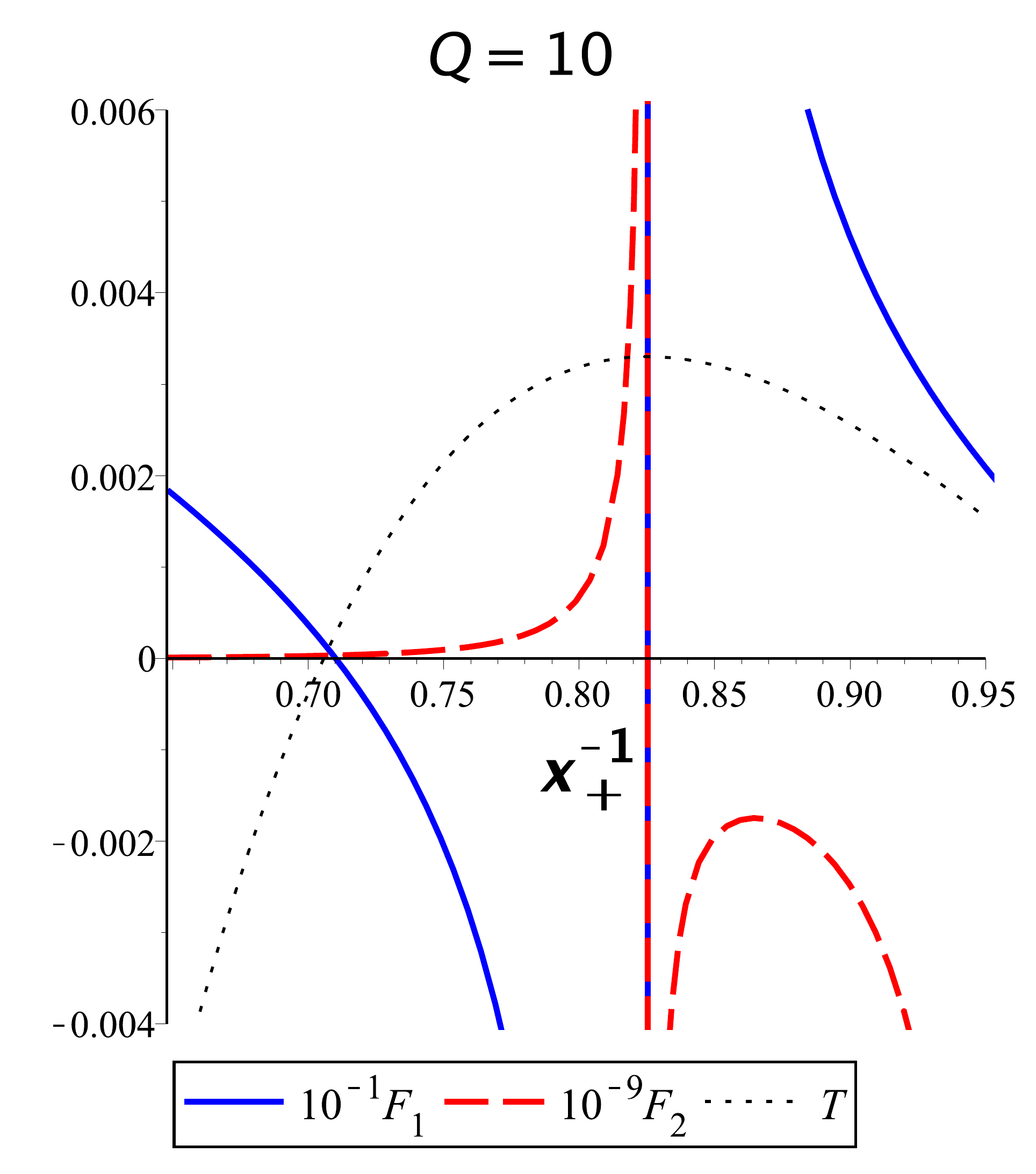} \qquad
	\includegraphics[width=5.6 cm]{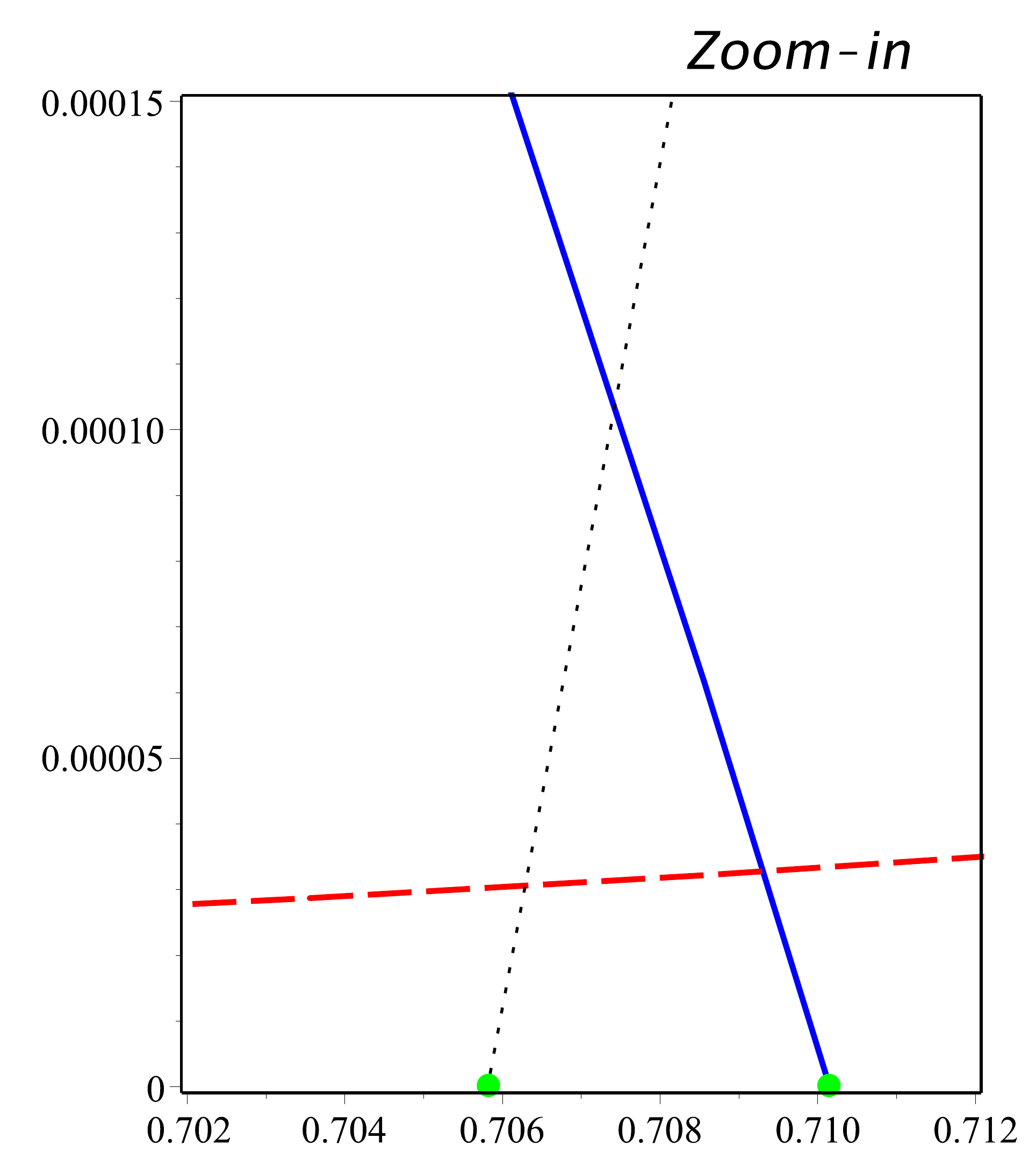}
	\caption{Second derivatives of the thermodynamic potential for $\gamma=1$, $\alpha=10$ and $Q=10$. The region for thermodynamic stability exists for any $Q$.}
	\label{resp44}
\end{figure}

\begin{figure}[H]
	\centering
	\includegraphics[width=11 cm]{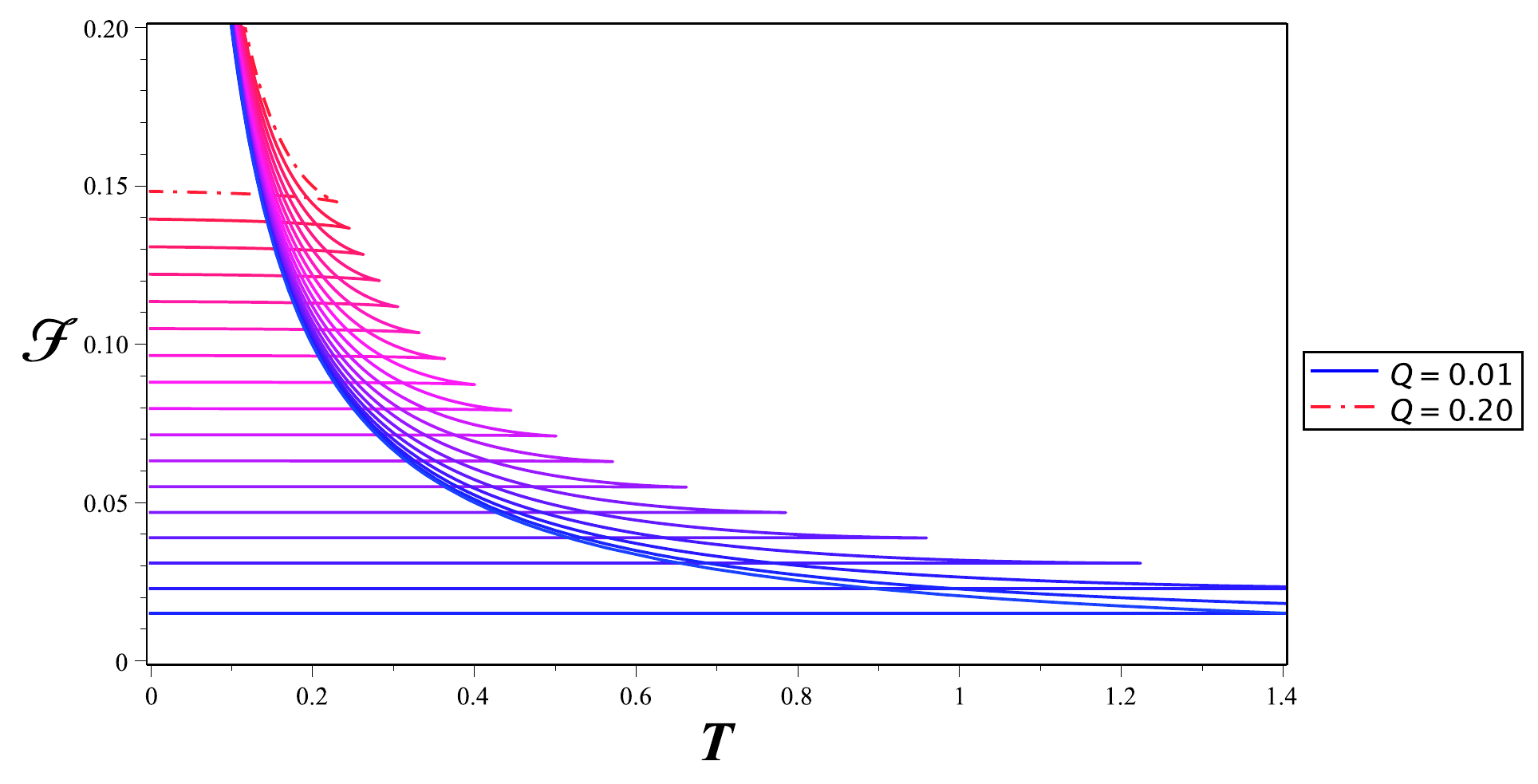}
	\caption{Thermodynamic potential $\mathcal{F}$ vs $T$, for $\gamma=1$ and $\alpha=10$. The sector with negative concavity (that is, $C_Q>0$) exists for any $Q$.}
	\label{FT}
\end{figure}

Let us now present a similar analysis for the canonical ensemble. In order to make explicit the dependence of the relevant thermodynamic quantities on $Q$, we are going to use the equation $q=-\eta Q$ to eliminate $q$ from the horizon equation $f(x_+)=0$ and then to solve for the positive root $\eta=\eta(x_+,Q)$.
%
%
This allows to write down the following parametric expressions:
\begin{align}
M&=\frac{12\eta^4Q^2-\alpha}{12\eta^3} ,\quad 
\Phi=\frac{\eta\,(x_{+}-1)Q}{x_+},\quad
S=\frac{\pi x_+}{\eta^2(x_{+}-1)^2}, \\
T&=\frac{\(x_{+}-1\)^2}{8\pi\eta x_+}
\(-\alpha-2\eta^2\,\frac{x_{+}+1}{x_{+}-1}
+4\eta^4Q^2\,\frac{x_{+}+2}{x_+}\)
\end{align}
Consequently, the corresponding thermodynamic potential can be parametrically expressed in the following compact form:
\begin{equation}
\mathcal{F}(x_+,Q)=
\frac{\alpha}{24\eta^3}
+Q^2\eta\,\frac{x_{+}-2}{2x_{+}}
+\frac{1}{4\eta}\,\frac{x_{+}+1}{x_{+}-1}
\end{equation}
To investigate the local thermodynamic stability, we shall verify the inequalities $\epsilon_T>0$ and $C_Q>0$, which, in terms of the second derivative of the $\mathcal{F}$, are equivalent to the conditions
\begin{equation}
F_1:=(\partial^2\mathcal{F}/\partial Q^2)_T>0,
\qquad
F_2:=-(\partial^2\mathcal{F}/\pa T^2)_Q>0
\end{equation}
respectively.\footnote{The third condition for stability against mixed fluctuations, $\epsilon_TC_Q>0$, follows from $\epsilon_T>0$ and $C_Q>0$ and thus does not need to be imposed as a independent condition.} The graphics  on the right hand side of the Figs. \ref{resp4} and \ref{resp44} show a zoomed-in region in parameter spacer for which the black holes are stable. The thermdoynamic stability occurs in the sector $\epsilon_T>0$ located in the for which $\frac{1}{\sqrt{2}}<\Phi$ (but below the corresponding value for the extremal black hole, otherwise it is a naked singularity), as can be seen from the equation of state depicted in Fig. \ref{state_hairy1} and, therefore, in agreement with our results in grand canonical ensemble. 
In Fig \ref{FT}, it was depicted the thermodynamic potential as a function of temperature, indicating that there are black holes with $C_Q>0$ for any $Q$. This is, again, consistent with the fact that, when $\frac{1}{\sqrt{2}}<\Phi$, it includes configurations for all $Q$ (Fig. \ref{state_hairy1}). However, the stable black hole, as can be seen, exists provided both $\epsilon_T>0$ and $C_Q>0$, therefore, not all the black holes with $C_Q>0$ in Fig. \ref{FT} are thermodynamically stable.

\subsection{Negative branch with $\gamma=1$ and $\alpha\neq 0$}
We turn now to the negative branch and investigate the local thermodynamic stability when $\phi<0$ or, equivalently, $x\in (0,1)$ and with $\gamma=1$ and $\alpha>0$, presented in section (\ref{sec:sols}). We show, without presenting all the  details (the steps are basically the same like for the analysis of the positive branch), that no thermodynamically stable black holes exist in this case.

\subsubsection{Grand canonical ensemble}
\label{gcnb}
Below, in Fig. \ref{state1} it is shown the equation of state $Q$---$\Phi$ at $T$ and $S$ fixed, respectively. Compared with the positive branch, the upper part of the graphic does not contain a region where $\epsilon_T>0$. 
\begin{figure}[H]
	\centering
	\includegraphics[width=7 cm]{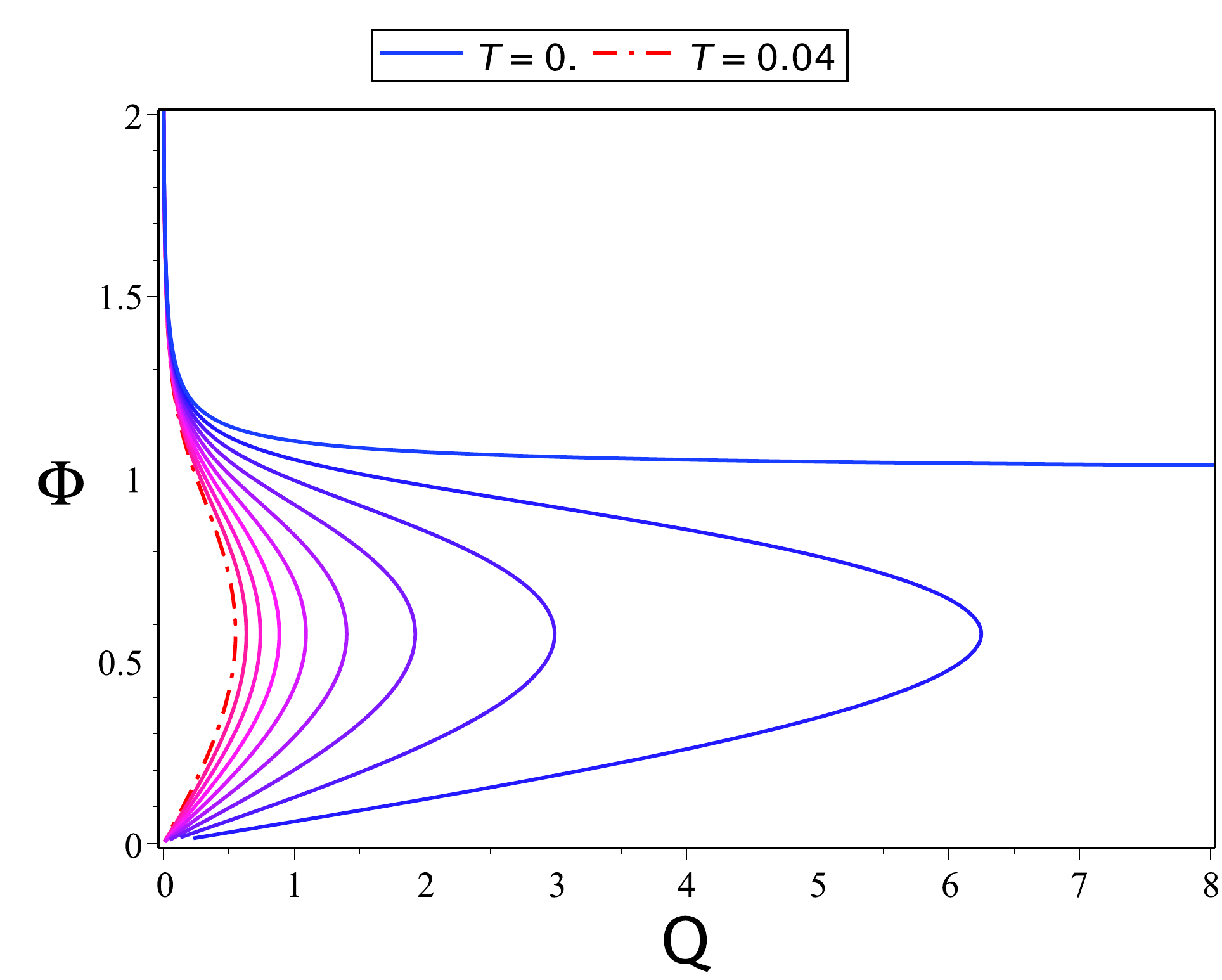}\quad
	\includegraphics[width=7 cm]{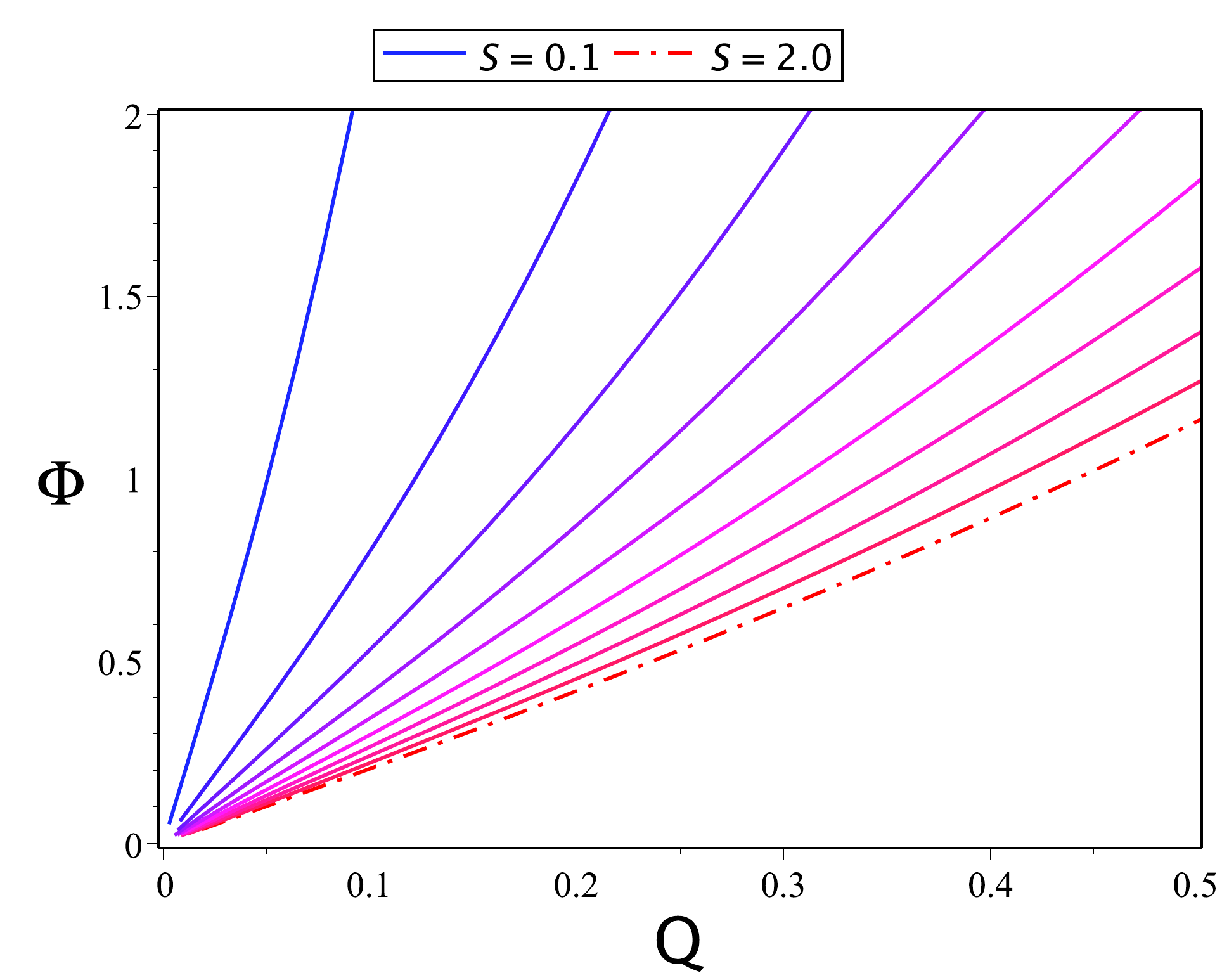}
	\caption{Left Hand Side: Equation of state in negative branch, for $\gamma=1$ and $\alpha=10$. Right Hand Side: isentropic curves $\Phi$---$Q$.}
	\label{state1}
\end{figure}

\begin{figure}[H]
	\centering
	\includegraphics[width=5 cm]{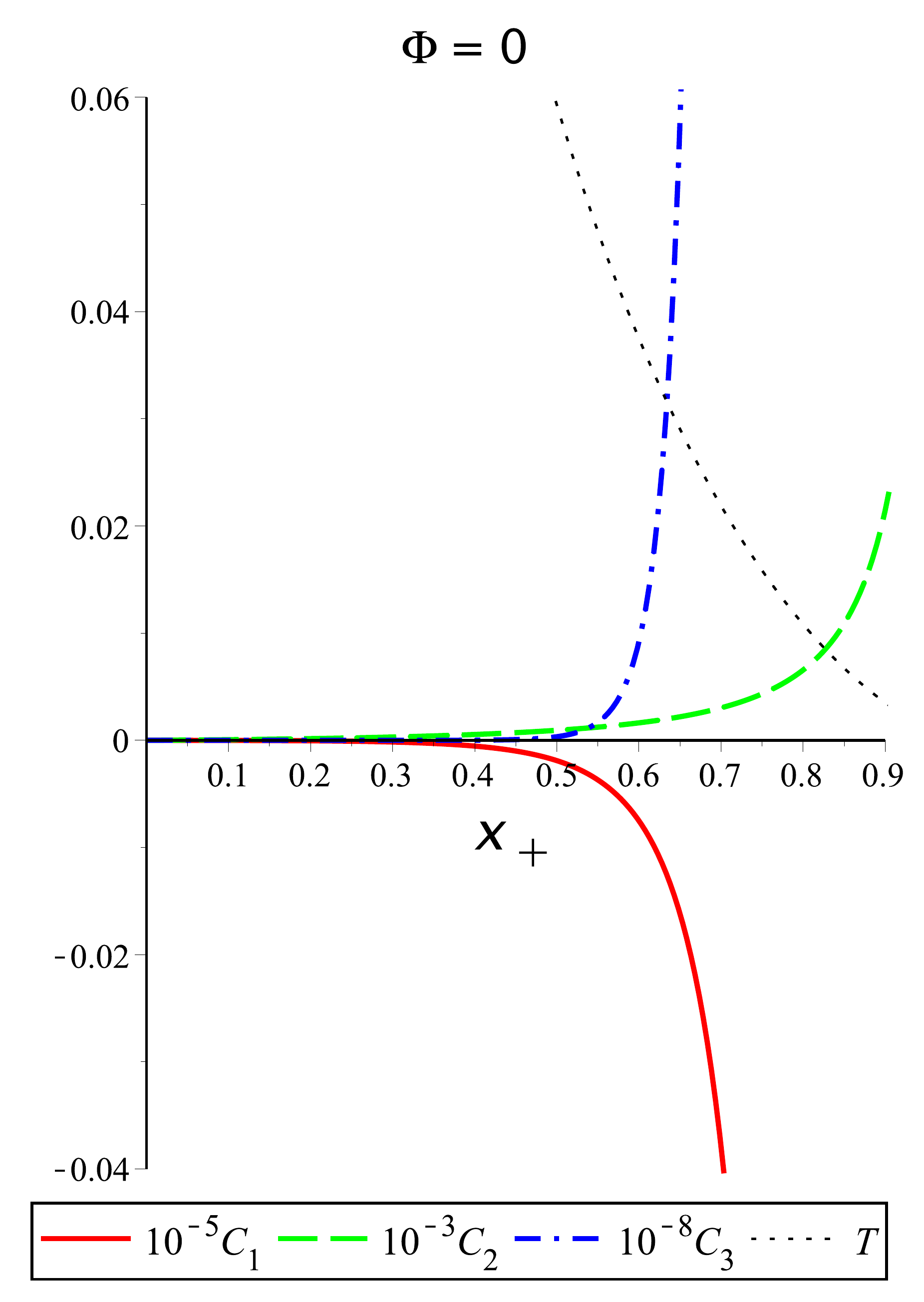}
	\includegraphics[width=5 cm]{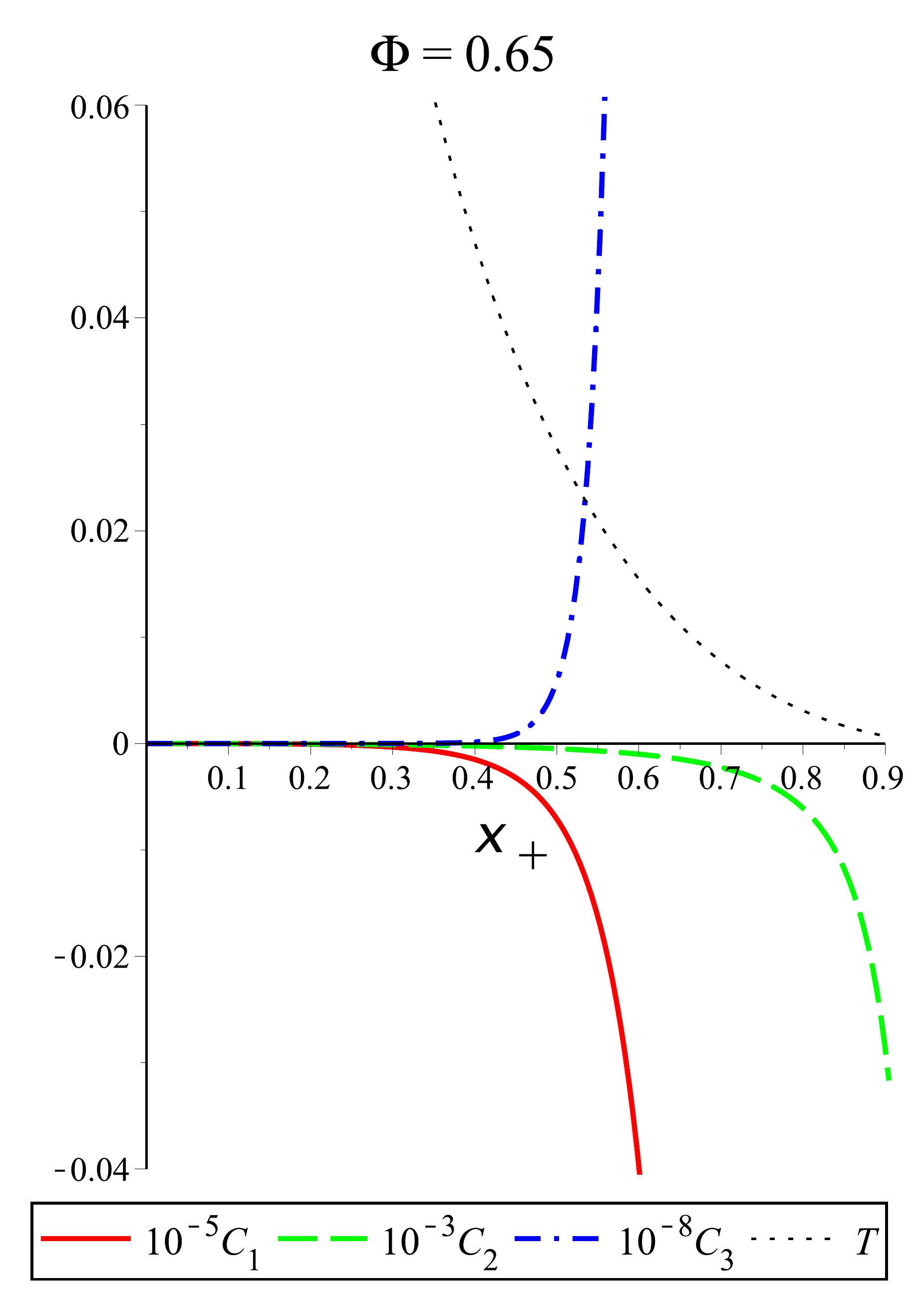}
	\includegraphics[width=5 cm]{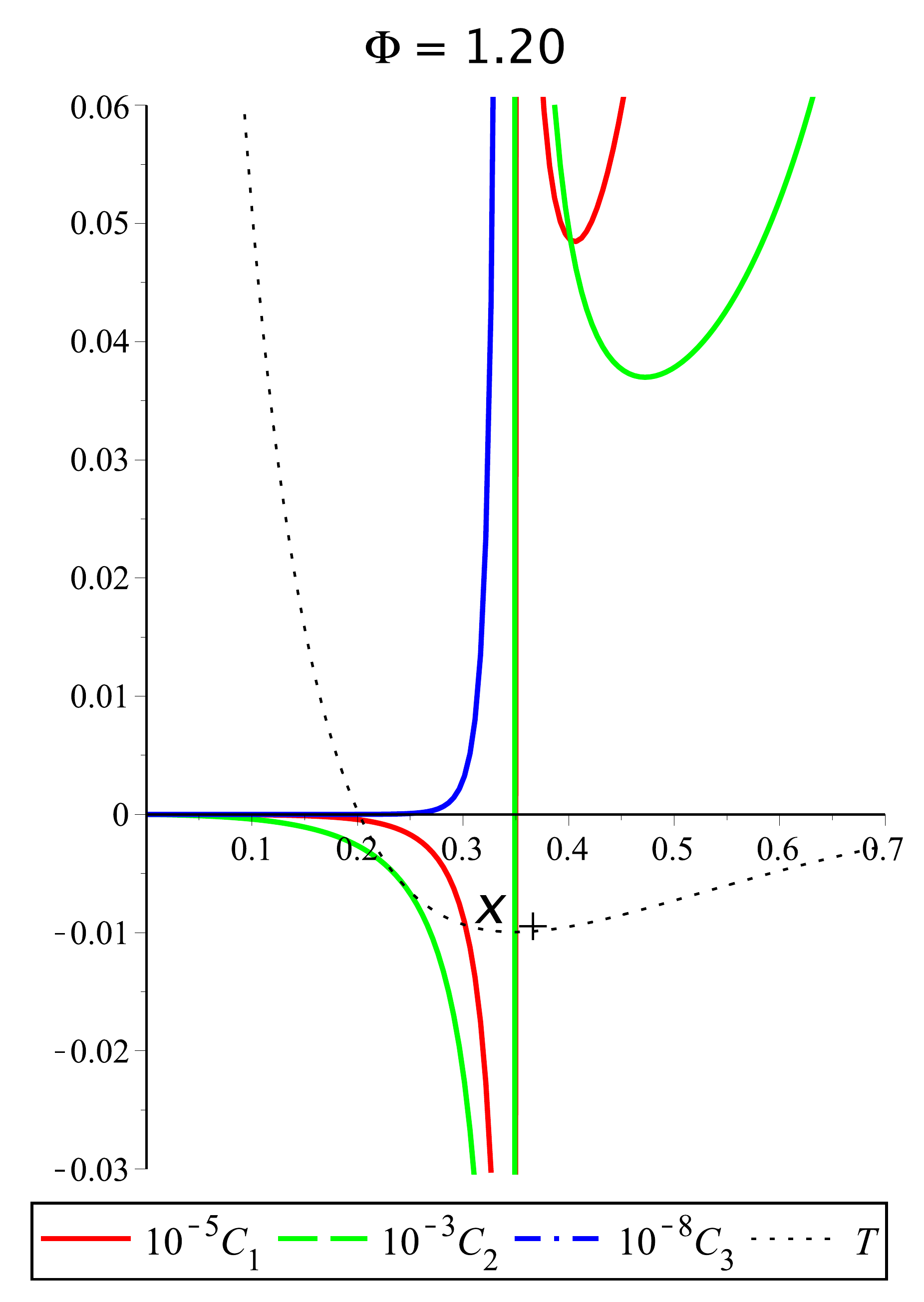}
	\caption{Response functions in terms of derivatives of $\mathcal{G}$, for negative branch in grand canonical ensemble, $\gamma=1$ and $\alpha=10$. Black dotted line represents temperature, the red curve represents $10^{-5}C_1$, the green one represents $10^{-4}C_2$ and blue curve $10^{-8}C_3$.}
	\label{resp1}
\end{figure}

The equation of state is rather similar with the one of RN black hole, in the sense that there is only one region with $\epsilon_T>0$ corresponding to the lower part of the graphic. On the other hand, one can explicitly show that, as in the previous section for positive branch, the permittivity at $S=const.$ is always positive, $\epsilon_S>0$. It turns out that the response functions do not share positive values in any physical region, that is, $C_\Phi<0$, as shown in Fig. \ref{resp1} --- we use the same conventions as in the case of the grand canonical in positive branch, namely the definitions (\ref{cc2}) and (\ref{cc3}).
Despite $\epsilon_S>0$ (blue curve), as we know from isentropic curves in Fig. \ref{state1}, there is no physical region where $C_\Phi>0$. Therefore, no thermodynamically stable black holes are found in negative branch. Notice that for $\Phi\approx 1/\sqrt{3}$, $\epsilon_T$ becomes negative, as clearly seen from equation of state in Fig. \ref{state1}. 

\begin{figure}[H]
	\centering
	\includegraphics[width=10 cm]{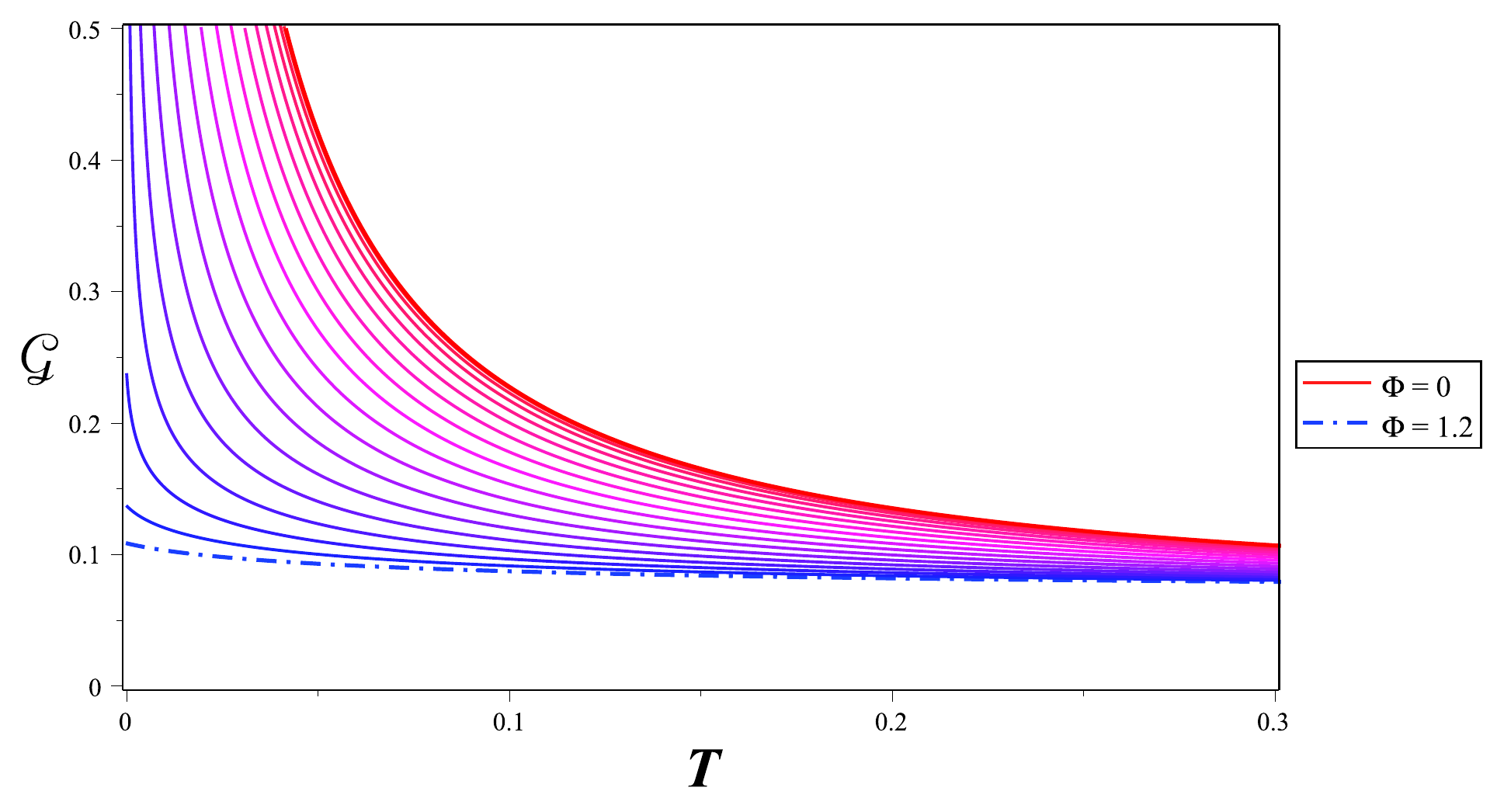}
	\caption{The concavity of free energy is positive definite and thus the heat capacity in grand canonical ensemble is negative, for $\gamma=1$ and $\alpha=10$.}
	\label{gtcan}
\end{figure}

The fact that $C_\Phi$ is always negative (in physical region) can be observed in the plot of the free energy vs temperature, in Fig. \ref{gtcan}.

\subsubsection{Canonical ensemble}
\label{cnb}

\begin{figure}[H]
	\centering
	\includegraphics[width=6 cm]{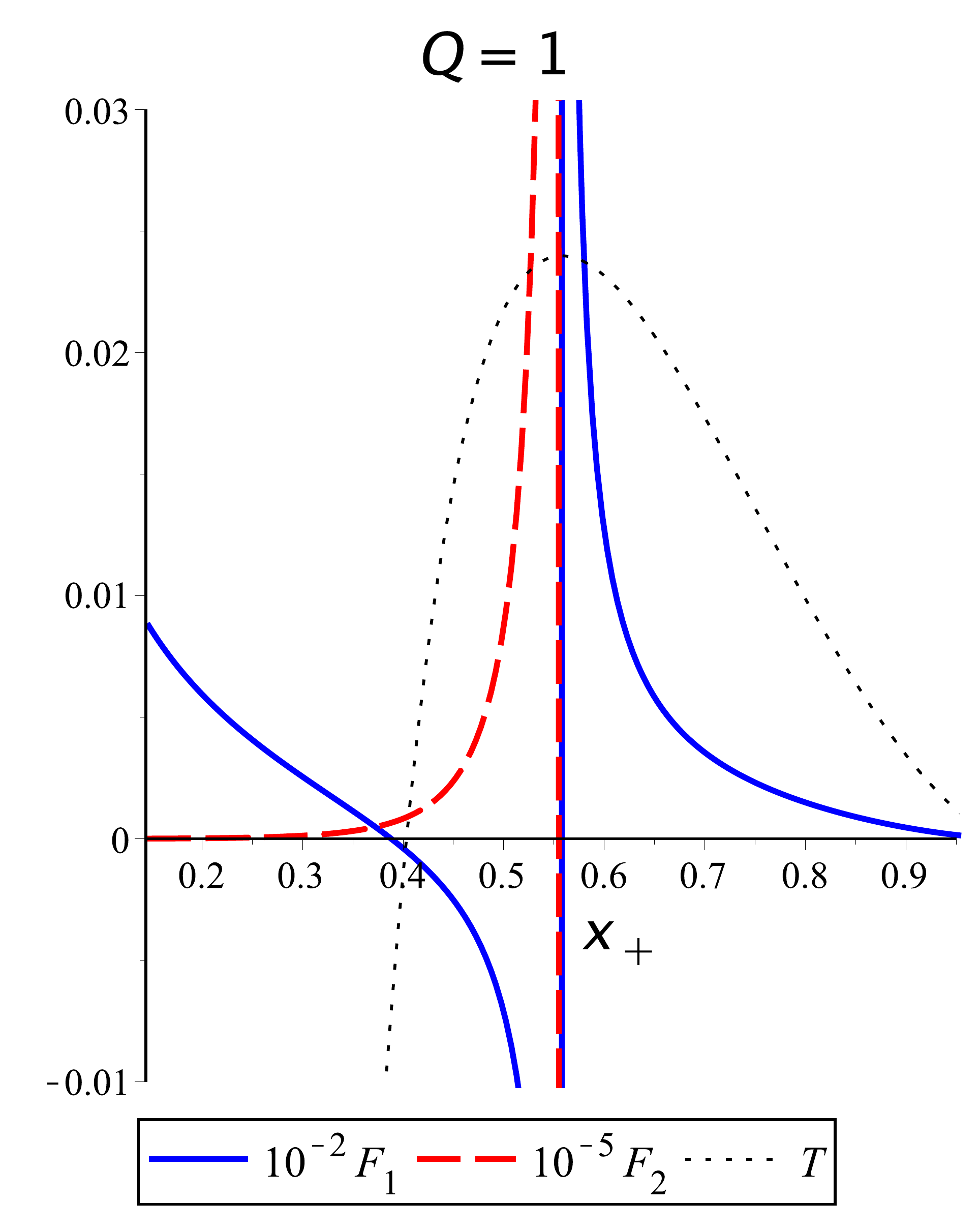} \quad
	\includegraphics[width=6 cm]{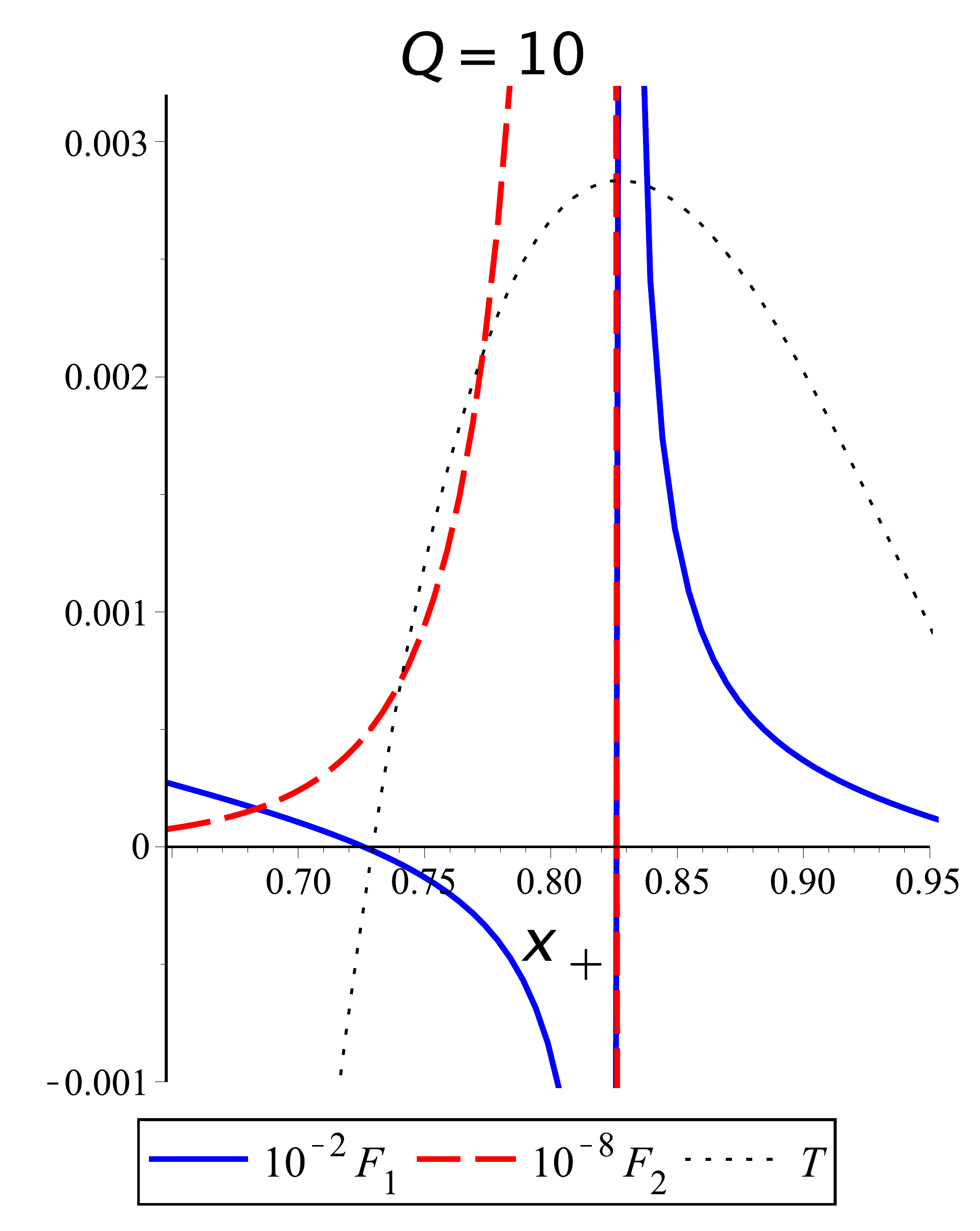}
	\caption{Second derivatives of thermodynamic potential in negative branch in canonical ensemble, for $\gamma=1$ and $\alpha=10$. Black dotted line represents the temperature; blue one represents $F_1:=(\pa^2\mathcal{F}/\pa Q^2)_T$ and the red one $F_2:=-(\pa^2\mathcal{F}/\pa T^2)_Q$. The simultaneous positivity of $F_1$ and $F_2$ indicates thermodynamic stability, but in negative branch they have opposite sign when $T>0$.}
	\label{resp3}
\end{figure}

The response functions are depicted in Fig. \ref{resp3}, where we observe that, in agreement with the grand canonical in negative branch, there is no physical region where simultaneously $\epsilon_T>0$ and $C_Q>0$. Instead, the product $C_Q\epsilon_T$ is negative inside for $T>0$.

As expected from the analysis in grand canonical ensemble, there are no thermodynamically stable black holes in canonical ensemble for the negative branch; the zero of $(\pa^2\mathcal{F}/\pa Q^2)_T$ is located in the $T<0$ region or, in other words, the extremal black hole is unstable and, thus, $\epsilon_T$ does not take positive values inside the region where $C_Q>0$.


\section{Discussion}
\label{disc}
Scalar fields play a central role in cosmology and particle physics and arise naturally
in the high energy physics unification theories. It is then important to understand  generic properties of gravity theories coupled to scalars (and other matter fields),  particularly the role played by the scalars to black hole physics.\footnote{Some recent interesting applications can be found in, e.g., \cite{Hirschmann:2017psw,Jai-akson:2017ldo,Cardenas:2017chu,McCarthy:2018zze,Brihaye:2018woc, Bzowski:2018aiq}.} In  this  work,  we  have  considered  thermodynamic properties of a family of exact asymptotically flat hairy black holes, with the goal of shedding light on their thermodynamic stability. In our investigations, we have been directly motivated by the results of \cite{Nucamendi:1995ex}, where it was conjectured the existence of such black hole solutions in theories with a non-trivial scalar potential that vanishes at the boundary\footnote{The scalar potential is blowing up at the singularity, $x=\{0,\infty\}$, but this should not be a matter of concern because the singularity is shielded by a horizon. At the singularity one expects that the quantum gravity effects become important and so the theory we have considered should be interpreted as an effective theory.} and \cite{Anabalon:2013qua} where exact regular hairy black hole solutions were obtained. 

Due to its intimate connection with the partition function, the Euclidean path integral formalism of quantum gravity is widely used when studying black hole thermodynamics. We have presented a complete analysis using the boundary terms required in the action functional of general relativity and prove that, for some values of the parameters, these black holes are thermodynamically stable in both, canonical and grand canonical ensembles. This result could come as a surprise because generally, in flat spacetime of different dimensions, it is known that the black holes/rings are not thermodynamically stable \cite{Hawking:1976de,Monteiro:2009tc, Dias:2010eu, Astefanesei:2010bm}.\footnote{However, there exist examples of thermodynamically stable black holes in theories with higher derivative terms \cite{Myers:1988ze, Bueno:2016lrh, Bueno:2017qce}.} It is possible to make a black hole thermodynamically stable by introducing a negative cosmological constant $\Lambda$ and considering asymptotically AdS black holes or they can also be stabilized by putting them in a finite volume cavity. However, when scalar fields are present in the theory, it seems that the existence of the scalar field self-interaction is the key ingredient for the thermodynamic stability.

It is known that, when the dilaton potential vanishes, one can also vary the asymptotic value of the scalar field, $\phi_\infty$. In this case, it was claimed that the first law of thermodynamics should be modified by adding a contribution coming from the scalar charge \cite{Gibbons:1996af} that can be explicitly verified for the exact hairy black hole solutions \cite{Gibbons:1987ps,Strominger:1991,Kallosh:1992ii}. Recently in \cite{Astefanesei:2018vga}, by considering the correct variational principle, it was shown that the quasilocal energy does not match the $ADM$ mass (obtained from expanding the $g_{tt}$ component of the metric). Once the correct definition of (quasilocal) gravitational energy is used, it was shown in \cite{Astefanesei:2018vga} that the first law preserves its usual form without including the extra contribution coming from the variation of the asymptotic value of the dilaton (see, also, \cite{Astefanesei:2006sy, Hajian:2016iyp}).  We should contrast this case with the hairy black holes in theories with a dilaton potential. Particularly, to obtain an asymptotically flat spacetime there should be imposed an important constraint on the scalar potential, namely to vanish at the boundary.  That is, the asymptotic value of the dilaton should be fixed, otherwise its variation is going to change the asymptotics of the spacetime. Therefore, the issue of the appearance of the scalar charge in the first law does not appear in our case. 

In the work of Brown and York \cite{Brown:1992br}, it was shown that the quasilocal stress tensor is covariantly conserved only if the asymptotic fall-off of the matter fields is fast enough, which is also our case: at the boundary when $x=1$, the scalar field is $\phi\rightarrow 0$, which implies that the the potential vanishes. For example, the hairy black hole in the theory with $\gamma=\sqrt{3}$ has the following quasilocal stress tensor
\begin{align}
\tau_{tt}
&=-\frac{8\alpha+3\(1-2q^{2}\)\eta^{2}}{3\eta\kappa}\,(x-1)^{2}
+\mathcal{O}\[(x-1)^{3}\] \\
\tau_{\theta\theta}
&=\frac{\tau_{\phi\phi}}{\sin^2\theta}
=\frac{\[8\alpha+3\(1-2q^{2}\)\eta^2\]^{2}-9\eta^{4}(4q^{2}-3)}{72\kappa\eta^{5}}\,(x-1)+\mathcal{O}[(x-1)^{2}] 
\end{align}
which is, indeed, covariantly conserved. This contrasts with the situation when the solution is not regular due to the presence of conical singularity in the boundary \cite{Astefanesei:2009mc}. 

Another important observation is about the ground state of the theory. Remarkably, the counterterm method provides a regularization method that yields an intrinsic
definition of the action without the necessity of using a reference background. However, it is important to obtain the soliton-like solutions (at zero temperature). We leave a detailed analysis of this point for future work, but, as in the case of charged black holes in AdS \cite{Chamblin:1999tk}, one can, in principle, consider the existence of the extremal black hole in the canonical ensemble (at fixed $Q$) as the reference background. The problem that appears for hairy black holes is that the extremal limit is not always well defined like in the case of RN black hole. This is related to the attractor mechanism \cite{Ferrara:1995ih, Strominger:1996kf, Ferrara:1996dd} and we would like to comment now on this subtle aspect of the theory. There are two different methods to study the near horizon data of an extremal black hole, the effective potential \cite{Goldstein:2005hq} and entropy function method \cite{Sen:2005wa, Astefanesei:2006dd}. When the dilaton potential vanishes, in theories with one electric field, the effective potential can not have an extremum at the horizon, which indicates that the extremal black hole does not exist. This can be also obtained directly by computing the geometric invariants at the inner horizon and prove that some of them are blowing up. However, there is a drastic change when the dilaton potential is non-trivial. That is, due to the competition between the effective potential and dilaton potential, there could exist a well defined extremal limit (on the Lorentzian section) for some values of the parameters of the dilaton potential. In this case the effective potential method fails to work, but, instead, one can use the entropy function formalism. An analysis of the case we are interested in was done in \cite{Anabalon:2013qua} (see, also, \cite{Anabalon:2013sra}) and, indeed, since there exist regular extremal hairy black hole solutions and so the canonical ensemble is well defined.

\begin{figure}[H]
	\centering
	\includegraphics[width=7 cm]{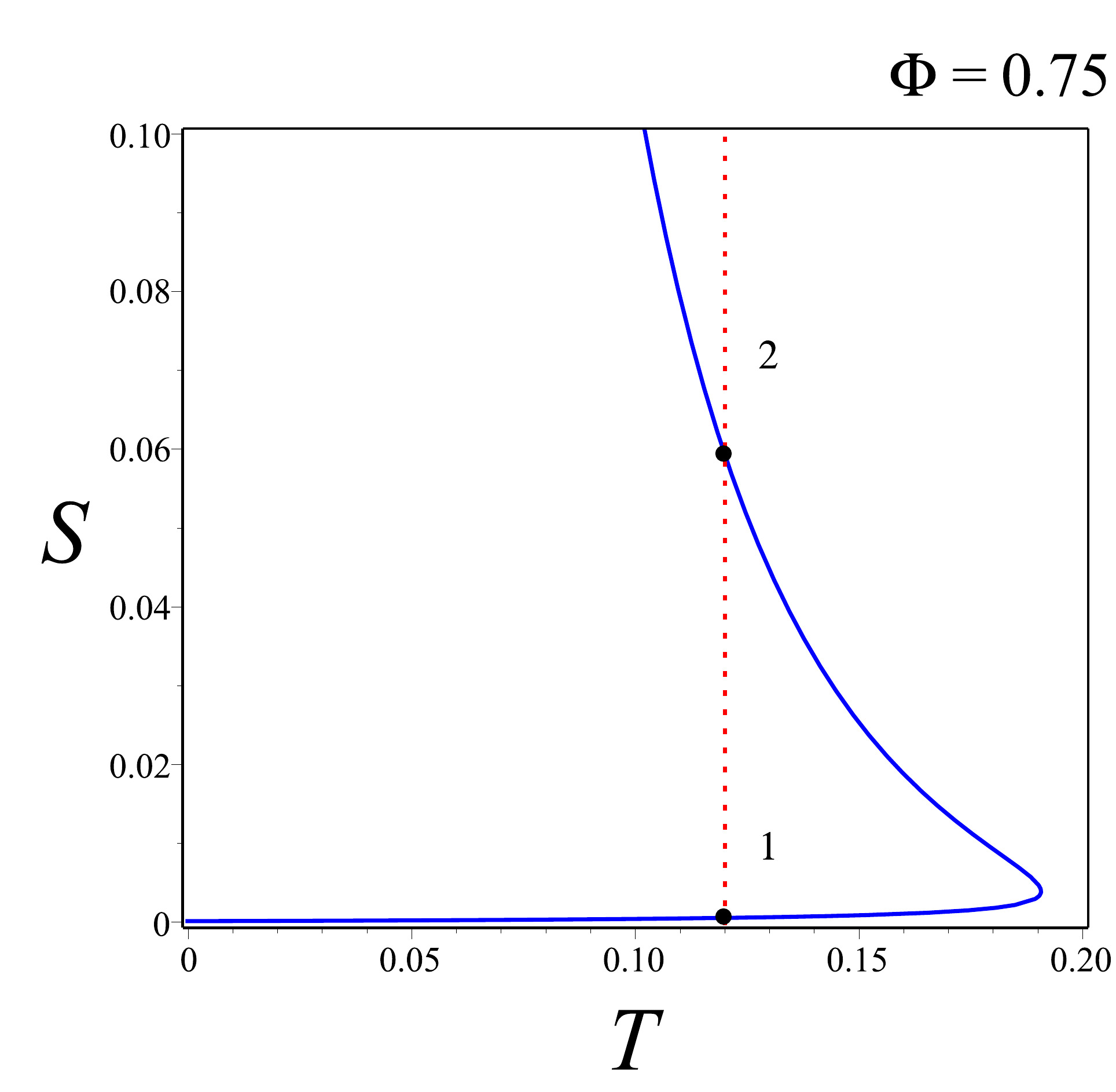}\quad
	\includegraphics[width=7 cm]{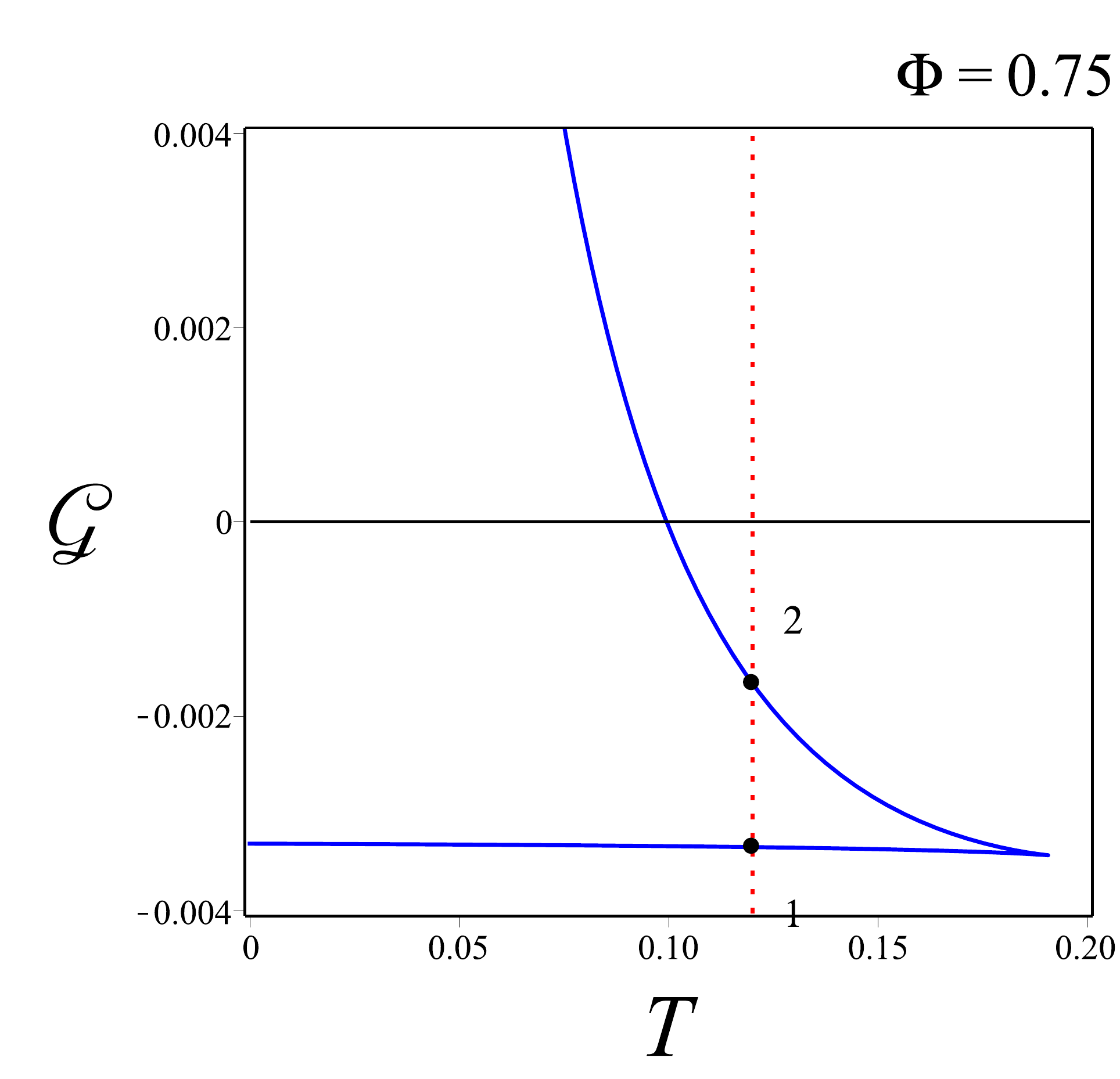}
	\caption{Plot for $\Phi=0.75$ and $T=0.06$ (vertical red doted line) to illustrate which configuration is stable. It turns out that the configuration $1$, as indicated on the plot, is the stable one.}
	\label{stable1}
\end{figure}

Now, we would like to discuss in more detail our main result from section (\ref{main}). Let us compare the thermodynamically stable black holes in flat space, which exist for $\frac{1}{\sqrt{2}}<\Phi$ (the value of the conjugate potential is smaller than $1$ for the extremal hairy black holes, but when$\Phi \rightarrow 1$ the extremal RN is recovered ), with the stable black holes in AdS. In Fig. \ref{stable1}, we plot $S$ vs $T$ and $\mathcal{G}$ vs $T$ for asymptotically flat hairy black holes in order to identify, for a given $T$, which thermodynamic configuration is preferred. 
From the first plot in Fig. \ref{stable1}, we observe that stable black holes, for which $C_\Phi=T(\pa S/\pa T)_\Phi>0$, correspond to the configuration indicated as $1$. This black hole has less entropy than the other one at the same temperature, thus, since
$
S=-({\pa\mathcal{G}}/{\pa T})_\Phi
$,
this can be identified in the second plot as the one with lower absolute value of the slope (indicated as 1, too). Another way to understand this is by investigating the second derivative of the thermodynamic potential. Since
\begin{equation}
\label{secderiv}
\(\frac{\pa S}{\pa T}\)_\Phi=-\(\frac{\pa^2\mathcal{G}}{\pa T^2}\)_\Phi
\end{equation}
the stable black hole, for which $(\pa S/\pa T)_\Phi>0$, should appear in the first plot as $({\pa^2\mathcal{G}}/{\pa T^2})_\Phi<0$, which corresponds to the configuration 1, because it has negative concavity. The identification has now been completed. 

\begin{figure}[H]
	\centering
	\includegraphics[width=7 cm]{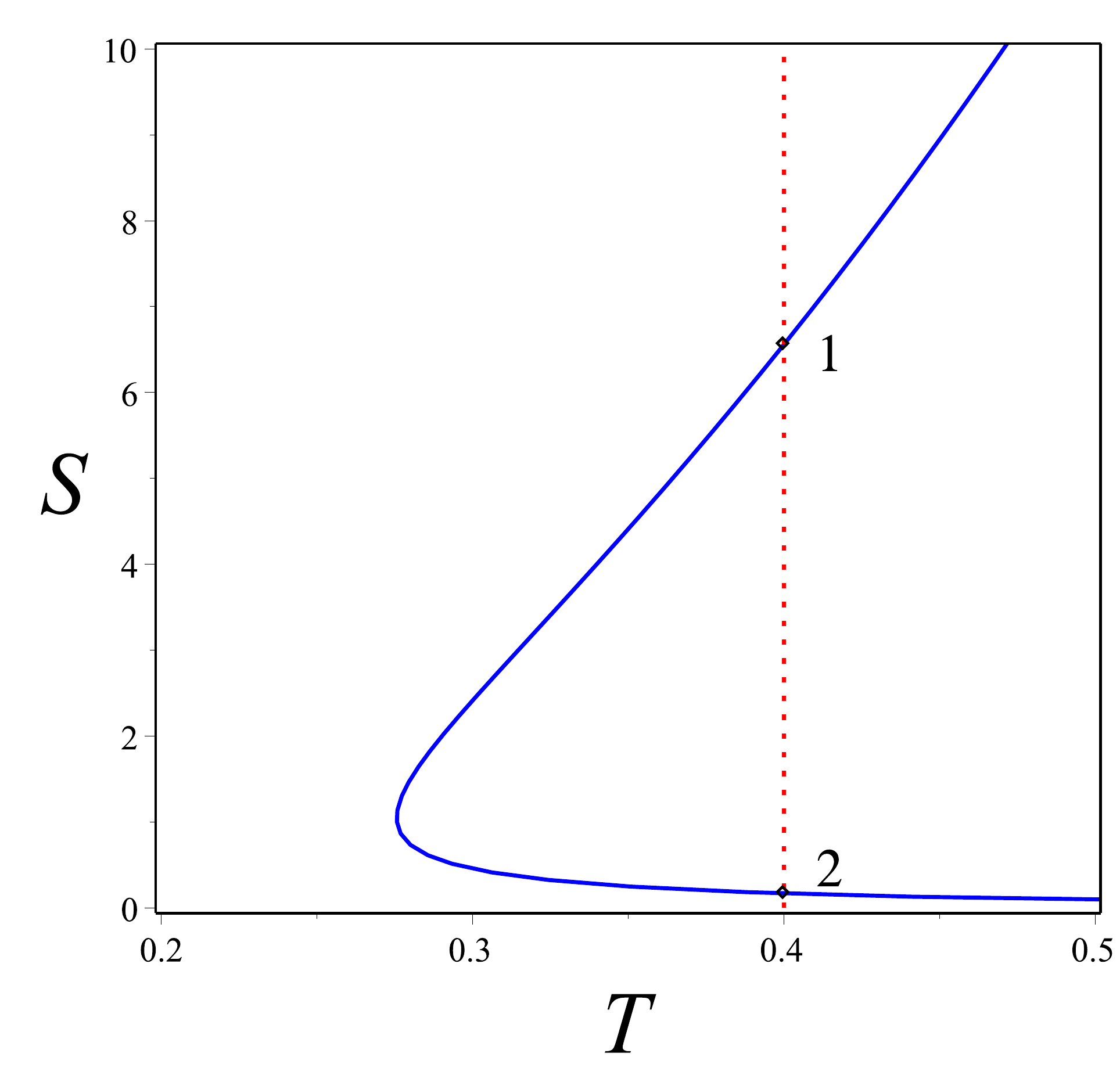} \quad
	\includegraphics[width=7 cm]{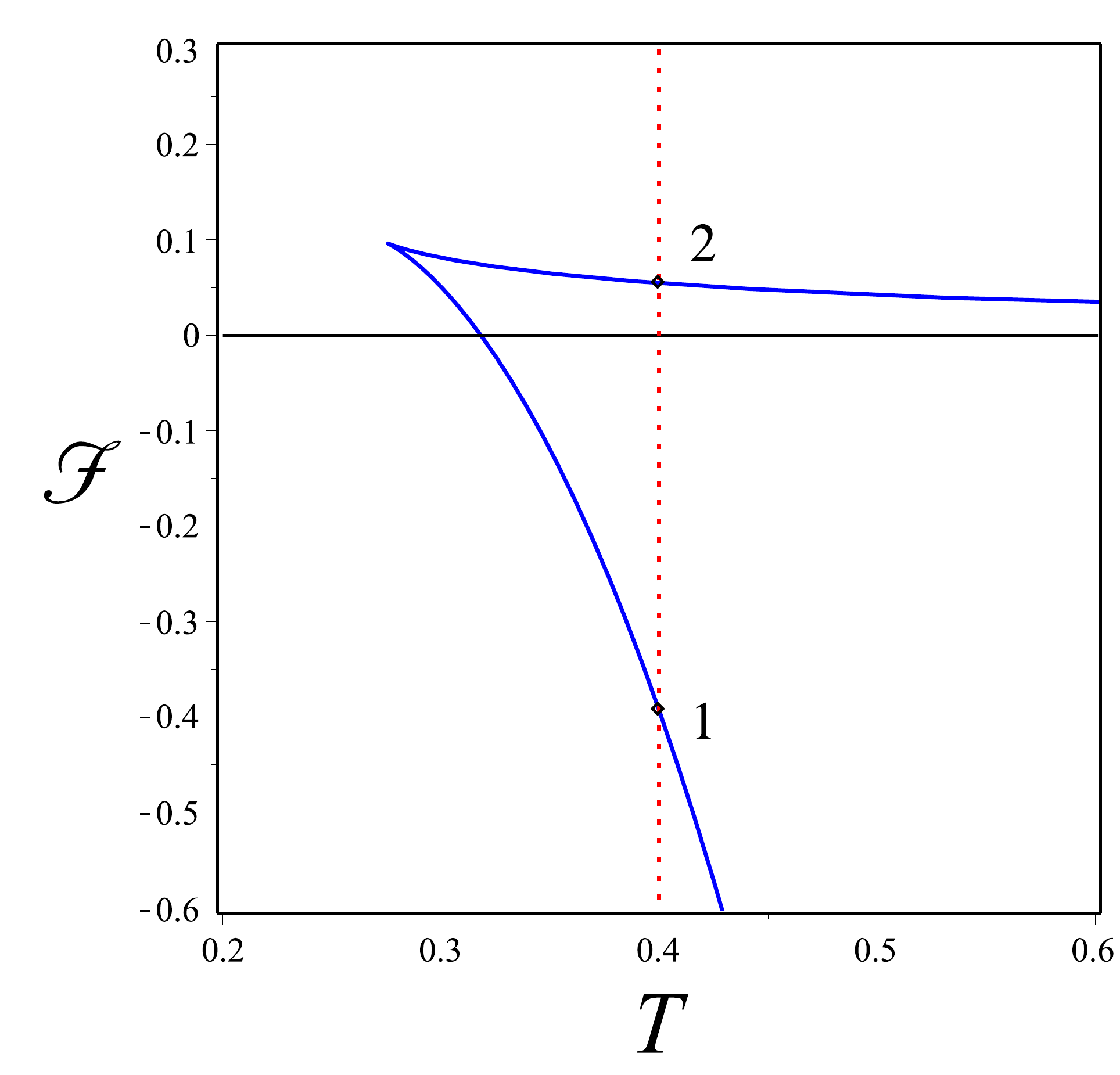}
	\caption{For Schwarzschild-AdS, the stable black hole is indicated as 1. In this case, the large black hole is the stable configurationm, contrary to the hairy case.}
	\label{stable2}
\end{figure}

Let us now turn our attention to Schwarzschild-AdS solution, when there also exist two black holes for the same temperature, so that we can compare with our results. In Fig. \ref{stable2}, we depict the free energy $\mathcal{F}=M-TS$ vs $T$ and $S$ vs $T$.
To identify the stable black hole note that, in the plot $S$ vs $T$, the positive slope correspond to  the configuration 1 at a fixed temperature. From the eq. (\ref{secderiv}), this one should be the one having negative concavity for the thermodynamic potential. Thus, in the second plot in Fig. \ref{stable2} it corresponds to the one with less free energy, indicated as the configuration 1 too. At first sight, it could be strange that in AdS the large black holes are stable and in flat space the small ones (comparing the black holes at the same temperature). However, there is a nice physical interpretation of this result. It is by now well known \cite{Hawking:1982dh} that AdS spacetime acts like a box and so, when the black hole horizon radius is comparable with the AdS radius $L$, they can be in stable thermal equilibrium. For hairy black holes in flat spacetime,  the self-interaction of the scalar field plays the role of the `box'. When the horizon radius is large, the dilaton potential takes smaller values (it vanishes at the boundary) and so the large black holes are not stable, while for small ones, the self interaction becomes relevant acting like a box allowing configurations in stable thermal equilibrium.

\begin{figure}[H]
	\centering
	\includegraphics[width=8 cm]{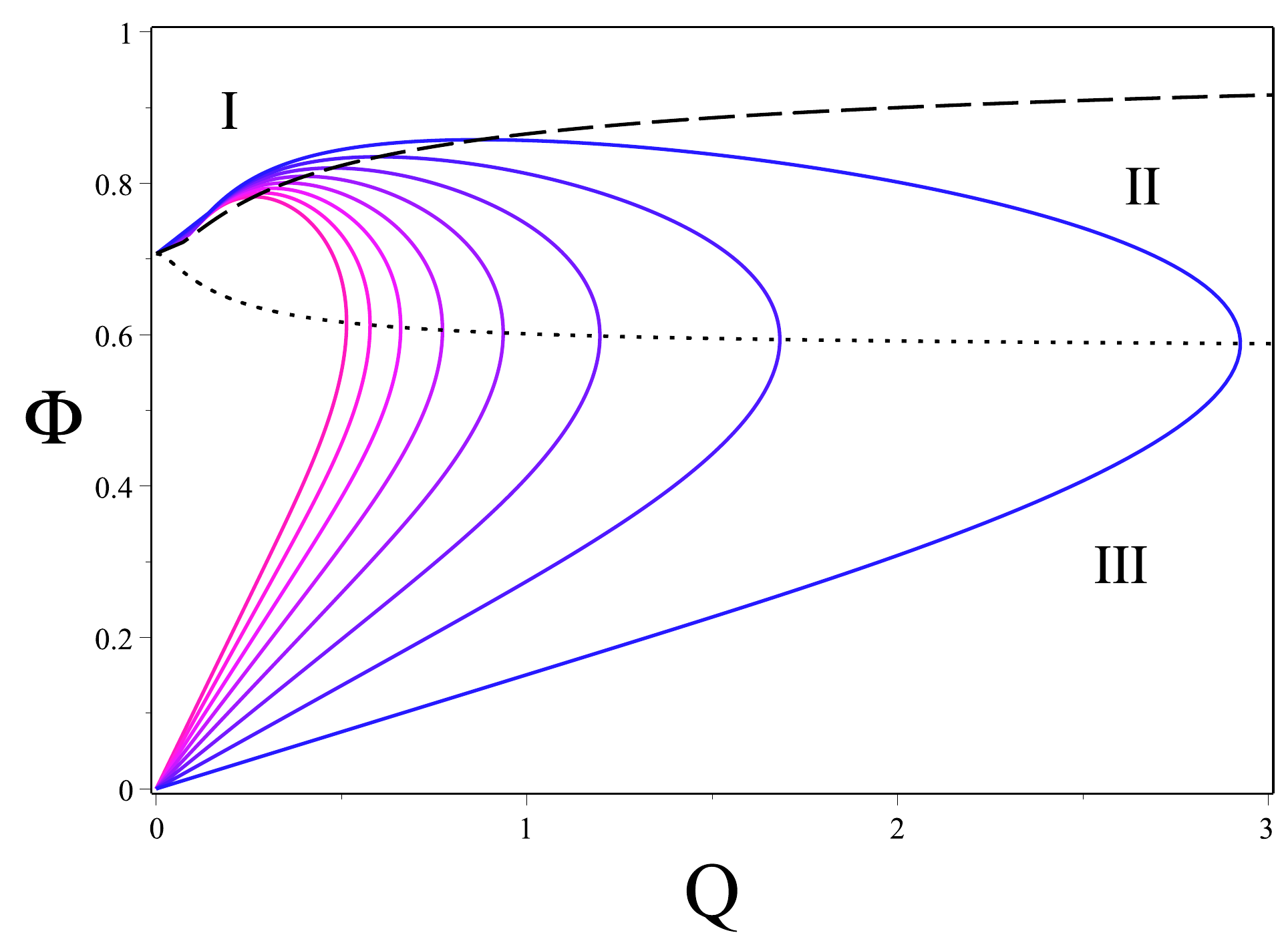}
	\caption{\textbf{Region I:} Here are plotted and isolated the stable configurations with $\epsilon_T>0$ and also $C_Q>0$. The relevant region, where these black holes exist, is bounded by the dashed black curve and extremal limit ($\Phi<1$ for the hairy case). \textbf{Region II:} This region for electrically unstable black holes, $\epsilon_T<0$, is bounded by the dashed curve and the dotted one (along which $\epsilon_T=0$). \textbf{Region III:} Electrically stable, $\epsilon_T>0$, but thermally unstable $C_Q<0$. In Reissner-Nordstr\"om, dotted curve appears as an horizontal line $\Phi=1/\sqrt{3}$ (this limit is recovered here only by large values of $Q$, namely $Q\gg \alpha^{-1/2}$). For this plot, we have fixed $\alpha=10$.}
	\label{StabilityRegions}
\end{figure}

Consider the equation of state, shown once again in Fig. \ref{StabilityRegions} and electric stability condition in canonical ensemble, $\epsilon_T>0$, now one can distinguish the relevant regions, as shown explicitly below.
The novel result, compared with RN black hole, is the existence of the Region I, where $\epsilon_T>0$.
To be more specific, let us compare the hairy black holes in the positive branch with the RN asymptotically flat black holes and, also, with the hairy solutions in the negative branch. Region II in all these cases is characterized by both $\epsilon_T<0$ and $C_Q>0$. However, only for the positive branch, the equation of state develops a new region (Region I) where the electric permittivity changes its sign, while $C_Q$ preserves the positivity, which is the region with stable black holes. 

\begin{figure}[H]
	\centering
	\includegraphics[width=7 cm]{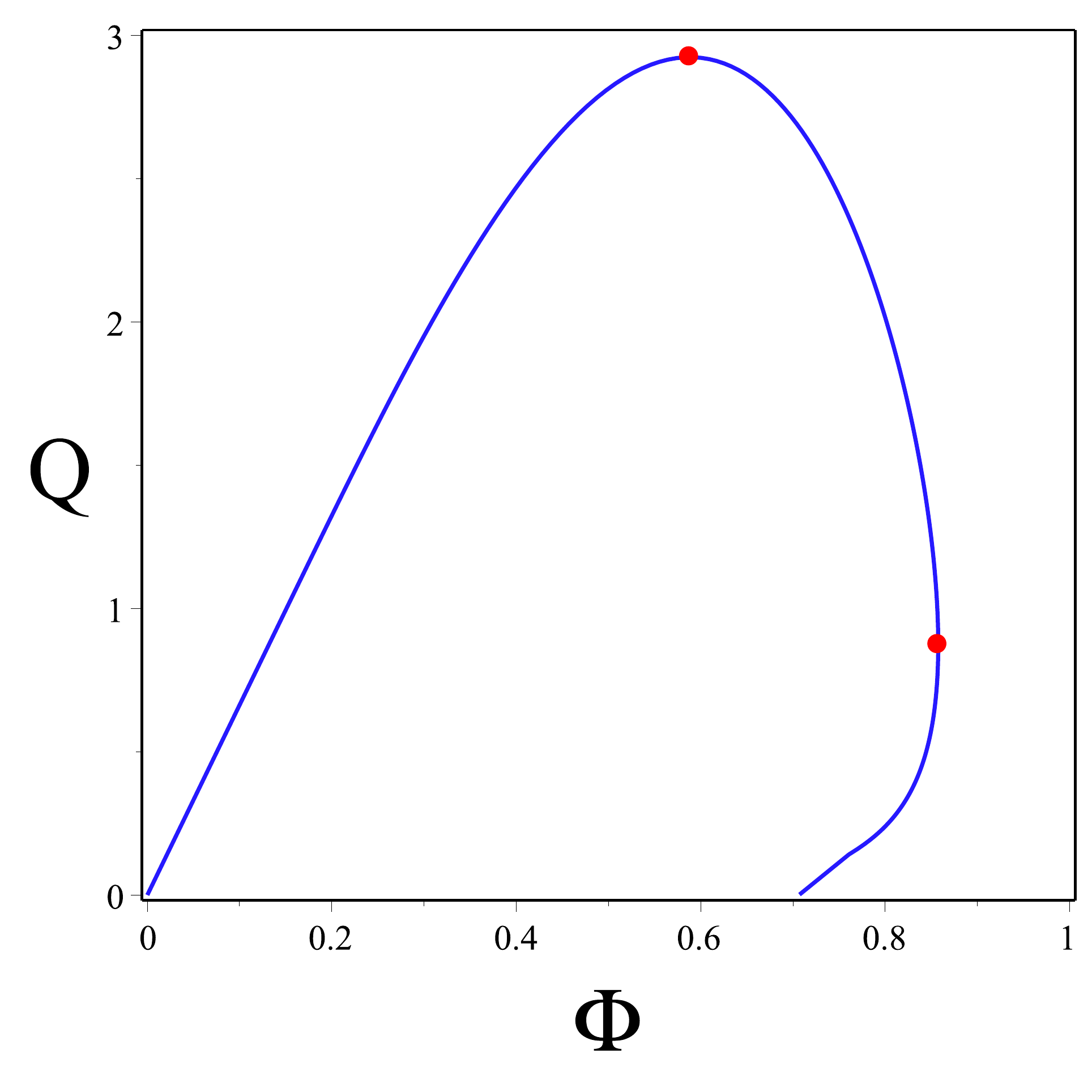}\quad
	\includegraphics[width=7 cm]{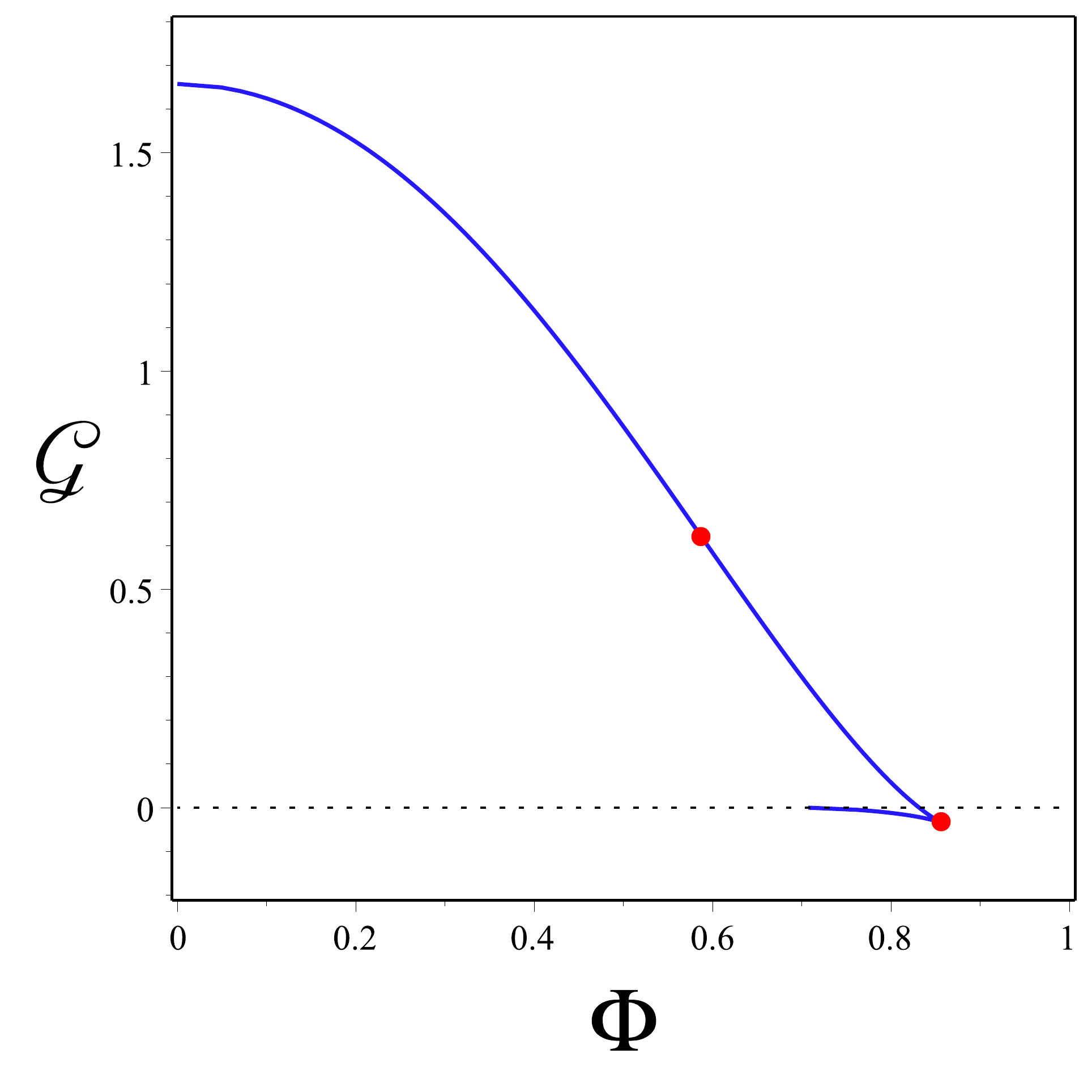}
	\caption{Plots $Q$ vs $\Phi$ (left hand side) and $\mathcal{G}$ vs $\Phi$ (right hand side) for the isotherm $T=0.012$. Red points indicates $\epsilon_T=0$ (the point at the left side in the plots) and $\epsilon_T\rightarrow\infty$.}
	\label{eSSPhi}
\end{figure}

As a consistency check that we have performed the computations correctly, let us discuss this result from a different prospective.  Since
\begin{equation}
dG=-SdT-Qd\Phi
\label{GCensm}
\end{equation}
by fixing then $T$, we have
$dG=-Qd\Phi$. Now, by integrating, one gets
\begin{equation}
\Delta G 
=-\int Q d\Phi
=-\int_{\Phi=\frac{1}{\sqrt{2}}}^{\Phi=\Phi_{m}} Q d\Phi
-\int_{\Phi=\Phi_m}^{\Phi=0} Q d\Phi
\end{equation}
where $\Phi_m$ is the maximum value that $\Phi$ reaches for a given fixed $T\neq 0$.
Therefore, Fig. \ref{StabilityRegions} provides information, up to a constant factor, of the free energy as a function of $\Phi$ and the comparison is made in Fig. \ref{eSSPhi}.
As mentioned before, electrical stability means negative concavity of the Gibbs free energy as a function of $\Phi$ that may be visualized in Fig. \ref{eSSPhi}. The red dots indicate $\epsilon_T=0$ and $\epsilon_T\rightarrow\infty$. Between $\Phi=0$ and the first red dot, the concavity $\mathcal{G}$ is negative (which means $\epsilon_T>0$). Between the red dots, $\epsilon_T<0$ and the concavity is positive. Finally, between the second red dot and $\Phi=1/\sqrt{2}$, the concavity turns positive again and $\epsilon_T>0$. Therefore, 
the plot of $\mathcal{G}$ vs $\Phi$ is consistent with the behaviour of the solution obtained from the equation of state.

The existence of thermodynamically stable asymptotically flat hairy black holes opens the possibility of investigating not only the phase diagram and possible phase transitions, but also to check the classical  stability (see, e.g., \cite{Gross:1982cv, Prestidge:1999uq, Gregory:2001bd, Hertog:2004bb, Gubser:2000ec, Gubser:2000mm, Reall:2001ag}) in this new context. 

\section*{Acknowledgments}
This work has been funded by the Fondecyt Regular Grant 1161418. The work of DC is supported by Fondecyt Postdoc Grant 3180185. RR was supported by the national Ph.D. scholarship Conicyt 21140024.

\newpage

\begin{appendix}
	
\section{Hairy black holes with $\alpha=0$}
\label{A}
	
	In this section, we briefly discuss the static solution of Einstein-Maxwell-dilaton theories with vanishing dilaton potential, for both values of the coupling coefficients, $\gamma=1$ and $\gamma=\sqrt{3}$. They are obtained by taking the limit $\alpha=0$ in theories described in section (\ref{sec:sols}) for which only the positive branch supports black hole configurations.
	
	\subsection{$\gamma=1$}
	
	From the horizon eq. (\ref{1gamma}) with $\alpha=0$, $x_+$ can be isolated and the thermodynamic quantities can be  written as follows:
	\begin{equation}
	M=\frac{1}{8\pi T}, \qquad
	S=\frac{1-32\pi^2Q^2T^2}{16\pi T^2},\qquad
	\Phi=4\pi QT
	\end{equation}
	They satisfy the first law $dM=TdS+\Phi dQ$ and the third equation is, in fact, the equation of state from which we can check that $\epsilon_T>0$.
	
	\subsubsection*{Grand Canonical}
	
	The electric permittivity at fixed entropy can be analytically obtained,
	\begin{equation}
	\epsilon_S=\frac{1}{4\pi T\(1-2\Phi^2\)}
	\end{equation}
	and it is positive for $1-2\Phi^2>0$. The free energy and temperature can be expressed as 
	\begin{equation}
	\mathcal{G}(T,\Phi)=\frac{1-2\Phi^2}{16\pi T^2},
	\qquad
	T=\sqrt{\frac{1-2\Phi^2}{16\pi S}}
	\end{equation}
	and we observed that $\Phi$ is restricted as $\Phi\leq 1/\sqrt{2}$ for the existence of regular solutions. This, in turn, implies that $\epsilon_S>0$. However, the heat capacity 
	\begin{equation}
	C_\Phi=-\frac{(1-2\Phi^2)}{8\pi T^2}
	\end{equation}
	is always negative that implies that there are no thermodynamically stable configurations in grand canonical ensemble.
	
	\subsubsection*{Canonical Ensemble}
	
	The electric permittivity at fixed $T$ can be directly obtained and it is positive,
	\begin{equation}
	\epsilon_T=\frac{1}{4\pi T}
	\end{equation}
	On the other hand, the thermodynamic potential is
	\begin{equation}
	\mathcal{F}(T,Q)=\frac{1+32\pi^2Q^2T^2}{16\pi T^2}
	\end{equation}
	and so the heat capacity is always negative,
	\begin{equation}
	C_Q=-\frac{1}{8\pi T^2}
	\end{equation}
	indicating the thermodynamic instability for all configurations.
	
	Therefore, there are no stable configuration in theory $\gamma=1$ when the self-interaction of the scalar field is turned off. 
	
	\subsection{$\gamma=\sqrt{3}$}
	
	In this theory, it is also straightforward to write down in a simple manner the thermodynamic quantities, by eliminating $x_+$ from the horizon eq. (\ref{3gamma}),
	\begin{align}
	S&={\frac {2\pi \sqrt {2} \left( {M}^{2}+M\sqrt {{M}^{2}+2{Q}^{2}}-{Q}^{2} \right) ^{3/2}}
		{M+\sqrt {{M}^{2}+2\,{Q}^{2}}}}
	,\quad
	T={\frac {\sqrt {2}}{8\pi \sqrt {{M}^{2}+M\sqrt {{M}^{2}+2{Q}^{2}}-{Q}^{2}}}}
	,\notag \\
	\Phi&={\frac {Q}{M+\sqrt {{M}^{2}+2{Q}^{2}}}}
	\label{quant4}
	\end{align}
	which satisfy the first law. In this case, the chemical potential is also restricted as $0<\Phi<1/\sqrt{2}$. This can be seen by solving $Q$ from $\Phi=\Phi(Q,M)$ in the third equation in (\ref{quant4}),
	\begin{equation}
	\label{charg3}
	Q=\frac {2M\Phi}{1-2\Phi^2}
	\end{equation}
	The equation of state can be obtained by using (\ref{charg3}) to eliminate $M$ in the equation for the temperature,
	\begin{equation}
	4\,\sqrt {-4\,{\Phi}^{2}+1}Q\pi \,T-\Phi=0
	\end{equation}
	One then gets that the electric permittivity at fixed temperature is
	\begin{equation}
	\epsilon_T=\frac{Q}{\Phi\(1-4\Phi^2 \)}
	\end{equation}
	that is positive. By inserting $M$ from (\ref{charg3}) into the expression for entropy, we can obtained the electric permittivity at fixed $S$,
	\begin{equation}
	S=\frac{\(1-4\Phi^2\)^{3/2}}{\Phi^2}\,\pi Q^2,
	\qquad
	\epsilon_S=\frac{\(1+2\Phi^2\)Q}{1-4\Phi^2}
	\end{equation}
	which is also positive. Let us now complete our analysis by computing the heat capacities in both ensembles.
	
	\subsubsection*{Grand Canonical}
	
	The thermodynamic potential is
	\begin{equation}
	\mathcal{G}(T,\Phi)=\frac{\sqrt{1-4\Phi^2}}{16\pi T}
	\end{equation}
	and then, heat capacity is
	\begin{equation}
	C_\Phi=-\frac{\sqrt{1-4\Phi^2}}{8\pi T^2}
	\end{equation}
	indicating that there is no stable configuration in grand canonical ensemble.
	
	\subsubsection*{Canonical ensemble}
	
	The thermodynamic potential is
	\begin{equation}
	\mathcal{F}(T,Q)=\frac{\sqrt{1+64\pi^2Q^2T^2}}
	{16\pi T}
	\end{equation}
	from which the following heat capacity at fixed $Q$ can be obtained:
	\begin{equation}
	C_Q=-\frac{1+96\pi^2Q^2T^2}
	{8\pi T^2\(1+64\pi^2Q^2T^2\)^{3/2}}
	\end{equation}
	This is always negative and, as expected, there also are no stable configurations in canonical ensemble.
	
	We conclude that the existence of the stable equilibrium configurations is related to the scalar field has self-interaction and when $\alpha=0$, the thermodynamically stable hairy black holes do not exist. Even if the coupling of a scalar field to the electromagnetic field improves electrical stability of black holes, the heat capacity is negative and they are thermally unstable.
	
	\section{Hairy black holes, $\gamma=\sqrt 3$, with $\alpha\neq 0$}
	\label{B}
	
	In this appendix, for completeness, we investigate the thermodynamic local stability of the solution $\gamma=\sqrt{3}$ presented in section (\ref{sec:sols}). Since the procedure is completely equivalent to that presented in section (\ref{sec:therm2}), here we shall only write down the  expressions of thermodynamic quantities and present the relevant results.
	
	\subsection{Grand Canonical (negative branch)}
	\label{gcnb2}
	
\begin{figure}[H]
		\centering
		\includegraphics[width=7.5 cm]{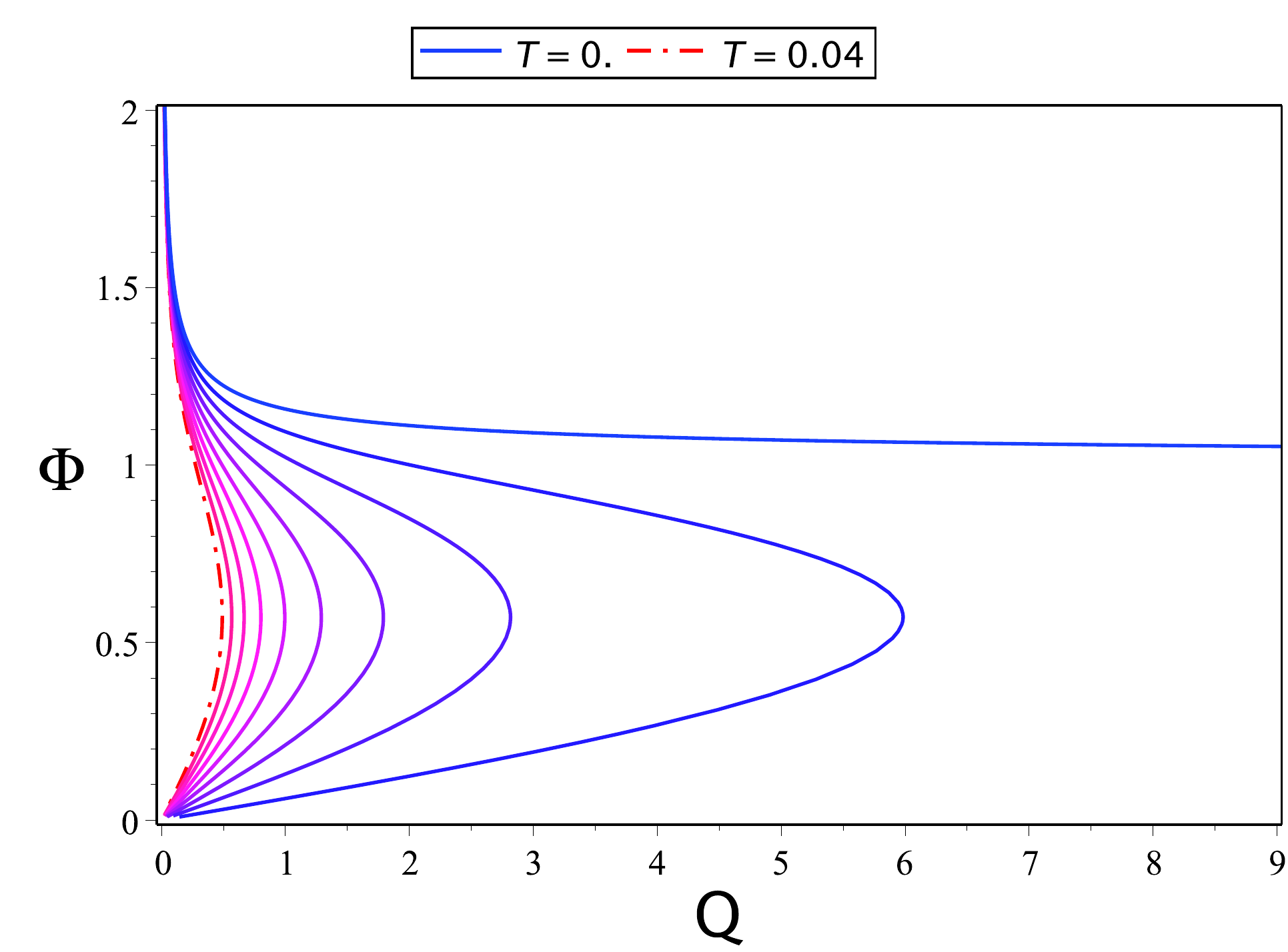}
		\includegraphics[width=7.5 cm]{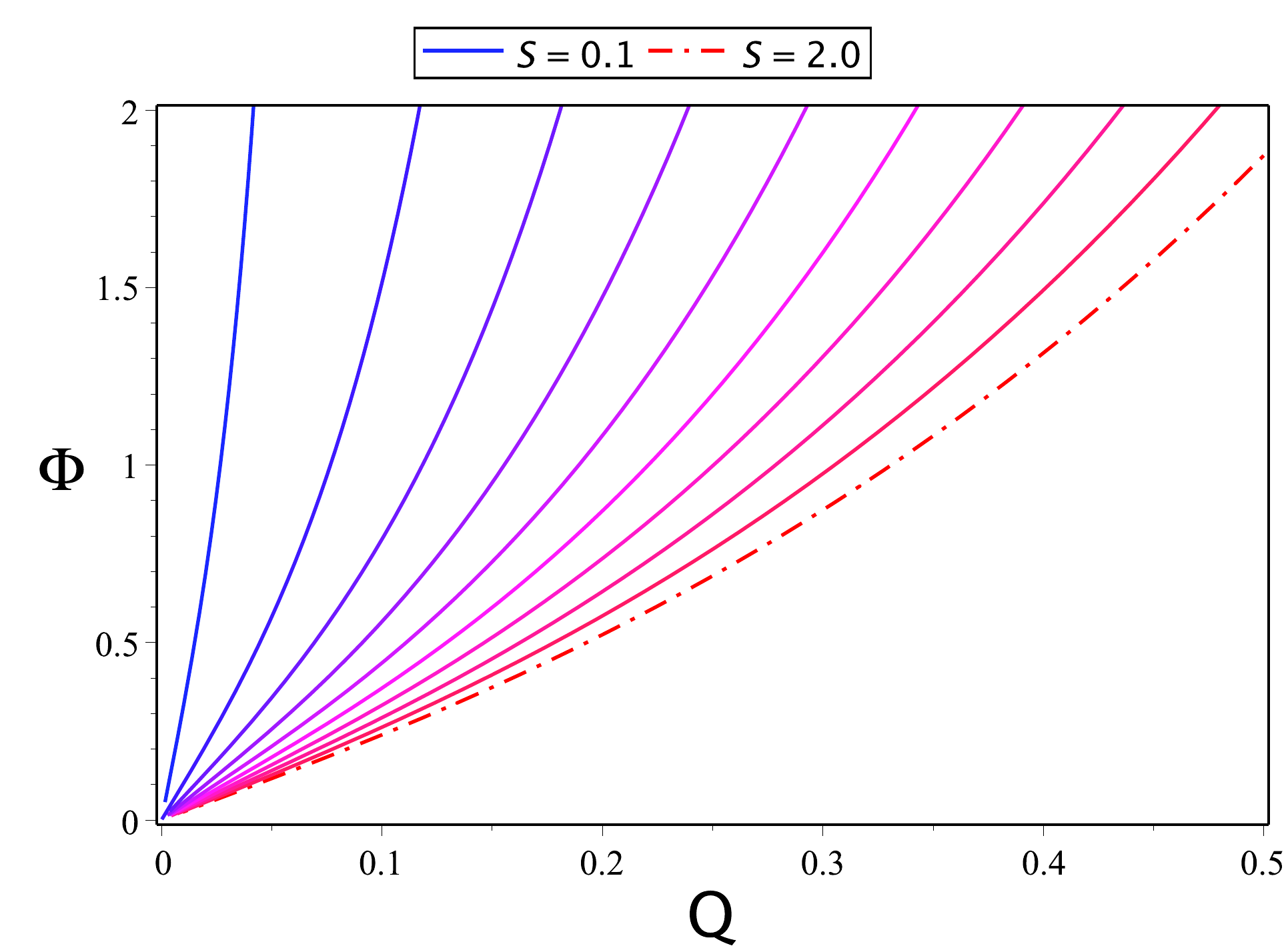}
		\caption{Left Hand Side: Equation of state, isotherms curves $Q$---$\Phi$, $\gamma=\sqrt{3}$ and $\alpha=10$, in negative branch. Right Hand Side: isentropic curves $Q$---$\Phi$.}
		\label{state3}
	\end{figure}

	The dilaton potential for which we have obtained the exact solutions for $\gamma=\sqrt 3$ is (\ref{dilaton2}). The equation of state can be studied parametrically by using the dependence of the radius of horizon and temperature, $Q=Q(x_+,T)$ and $\Phi=\Phi(x_+,T)$. It is useful to also have $Q=Q(x_+,S)$ and $\Phi=\Phi(x_+,S)$ so that we can get $\epsilon_S$. In Fig. \ref{state3}, they are graphically represented and have a similar behaviour as the one of $\gamma=1$.

	Next, we use $q=2x_+^2\Phi/(1-x_+^2)$ from eq. (\ref{quant2}) into the horizon equation $f(x_+)=0$ in order to get the positive root of $\eta=\eta(x_+,\Phi)$. Once done, we are able to express all the thermodynamic quantities as function of $x_+$ and $\Phi$. The thermodynamic potential
	\begin{equation}
	\mathcal{G}(x_+,\Phi)=-\frac{2\alpha}{3\eta^3}
	+\frac{2\Phi^2x_+^4}{\eta\(x_+^2-1\)^2}
	+\frac{1+x_+^2}{2\eta(1-x_+^2)}
	\end{equation}
	and heat capacity $C_\Phi$ are plotted below, in Fig. \ref{resp5},
	\begin{figure}[H]
		\centering
		\includegraphics[width=5 cm]{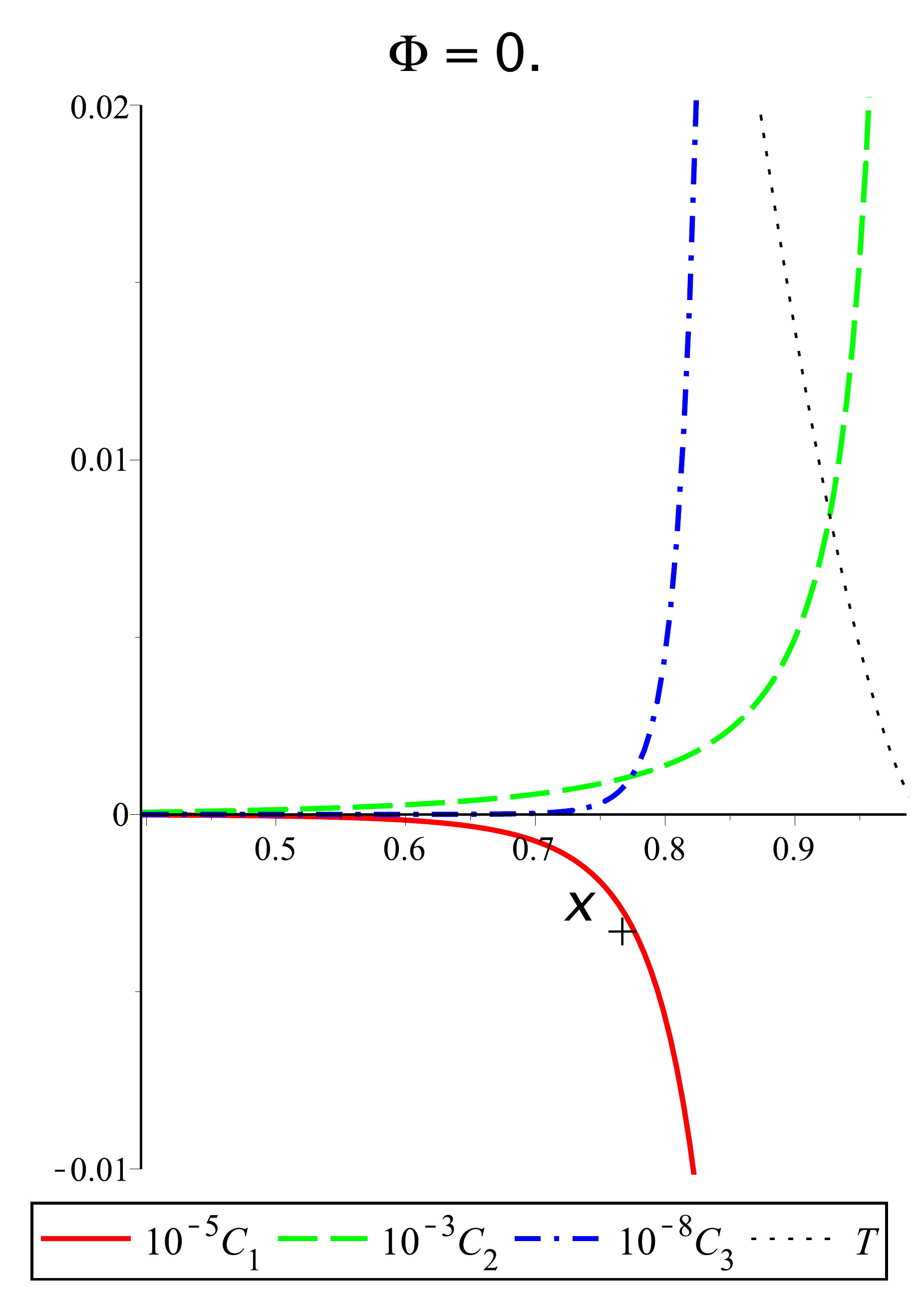}
		\includegraphics[width=5 cm]{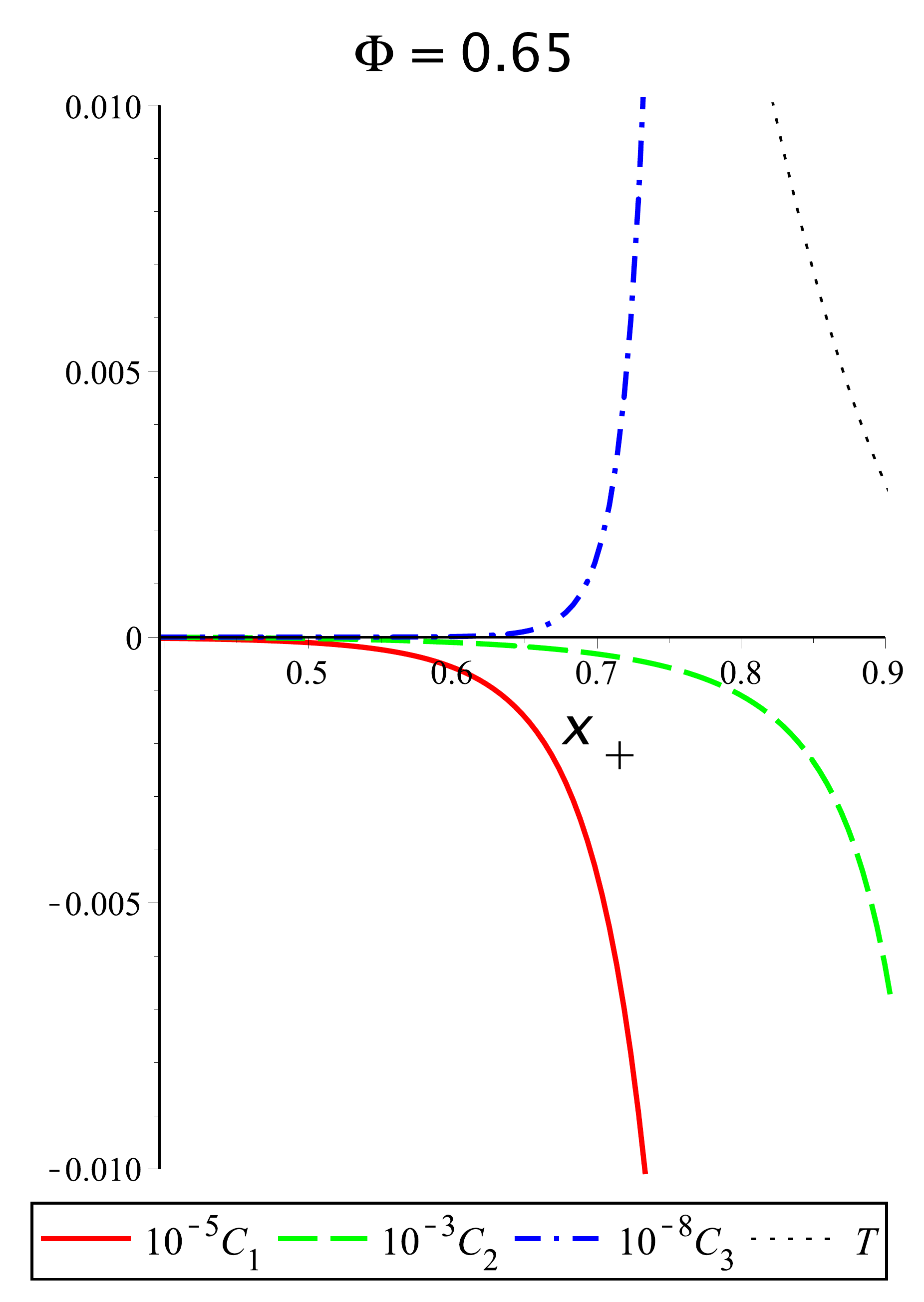}
		\includegraphics[width=5 cm]{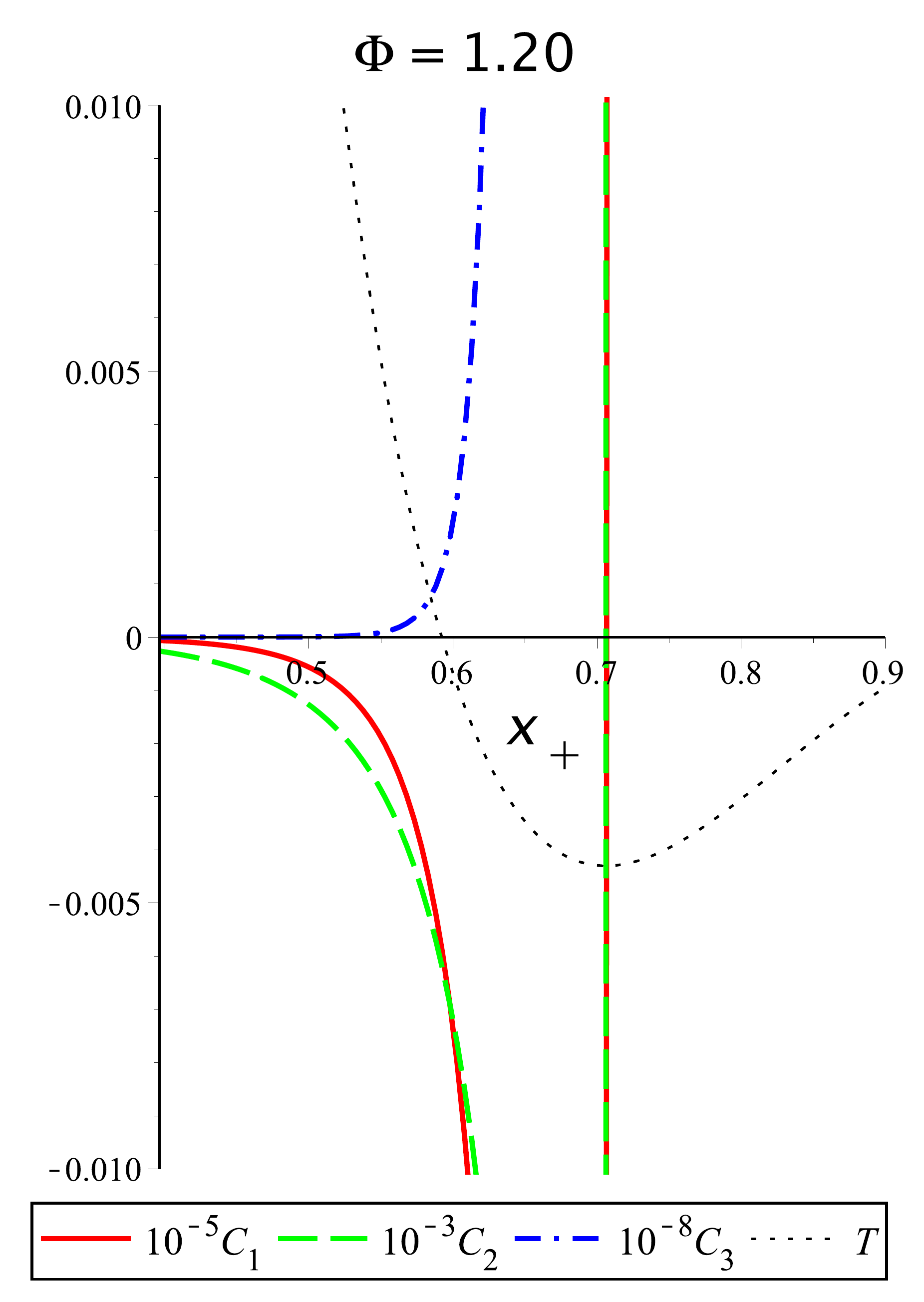}
		\caption{Response functions in terms of derivatives of $\mathcal{G}$, for negative branch in grand canonical ensemble, $\gamma=\sqrt{3}$. It was considered $\alpha=10$. Black dotted line represents $T$, the red curve represents $10^{-5}C_1$; the green one represents $10^{-2}C_2$ and blue curve $10^{-6}C_3$.}
		\label{resp5}
	\end{figure}

	Since there is no region in parameter space where both $\epsilon_S$ and $C_\Phi$ are simultaneously positive, the thermodynamically stable hairy  black holes do no exist in the negative branch. The response functions have the same skematic behaviour than the RN black hole, they have opposite sign for every configuration.
	
	\subsection{Grand Canonical (positive branch)}
	
\begin{figure}[H]
		\centering
		\includegraphics[width=7 cm]{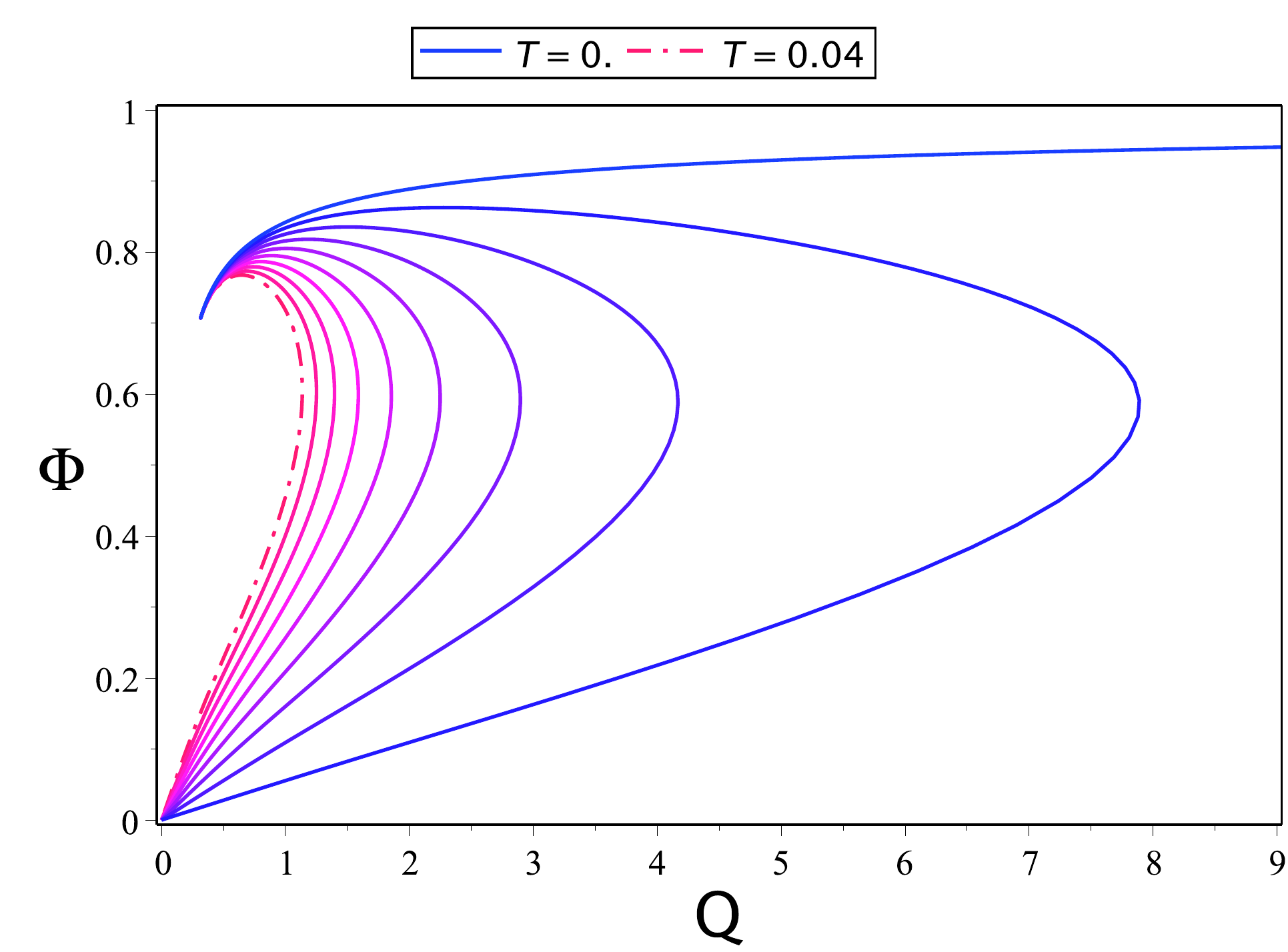}\quad
		\includegraphics[width=7 cm]{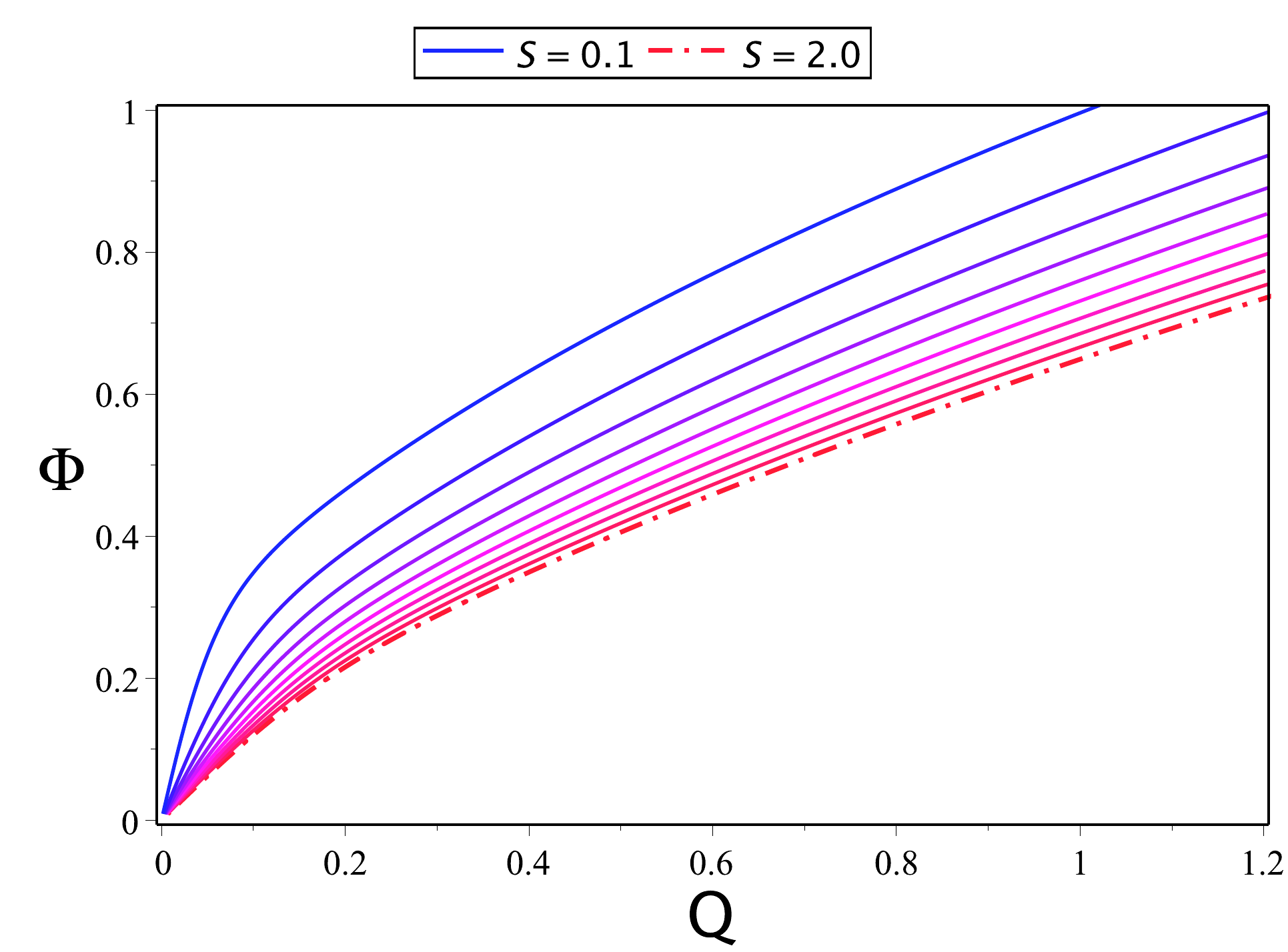}
		\caption{Left Hand Side: Equation of state in the positive branch, $\gamma=\sqrt{3}$ and $\alpha=10$. Right Hand Side: $\Phi$ vs $Q$ at entropy fixed.}
		\label{state4}
	\end{figure}

\begin{figure}[H]
		\centering
		\includegraphics[width=5 cm]{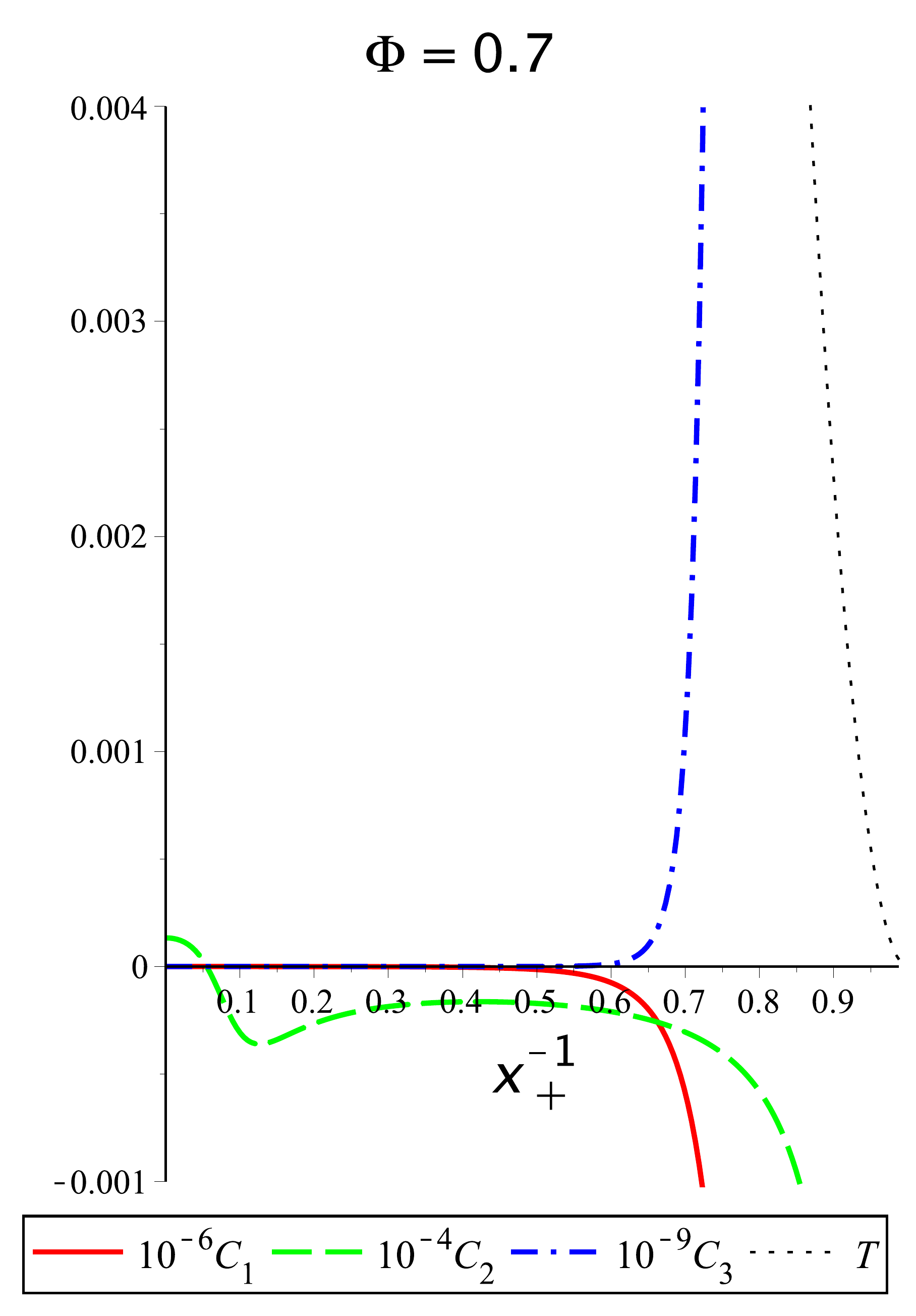}
		\includegraphics[width=5 cm]{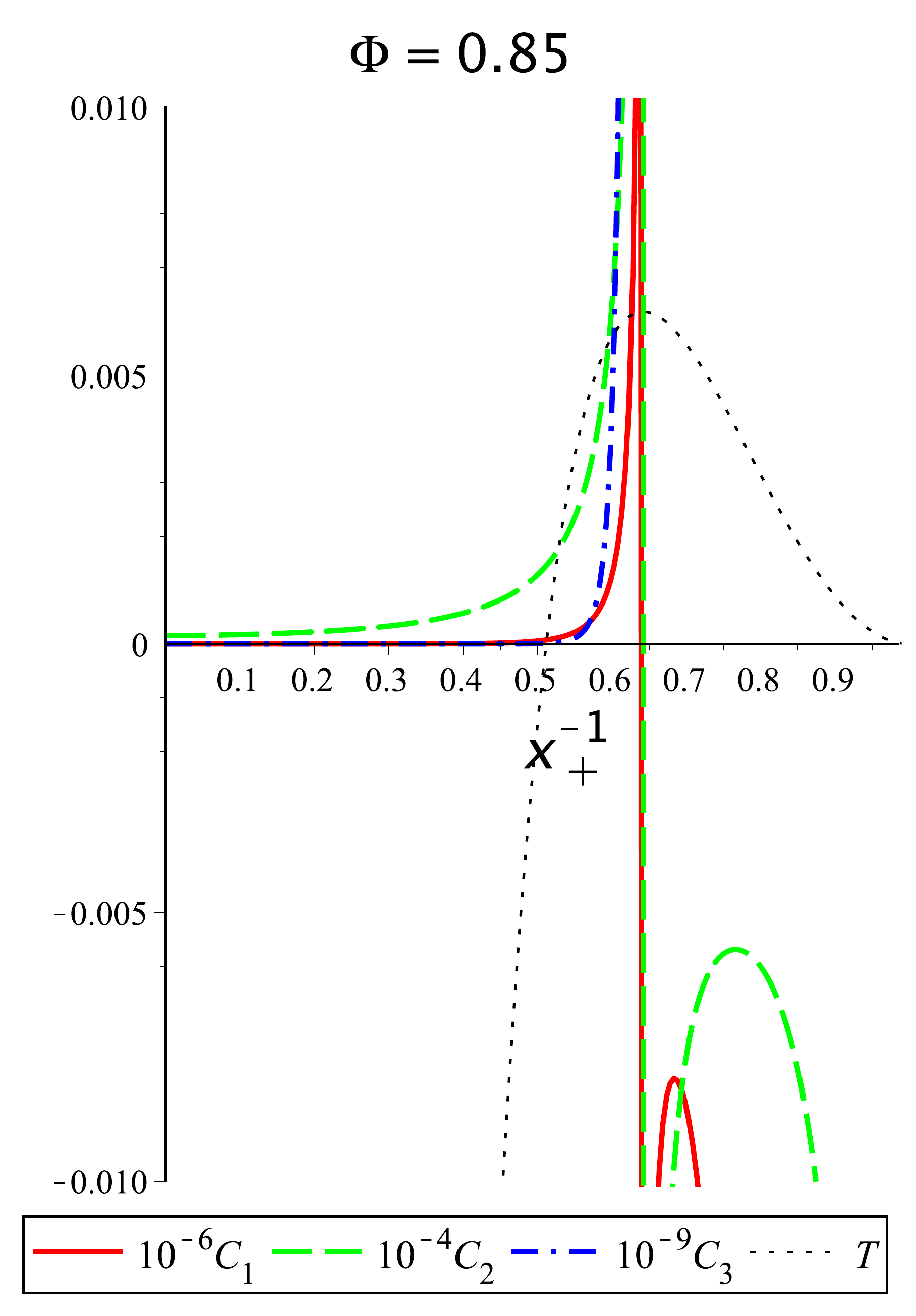}
		\includegraphics[width=5 cm]{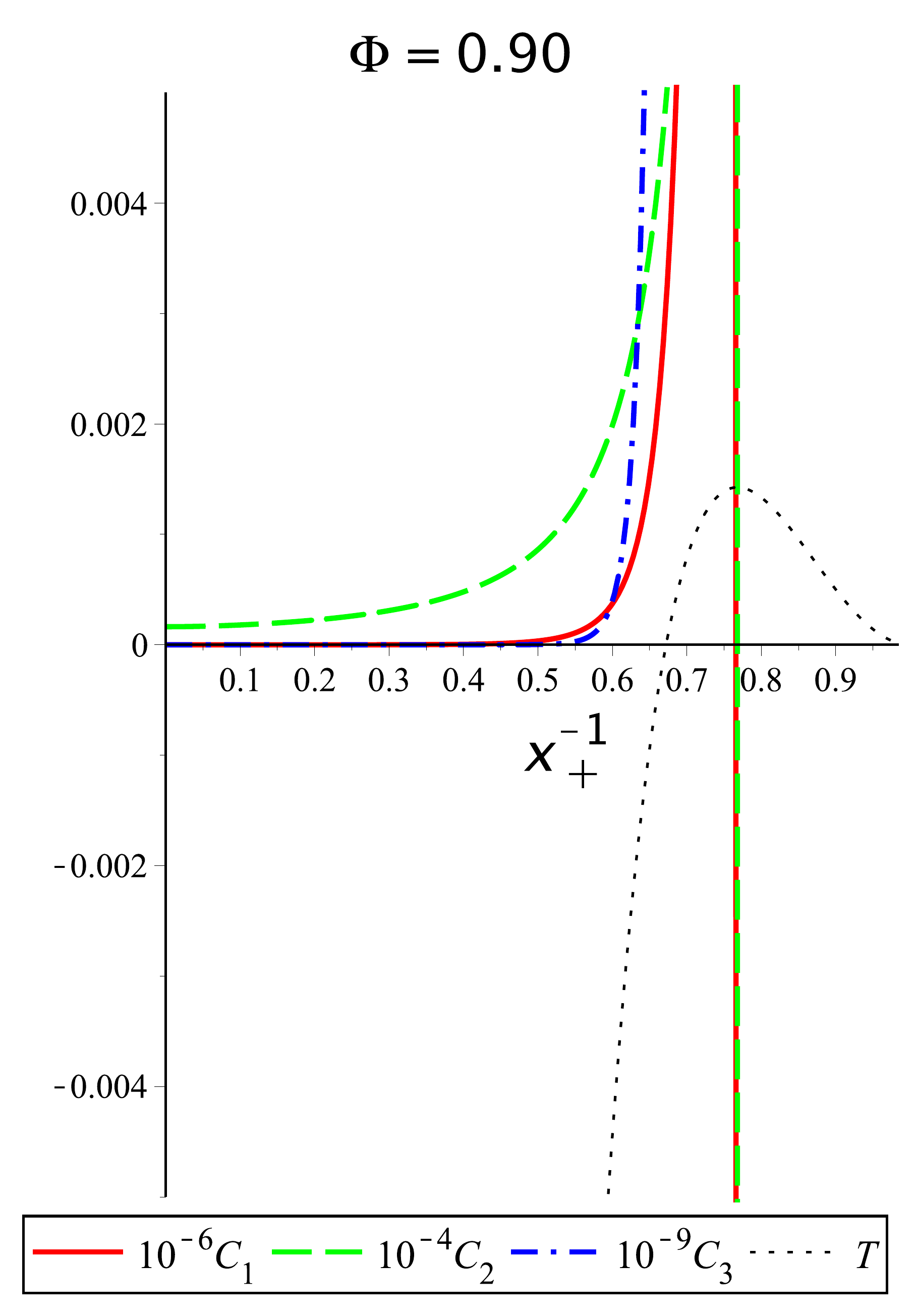}
		\caption{Response functions in terms of derivatives of $\mathcal{G}$, for positive branch in grand canonical ensemble, $\gamma=\sqrt{3}$ and $\alpha=10$.}
		\label{resp6}
	\end{figure}

	The equation of state and also  $\Phi$---$Q$ at $S$ fixed are represented graphically in Fig. \ref{state4}. The relevant response functions are plotted in Fig. \ref{resp6}.
	
	The thermodynamic potential can be expressed as
	\begin{equation}
	\mathcal{G}(x_+,\Phi)=
	\frac{2\alpha}{3\eta^3}
	-\frac{2x_+^4\Phi^4}{\(x_+^2-1\)^2\eta}
	+\frac{x_+^2+1}{2\eta(x_+^2-1)}
	\end{equation}
	where, as usual, $\eta=\eta(x_+,\Phi)$ is obtained from horizon equation.
	
	One interesting aspect of this theory in positive branch is that each isotherm starts at $Q=0$ and $\Phi=0$ and stops at $Q= 1/\sqrt{\alpha}$ and $\Phi=1/\sqrt{2}$ (this can be proven in the same way as in the case $\gamma=1$)
	
\begin{figure}[H]
		\centering
		\includegraphics[width=13 cm]{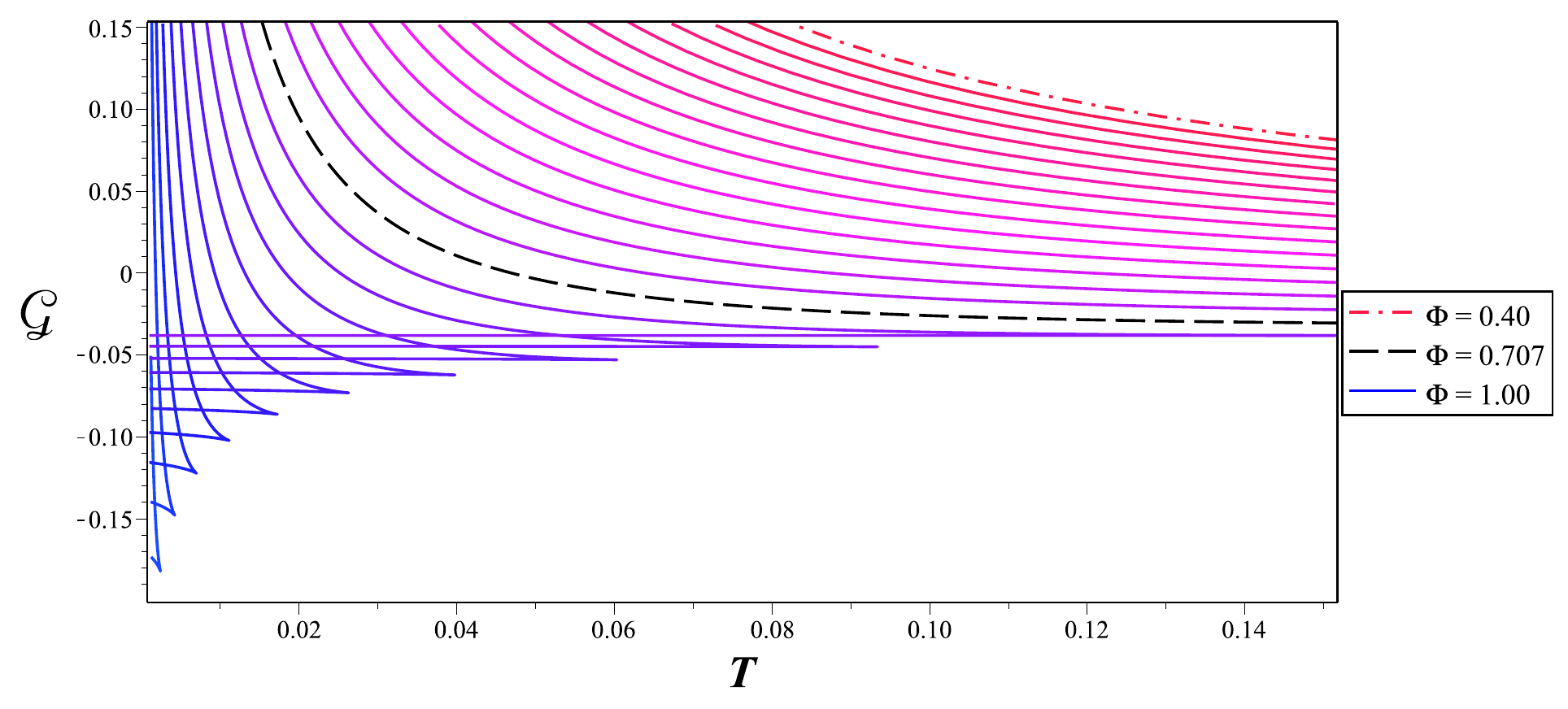}
		\caption{Free energy $\mathcal{G}$ vs $T$, for $\gamma=\sqrt{3}$ and $\alpha=10$, in positive branch.}
		\label{GT2}
	\end{figure}

	Just as in the case $\gamma=1$, there exist thermodynamically stable black holes only when 
	\begin{equation}
	\frac{1}{\sqrt{2}}<\Phi
	\end{equation}
	Also, these stable black holes have negative free energy, as can be seen in Fig. \ref{GT2}.
	
	
	\newpage
	\subsection{Canonical Ensemble (negative branch)}
	\label{cnb2}
	
	The thermodynamic potential can be written as
	\begin{equation}
	\mathcal{F}(x_+,Q)=
	-\frac{2\alpha}{3\eta^3}
	+\frac{\eta Q^2}{2x_{+}^2}
	+\frac{x_{+}^2+1}{2\eta(1-x_{+}^2)}
	\end{equation}
	from which we can get all the response functions.
	\begin{figure}[H]
		\centering
		\includegraphics[width=6 cm]{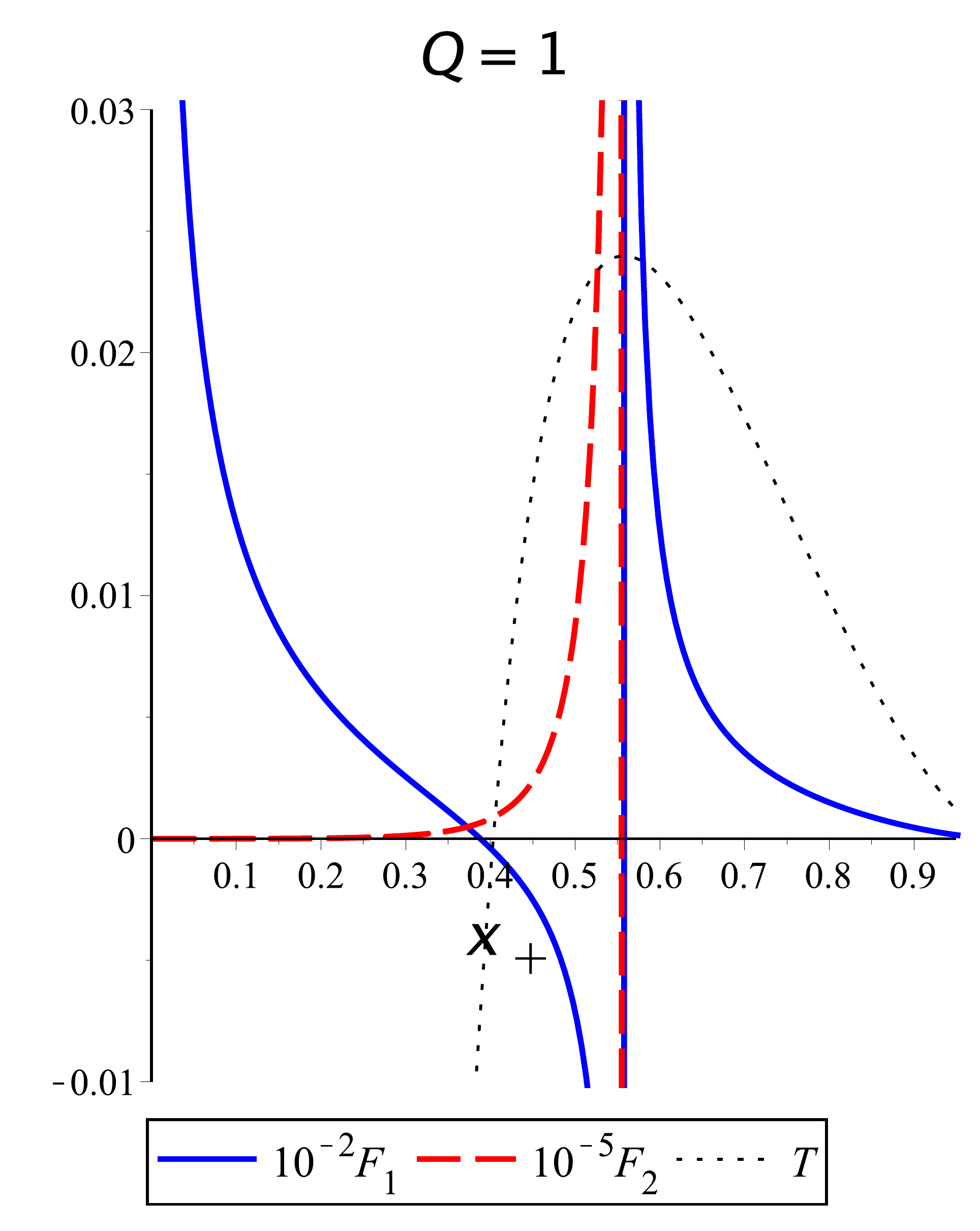}\qquad
		\includegraphics[width=6 cm]{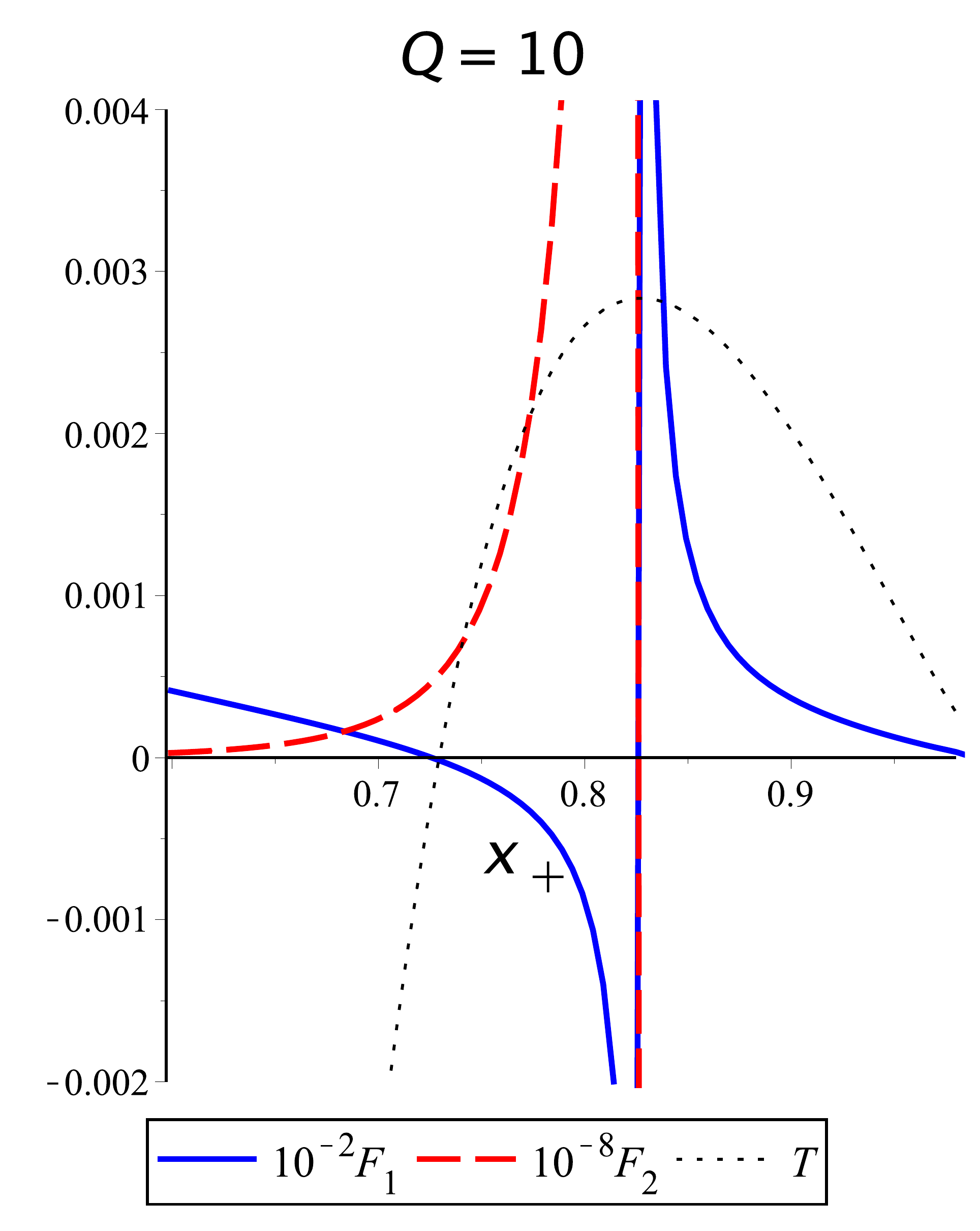}
		\caption{Second derivatives for negative branch in canonical ensemble, $\gamma=\sqrt{3}$ and $\alpha=10$. Black dotted line represents temperature, while $F_1:=\left(\pa^2\mathcal{F}/\pa{Q}^2\right)_T$ and $F_2:=\left(\pa^2\mathcal{F}/\pa{T}^2\right)_Q$.}
		\label{resp7}
	\end{figure}
	From Fig. \ref{resp7}, it can be observed that $C_Q$ and $\epsilon_T$ are not simultaneously postive definite in any physical region. In fact, the extremal black hole is electrically unstable $\left(\pa^2\mathcal{F}/\pa{Q}^2\right)_T<0$ and thus, between $T=0$ and $T_{max}$, for a given $Q$, the response functions have opposite signs. This result is expected because of the lack of thermodynamic stability in grand canonical ensemble.
	
	\subsection{Canonical Ensemble (positive branch)}
	\label{cpb2}
	
\begin{figure}[H]
		\centering
		\includegraphics[width=6.2 cm]{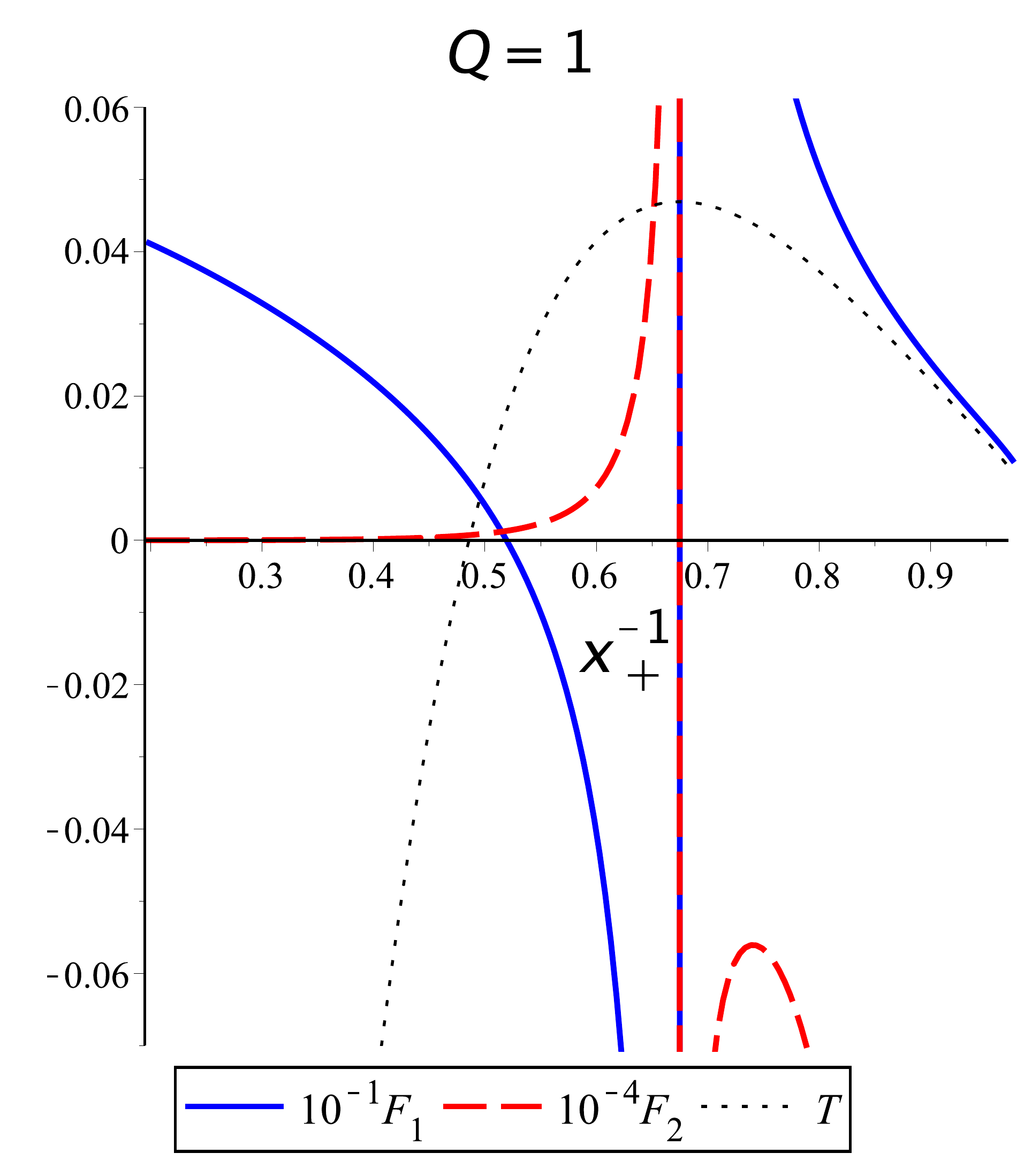}\qquad
		\includegraphics[width=6.2 cm]{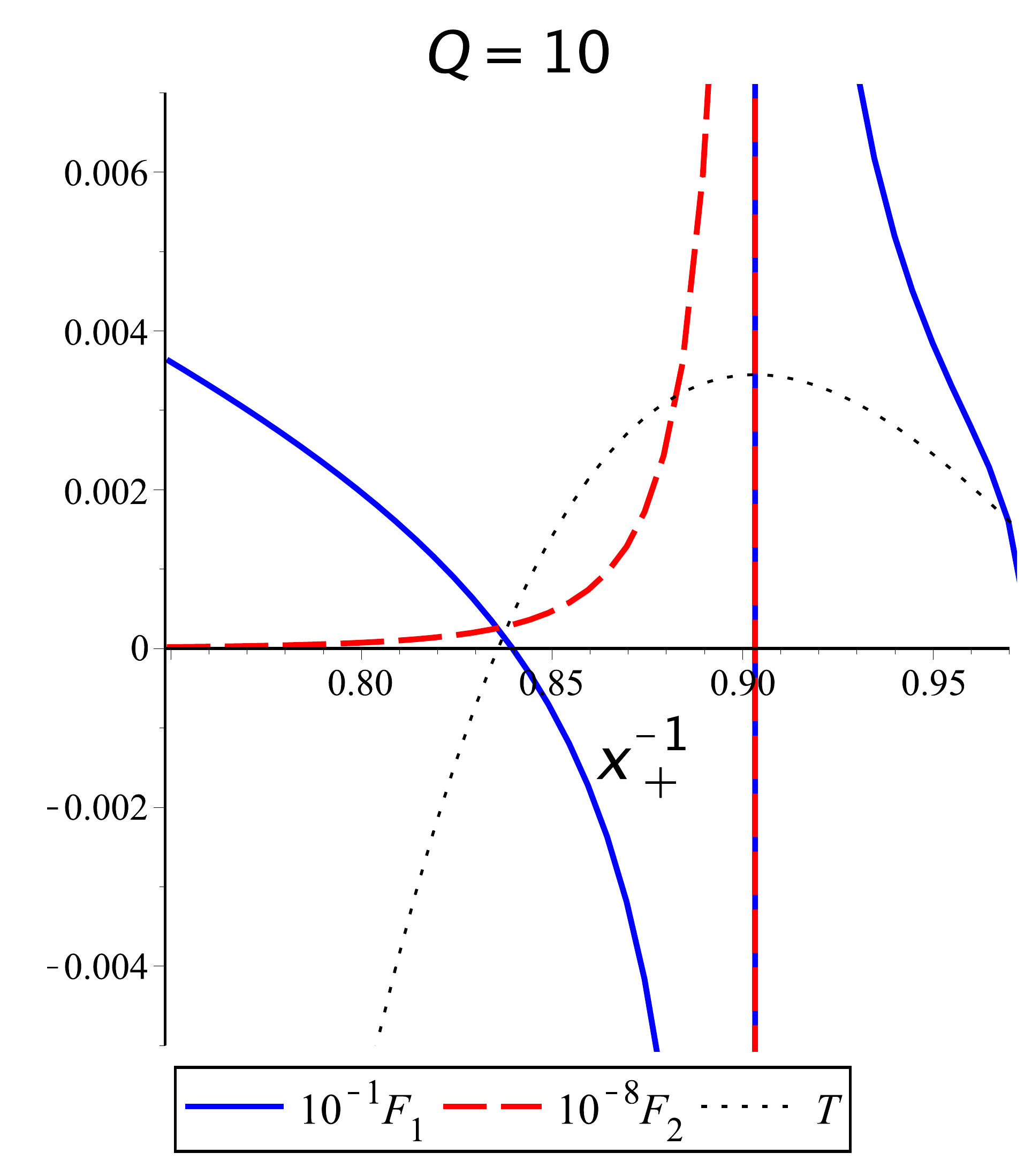}
		\caption{Second derivatives for positive branch in canonical ensemble, $\gamma=\sqrt{3}$ and $\alpha=10$. Black dotted line represents temperature, while $F_1:=\left(\pa^2\mathcal{F}/\pa{Q}^2\right)_{T}$ and $F_2:=\left(\pa^2\mathcal{F}/\pa{T}^2\right)_{Q}$.}
		\label{resp8}
	\end{figure}

	Finally, we show that there exist thermodynamically stable hairy black holes in the positive branch, which is consistent with our findings for the grand canonical ensemble. After a Legendre transform, the thermodynamic potential in canonical ensemble can be expressed as
	\begin{equation}
	\mathcal{F}(x_+,Q)=\frac{2\alpha}{3\eta^3}
	-\frac{\eta Q^2}{2x_+^2}
	+\frac{x_+^2+1}{2\eta\(x_+^2-1\)}
	\end{equation}
	where $\eta=\eta(x_+,Q)$ is obtained from horizon equation. The response function in terms of the second derivatives are graphically represented in Fig. \ref{resp8}. As can be seen, only black holes with $Q>1/\sqrt{\alpha}$ can be thermodynamically stable, since $\epsilon_T>0$ and $C_Q>0$.
	
	The main conclusion is that, even for the case $\gamma=\sqrt 3$, the self-interaction of the scalar field thermodynamically stabilizes the hairy black holes.
	
	\begin{figure}[H]
		\centering
		\includegraphics[width=13 cm]{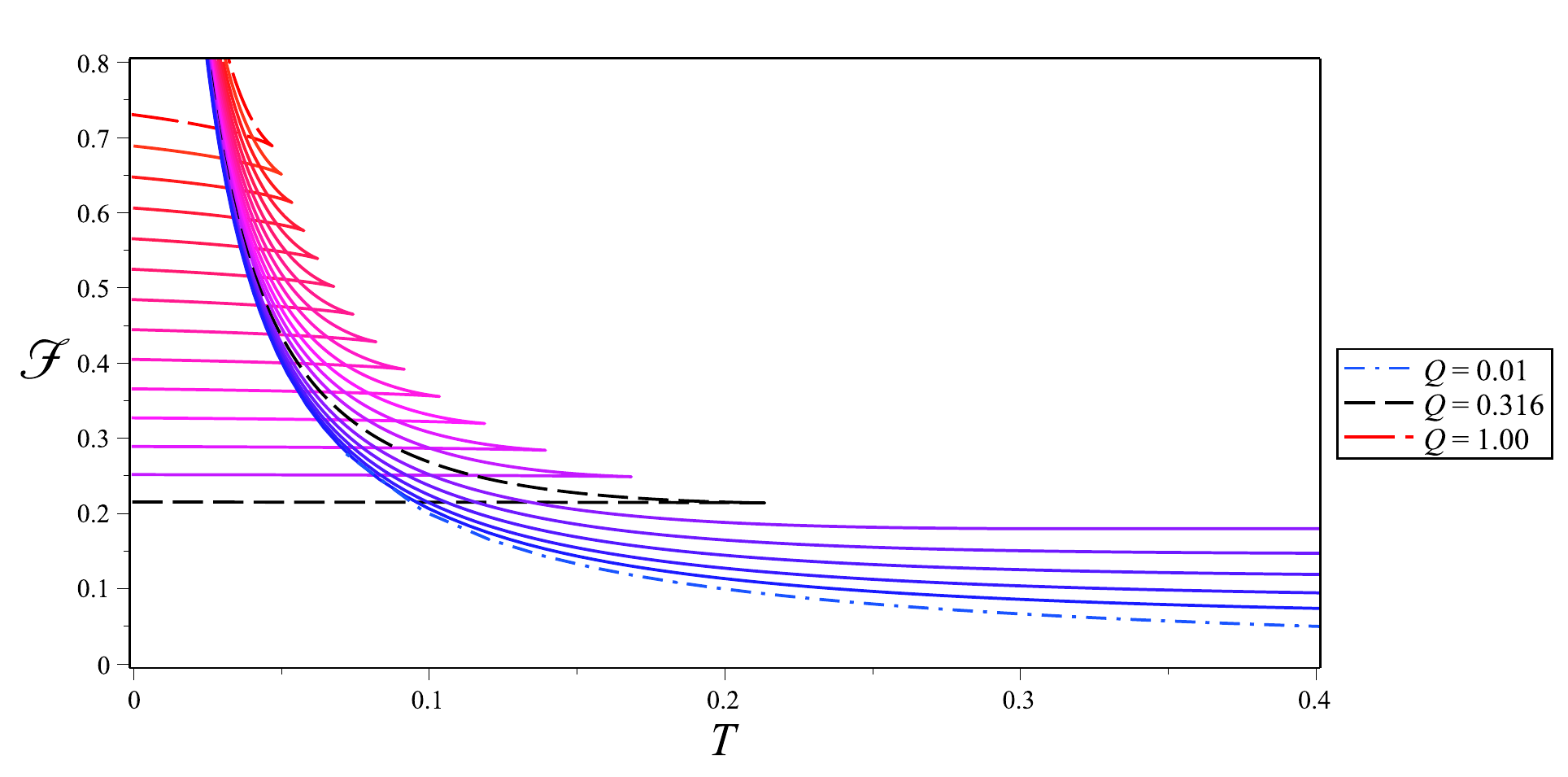}
		\caption{Thermodynamic potential $\mathcal{F}$ vs $T$, for $\gamma=\sqrt{3}$. The sector with negative concavity ($C_Q>0$) exists for $Q>1/\sqrt{\alpha}$. For the particular case considered here, $\alpha=10$, stable black holes appear for $Q\gtrsim  0.316$, as shown.}
		\label{FT2}
	\end{figure}
	
\end{appendix}

\newpage


\end{document}